\documentclass[a4paper,11pt]{article}
\usepackage{a4wide,amsmath,amssymb,bm,bbm}
\usepackage[makeroom]{cancel}
\usepackage[english]{babel}
\usepackage[utf8]{inputenc}

\usepackage{hyperref,enumerate}
\hypersetup{
    colorlinks,
    linkcolor={black},
    citecolor={black},
    urlcolor={black}
}

\DeclareMathOperator{\tr}{tr}
\makeatletter

\@addtoreset{equation}{section} \makeatother

\begin{document}
\setcounter{page}{0}

\hfill
\vspace{30pt}

\begin{center}
{\huge{\bf {Tree-level $R^4$ correction from $O(d,d)$:\\ NS-NS five-point terms}}}

\vspace{80pt}

Linus Wulff

\vspace{15pt}

\small {\it Department of Theoretical Physics and Astrophysics, Faculty of Science, Masaryk University\\ 611 37 Brno, Czech Republic}
\\
\vspace{12pt}
\texttt{wulff@physics.muni.cz}\\

\vspace{80pt}

{\bf Abstract}
\end{center}
\noindent
The tree-level string effective action reduced from $D$ to $D-d$ dimensions possesses a continuous $O(d,d)$ symmetry, closely related to T-duality. A necessary condition for a higher derivative correction to preserve this symmetry is that certain $O(d,d)$ violating terms which appear in the dimensional reduction have to cancel out. We use this idea to complete the quartic Riemann correction with all terms involving five NS-NS sector fields. The resulting Lagrangian is considerably simpler than expressions that have previously appeared in the literature.

\clearpage
\tableofcontents

\section{Introduction}
The first correction to the tree-level effective action for the type II string has been known for a long time to involve a quartic Riemann invariant at order $\alpha'^3$. More explicitly it takes the form \cite{Gross:1986iv,Grisaru:1986vi,Freeman:1986zh}
\begin{equation}
\frac{\zeta(3)\alpha'^3}{2^{11}}\int d^Dx\sqrt{-g}\,e^{-2\Phi}L\,,\qquad L=\frac{1}{4!}t_8t_8R^4+\ldots
\label{eq:alpha3correction}
\end{equation}
In fact, a similar correction appears also for the heterotic string \cite{Cai:1986sa,Gross:1986mw} and the bosonic string \cite{Jack:1988sw} ($D$ denotes the critical dimension). This $R^4$-term is known to be accompanied by a large number of terms, but the full expression is not known. Here we confine ourselves to the NS-NS sector only. In that case we know from four-point string scattering amplitude calculations that the Riemann tensor should actually be replaced by the torsionful Riemann tensor\footnote{This is the Riemann tensor computed from the torsionful spin connection $\omega^{(-)}=\omega-\frac12H$ (it is related to the one compute from $\omega^{(+)}=\omega+\frac12H$ by sending $H\rightarrow-H$ or, equivalently, exchanging the first two and last two indices).}
\begin{equation}
\mathcal R^{ab}{}_{cd}=\mathcal R_S^{ab}{}_{cd}-(\nabla H)^{ab}{}_{cd}\,,
\label{eq:Rcal}
\end{equation}
where
\begin{equation}
\mathcal R_S^{ab}{}_{cd}=R_S^{ab}{}_{cd}+\frac12H^{[a}{}_{ce}H^{b]e}{}_{d}\,,
\end{equation}
is symmetric under exchanging the first two and the last two indices, while
\begin{equation}
(\nabla H)^{ab}{}_{cd}=\nabla^{[a}H^{b]}{}_{cd}
\end{equation}
is anti-symmetric. Furthermore, it is known from five-point amplitude calculations that there is also a $\varepsilon_8\varepsilon_8\mathcal R^4$ term, as well as certain $H^2\mathcal R^3$ terms \cite{Liu:2019ses}. Terms involving the derivative of the dilaton and terms quartic in the $B$-field have so far not been computed from amplitudes. In principle one could simply continue and compute all the remaining terms from scattering amplitudes. This would however be very involved for two reasons -- (1) one would have to go up to the eight-point amplitude (to get the $H^8$ terms) and (2) one has to match this to a field theory calculation involving an ansatz with an enormous number of terms. This approach does not seem very practical. Another possible approach is to try to use supersymmetry to find the remaining terms. We will not follow this approach here, instead we refer to \cite{Ozkan:2024euj} for a recent review of the state of the art. A third possible approach is to try to use duality symmetries to fix the remaining terms. This is the approach we will take here.

The most straightforward way to do this is to take the most general ansatz for the $D$-dimensional action, reduce it on $S^1$, and require the result to be invariant under T-duality. This calculation has been carried out by Garousi in a series of papers. Due to the ansatz involving of the order of 1000 terms \cite{Garousi:2020mqn} the calculation was done using symbolic software. He found that the full NS-NS sector Lagrangian is uniquely fixed by the requirement of T-duality \cite{Garousi:2020gio}. While this result is very nice, there is one serious drawback -- the resulting Lagrangian contains over 400 terms! While it turns out to be possible to use the previously known structures to simplify the result \cite{Garousi:2020lof,Garousi:2022ghs}, the Lagrangian is still too unwieldy, involving just under 200 terms. There is good reason to think that the result can be cast in a much better form, but it is very hard to find the right form without some additional hints for what structures should appear. For this reason we will follow a different approach, where it is easier to recognize which structures must appear along the way.

We will use the fact that the tree-level string effective action reduced to $D-d$ dimensions has a continuous $O(d,d)$ symmetry \cite{Meissner:1991zj,Meissner:1991ge,Sen:1991zi}. This symmetry is closely related to T-duality. The presence of this symmetry in the reduced theory gives very strong constraints on the form of the unreduced theory. However, carrying out the full dimensional reduction and imposing the $O(d,d)$ symmetry is even more complicated (for general $d$) than requiring T-duality of the $S^1$ reduction. But there is a way to simplify the calculations, which was introduced in \cite{Wulff:2021fhr}.\footnote{Other approaches to simplifying the use of duality symmetries to constrain higher derivative corrections include \cite{Godazgar:2013bja} and \cite{Baron:2022but,Baron:2023qkx}. These have so far only been applied at the first order in $\alpha'$. Other works, e.g. \cite{Codina:2020kvj,Codina:2021cxh,David:2021jqn}, have considered a ``cosmological reduction'' to one dimension, but in that case the constraints are not strong enough to fix all terms in the action. Finally, approaches which attempt to make the duality symmetries manifest from the beginning appear to fail at order $\alpha'^3$ \cite{Hronek:2020xxi}.} The idea is to implement only a necessary condition for $O(d,d)$ symmetry. We consider only the terms quadratic in the KK vectors that arise in the dimensional reduction, with no dependence on the scalars (except the dilaton). Dimensional reduction of the NS-NS sector fields gives rise to two vector fields, one coming from the metric and one from the B-field. Taking their sum and difference we have the two fields $A_a{}^{c'}$ and $\hat A_b{}^{d'}$, where unprimed/primed indices are external/internal. Since any internal indices must be contracted in the effective action there are three possible contractions of these that can appear in the reduced action: $A_a\cdot A_b$, $\hat A_a\cdot\hat A_b$ and $A_a\cdot\hat A_b$ (actually, due to gauge invariance their fields strengths appear instead of $A$ and $\hat A$). Now, the point is that the KK vectors combine to a single $O(d,d)$ vector, with respect to the internal index, but there are only \emph{two} quadratic $O(d,d)$ invariants: the contraction with the constant $O(d,d)$ invariant metric and the contraction with the so-called generalized metric. One finds that the contractions $A_a\cdot A_b$ and $\hat A_a\cdot\hat A_b$ respect the $O(d,d)$ symmetry, while $A_a\cdot\hat A_b$ violates it. Therefore a necessary condition for $O(d,d)$ invariance of the reduced action is that all terms proportional to $A_a\cdot\hat A_b$ cancel out. This condition turns out to be extremely strong. It fixes the form of the correction essentially completely.\footnote{Terms involving only the derivative of the dilaton turn out not to be fixed. Such terms are not expected and indeed they are known not to be present for the correction in question.}

In \cite{Wulff:2021fhr} this condition was used to complete the $R^4$ terms up to fifth order in fields, up to dilaton terms. The $H^2\mathcal R^3$ terms were shown to agree with amplitude calculations \cite{Liu:2019ses}. They also agree with the results of Garousi \cite{Garousi:2022ghs}. Here we revisit this calculation in order to (1) include also the dilaton terms (2) determine a better form for the $H^4\mathcal R$ terms, whose form in \cite{Wulff:2021fhr}, while simpler than the expressions found by Garousi, did not have a particularly natural form. Taking into account a similar calculation at order $\alpha'^2$ \cite{Wulff:2024ips}, we are able to reduce the $H^4\mathcal R$ terms by almost a factor of two compared to \cite{Wulff:2021fhr}, as well as find a natural structure for at least some of these terms. The result is a complete form for the completion of $R^4$ up to fifth order in NS-NS fields, which is by far the simplest form known to date. Ultimately this result should of course be completed to include the higher order terms, which have the form $H^4\mathcal R^2$, $H^6\mathcal R$ and $H^8$, either by the same approach or by comparing to the all order results of Garousi. Both approaches involve long calculations, but the latter may easily be done with the aid of a computer. However, there is good reason to expect that these higher order terms will not be too complicated as there are not as many terms with these structures and also given the result at $\alpha'^2$ in \cite{Wulff:2024ips}, where most of the higher order terms could be removed (see also the results of \cite{Garousi:2022ghs}). Finally, the RR fields also need to be included to have the complete correction in the case of the type II string, for some progress in this direction see \cite{Liu:2022bfg}.

Some words of clarification are in order here. Firstly, note that we are not trying to prove the full $O(d,d)$ invariance of the reduced action. This was already (indirectly) proven by Sen in \cite{Sen:1991zi}. Instead, we are imposing only a \emph{necessary} condition for $O(d,d)$.\footnote{Using our results it would be straightforward (but very tedious) to prove directly the full $O(d,d)$ invariance. To do this one of course has to carry out the full dimensional reduction, without truncating the scalars. The argument at the beginning of section 4 of \cite{Wulff:2021fhr} shows that this is guaranteed to work.} Another important point concerns the uniqueness of the result. As already mentioned the result is not unique since certain dilaton terms, e.g. $((\nabla_a^{(+)}\partial_b\Phi)^2)^2$, can be added without affecting the results. Besides these terms, which involve only the dilaton and its derivatives, no other non-uniqueness is encountered in the calculations. But we did not prove that there is no other source of non-uniqueness, since this would involve lengthy calculations (and anyway seems very unlikely). Instead we rely on the results of Garousi mentioned above, which show that there is a unique Lagrangian compatible with the $O(1,1)$ symmetry of the circle reduction. The only thing that remains to do is then to match our $H^4\mathcal R$ terms to those of Garousi (all other terms already match), which would be straightforward with the aid of a computer. There is very little doubt that the two will match (up to field redefinitions). Indeed, at order $\alpha'^2$ the two results match perfectly, as shown in \cite{Wulff:2024ips}. The point of the present paper is not so much to rederive the results of Garousi in a different way, but rather to find a more suitable form for those results.

The calculations in this paper are essentially straightforward, but in places very long due to a large number of integrations by parts and use of Bianchi identities needed. We give a lot of detail for most steps of the calculation in the bulk of the paper. However, most readers will be mainly interested in the end result and for this reason we summarize it here.

\subsection{Summary of results}
For ease of presentation we have divided the Lagrangian into seven pieces
\begin{equation}
L=L_1+L_2+L_3+L_4+L_5+L_6+L_7\,,
\label{eq:L}
\end{equation}
where
\begin{equation}
\begin{split}
L_1
&=
\frac{1}{4!}t_8t_8\mathcal R^4
-\frac{1}{2(3!)}t_8t_8H^2\mathcal R^3
+\frac12\tilde t_8\tilde t_8H^2(\nabla H)^2\mathcal R
+\frac{1}{4(4!)}(\varepsilon_8\varepsilon_8)'\mathcal R^4
\\
&\quad
+\frac{1}{2(3!)^3}(\varepsilon_9\varepsilon_9)'H^2\mathcal R^3
-\frac{1}{2(3!)^2}H^2(\varepsilon_6\varepsilon_6)'\mathcal R^3
+\frac{1}{2(3!)^3}(\varepsilon_9\varepsilon_9)'[H^2]\mathcal R^3
\\
&\quad
-\frac{1}{2(3!)^2}(\varepsilon_9\varepsilon_9)'[H^2](\nabla H)^2\mathcal R
+2\varepsilon_4H^2\varepsilon_4\mathcal R^2\mathcal R
-4\varepsilon_4H^2\varepsilon_4(\nabla H)^2\mathcal R\,,
\end{split}
\label{eq:L1}
\end{equation}
contains the terms with (more or less) familiar structures. In particular it contains the quartic Riemann terms and all $H^2\mathcal R^3$ terms. With the exception of the $H^2(\varepsilon_6\varepsilon_6)'\mathcal R^3$ term these are the same as in \cite{Wulff:2021fhr}, but written more compactly, and they match with amplitude calculations as verified there. The former term was not visible in those calculations due to neglecting dilaton terms (related to this term via the equations of motion), it does match with \cite{Garousi:2022ghs}. We have also included $H^2(\nabla H)^2\mathcal R$ terms with similar structures. The tensor structures appearing here are defined as follows
\begin{equation}
t_8M_1M_2M_3M_4
=
8\tr(M_1M_2M_3M_4)
-2\tr(M_1M_2)\tr(M_3M_4)
+\mbox{cyclic}(2,3,4)\,,
\label{eq:t8}
\end{equation}
which is standard, but we also define another version
\begin{equation}
\begin{split}
\tilde t_8M_1M_2M_3M_4
&=
-8\tr(M_1M_2M_3M_4)
+8\tr(M_1M_3M_4M_2)
+8\tr(M_1M_4M_2M_3)
\\
&\quad
+2\tr(M_1M_2)\tr(M_3M_4)
-2\tr(M_1M_3)\tr(M_4M_2)
+2\tr(M_1M_4)\tr(M_2M_3)\,,
\end{split}
\label{eq:t8tilde}
\end{equation}
which differs from $t_8$ in the sign of the first, fourth and sixth term. The $\varepsilon\varepsilon$ structures are defined as
\begin{equation}
(\varepsilon_8\varepsilon_8)'\mathcal R^4=-8!\left(\mathcal R^4\right)'^{a_1\cdots a_8}{}_{[a_1\cdots a_8]}
\label{eq:e8e8R4}
\end{equation}
and
\begin{equation}
(\varepsilon_9\varepsilon_9)'H^2\mathcal R^3=-9!H^{a_1a_2a_3}H_{[a_1a_2a_3}\left(\mathcal R^3\right)'^{a_4\cdots a_9}{}_{a_4\cdots a_9]}\,,
\label{eq:e9e9H2R3}
\end{equation}
which is also standard except for the prime, which means that all terms with self-contraction, i.e. terms involving the (torsionful) Ricci tensor and scalar, are to be removed. Again, this difference is only seen when dilaton terms are included so it was not seen in \cite{Wulff:2021fhr} or \cite{Liu:2019ses}, but again this agrees with \cite{Garousi:2022ghs} where the dilaton terms were included. We similarly have
\begin{equation}
H^2(\varepsilon_6\varepsilon_6)'\mathcal R^3=-6!H^2\left(\mathcal R^3\right)'^{a_1\cdots a_6}{}_{[a_1\cdots a_6]}\,,
\label{eq:H2e6e6R3}
\end{equation}
where all the indices of the $H$s are contracted, i.e. $H^2=H_{abc}H^{abc}$. This term simply serves to remove the term with the $H$s completely contracted coming from $(\varepsilon_9\varepsilon_9)'H^2\mathcal R^3$. Finally, we have the term
\begin{equation}
(\varepsilon_9\varepsilon_9)'[H^2]\mathcal R^3=-\frac{6!^2}{6}H^{a_1a_2a_3}H_{a_4a_5a_6}\left(\mathcal R^3\right)'^{[a_4\cdots a_9]}{}_{[a_7a_8a_9a_1a_2a_3]}\,,
\label{eq:e9e9HHR3}
\end{equation}
where the square brackets mean that $H^2$ should be treated as a single object, i.e. any term with contractions of the two $H$s is also removed. This term simply doubles the corresponding term coming from $(\varepsilon_9\varepsilon_9)'H^2\mathcal R^3$ (curiously at order $\alpha'^2$ it has the opposite sign and the two contributions cancel rather than add up \cite{Wulff:2024ips}). We have also defined a term with a somewhat different structure
\begin{equation}
\varepsilon_4H^2\varepsilon_4\mathcal R^2\mathcal R
=
-4!H_a{}^{a_1a_2}H^{ba_3a_4}\mathcal R_{ce[a_1a_2}\mathcal R_{a_3a_4]}{}^{ed}\mathcal R_{bd}{}^{ac}\,,
\label{eq:e4}
\end{equation}
which was defined slightly differently, with an anti-symmetrization of five indices, in \cite{Wulff:2021fhr}.

The remaining terms in the Lagrangian consist of 25 terms of the form $H^2(\nabla H)^2\mathcal R$, which seem not to be expressible in terms of the above structures. Of course we can replace $\mathcal R$ by $R$, the usual Riemann tensor, in these terms, since they differ only by higher order terms which we neglected here. First we have the four terms ($H^2_{abcd}=H_{abe}H^e{}_{cd}$)
\begin{equation}
\begin{split}
L_2
&=
-4!(\nabla H)^{abef}(\nabla H)^{cd}{}_{ef}H_{k[ab}\nabla^2H^k{}_{cd]}
+4!\nabla^eH^{fab}\nabla_eH_f{}^{cd}H_{k[ab}\nabla^2H^k{}_{cd]}
\\
&\quad
-2(4!)H^2_{[abcd]}(\nabla H)^{abef}(\nabla H)_{ghef}\mathcal R_S^{cdgh}
-2(4!)H^2_{[abcd]}\nabla^eH^{fag}\nabla_eH_f{}^{bh}\mathcal R_S^{cd}{}_{gh}\,.
\end{split}
\end{equation}
The reason we have written them this way is that the first two terms turn out to be precisely the same as two terms that appear at order $\alpha'^2$ \cite{Wulff:2024ips}, except for an extra insertion of $\nabla^2$. The last two are also of the same form but with an extra insertion of $\mathcal R$. At the moment we don't have a deeper understanding of why these terms should appear. Note that using the fact that $\nabla^2H_{abc}=-3R^{de}{}_{[ab}H_{c]de}$, up to higher order terms and terms proportional to the equations of motion (which can be removed by a field redefinition), also the first two terms take the form $H^2(\nabla H)^2\mathcal R$. Which is the right way to write these terms will only be clear once the higher order terms are computed (when working to higher order in fields on has to replace $\nabla^2$ by $(\nabla_a-2\partial_a\Phi)\nabla^a$ in order not to generate extra dilaton terms).

The remaining terms have a less clear structure, but they all have at least two pairs of anti-symmetrized indices contracted, which we refer to loosely as at least two `traces'. First we have the terms
\begin{equation}
\begin{split}
L_3
&=
8\nabla_eH^{kcd}\nabla^eH_{fcd}H_{abk}\nabla^2H^{abf}
-8\nabla_eH_{kcd}H^{fcd}\nabla^eH_{abk}\nabla^2H^{abf}
\\
&\quad
+8\nabla^eH^2_{abcd}H^{agh}\nabla_eH_{fgh}\mathcal R_S^{bfcd}
-8H_{abf}H_{kcd}\nabla_eH^{abk}\nabla^eH^{fgh}\mathcal R_S^{cd}{}_{gh}\,.
\end{split}
\end{equation}
Their structure is again similar to terms that appear at order $\alpha'^2$, with an extra insertion of $\nabla^2$ or $\mathcal R$. Note that in these terms all derivatives are contracted among themselves. Next we have three terms with both derivatives contracted with $\mathcal R$
\begin{equation}
\begin{split}
L_4
&=
-16H^2_{abcd}\nabla_eH^{agh}\nabla_fH^c{}_{gh}\mathcal R_S^{bdef}
-4H^2_{abcd}\nabla_eH^{cgh}\nabla_fH^d{}_{gh}\mathcal R_S^{abef}
\\
&\quad
+8H_{abh}H_{kcd}\nabla_eH^{abg}\nabla_fH_g{}^{cd}\mathcal R_S^{hkef}\,.
\end{split}
\end{equation}
Then we have five terms whose structure is different
\begin{equation}
\begin{split}
L_5
&=
64H^{abc}H_{def}(\nabla H)_{ck}{}^{gd}(\nabla H)^{hk}{}_{ab}\mathcal R_S^{ef}{}_{gh}
-16H^{abc}H_{def}(\nabla H)^{dk}{}_{ab}(\nabla H)_{ckgh}\mathcal R_S^{efgh}
\\
&\quad
+16H_{abe}H_{kcd}\nabla^kH^{egh}(\nabla H)^c{}_{fgh}\mathcal R_S^{fdab}
+16H^2_{abcd}\nabla_fH^{abc}(\nabla H)^{efgh}\mathcal R_{Sdegh}
\\
&\quad
+4H_{abc}H^{def}\nabla_dH_k{}^{ab}(\nabla H)_{efgh}\mathcal R_S^{ckgh}\,,
\end{split}
\end{equation}
where the last two involve a `Chern-Simons like' combination $H_{ab[c}\nabla^eH_{d]}{}^{ab}$, which also appears at order $\alpha'^2$. Finally we have terms involving $H^2_{ab}=H_{acd}H_b{}^{cd}$
\begin{equation}
\begin{split}
L_6
&=
4H^2_{ef}\nabla_kH^{ecd}\nabla^fH^{kgh}\mathcal R_{Scdgh}
+8H^2_{ef}\nabla_kH^{eab}(\nabla H)_{abgh}\mathcal R_S^{kfgh}
\\
&\quad
-8\nabla^kH^2_{ef}H_{abk}\nabla^eH^a{}_{gh}\mathcal R_S^{bfgh}
-4\nabla_kH^2_{ef}H^{kcd}\nabla_gH_{hcd}\mathcal R_S^{eghf}
\end{split}
\end{equation}
and terms where two of the $H$s have all their indices contracted ($H^2=H_{abc}H^{abc}$)
\begin{equation}
\begin{split}
L_7
&=
\frac13\nabla^eH^2_{abcd}\nabla_eH^2\mathcal R_S^{abcd}
-\frac43H^2_{cd}\nabla_aH_{efg}\nabla_bH^{efg}\mathcal R_S^{acdb}
-\frac23H^2\nabla_cH_{agh}\nabla_dH_b{}^{gh}\mathcal R_S^{abcd}
\\
&\quad
+\frac43\nabla^hH_{efg}\nabla^kH^{efg}H_{kcd}H_{abh}\mathcal R_S^{cdab}
+\frac43H^2(\nabla H)_{abcd}(\nabla H)^{abgh}\mathcal R_S^{cd}{}_{gh}\,.
\end{split}
\end{equation}
Note that there are no terms involving the derivative of the dilaton, which is consistent with the results of Garousi.

This Lagrangian was derived from the requirement that the $O(d,d)$ violating terms quadratic in the KK vectors cancel out upon dimensional reduction. More precisely, they cancel out up to total derivative terms, which we drop throughout, and terms proportional to the equations of motion. The latter we keep track of up to first order in the $B$-field. They are given in (\ref{eq:k4eom}). The presence of these terms shows that (very complicated) field redefinitions are required in the reduced theory before it becomes compatible with $O(d,d)$ symmetry. Their form can be directly read off from (\ref{eq:k4eom}). An interesting question, which we plan to return to, is whether these field redefinitions can be lifted to $D$ dimensions.

The remainder of this paper is dedicated to the derivation of the Lagrangian presented above. In Sec. \ref{sec:KK} we briefly review the Kaluza-Klein reduction focusing on the $O(d,d)$-violating terms quadratic in the KK vectors. In Sec. \ref{sec:t8} we analyze the $O(d,d)$-violating terms coming from the terms with the $t_8$-structure. Their cancellation turns out to require many terms of different structures, whose dimensional reduction and simplification is deferred to the appendix.

\section{Kaluza-Klein reduction and \texorpdfstring{$O(d,d)$}{O(d,d)}}\label{sec:KK}
A nice discussion of Kaluza-Klein reduction and $O(d,d)$ symmetry can be found in \cite{Maharana:1992my}. Here we will mostly be interested in how $O(d,d)$ violating contractions of two KK vector fields can appear. We will summarize this briefly here and refer to \cite{Wulff:2024ips} for more details.

The starting point is the usual KK ansatz for the vielbein (internal indices are denoted with a prime)
\begin{equation}
\underline e_{\underline m}{}^{\underline a}
=
\left(
\begin{array}{cc}
	e_m{}^a & A^{(1)n'}_me_{n'}{}^{a'}\\
	0 & e_{m'}{}^{a'}
\end{array}
\right)
\end{equation}
and the $B$-field
\begin{equation}
\underline B_{\underline {mn}}
=
\left(
\begin{array}{cc}
	\hat B_{mn} & A^{(2)}_{mn'}+A^{(1)k'}_mB_{k'n'}\\
	-A^{(2)}_{nm'}-A^{(1)k'}_nB_{k'm'} & B_{m'n'}
\end{array}
\right)\,,
\end{equation}
where
\begin{equation}
\hat B_{mn}=B_{mn}-A_{[m}^{(1)m'}A_{n]m'}^{(2)}+A^{(1)k'}_mA^{(1)l'}_nB_{k'l'}\,.
\end{equation}

Besides the lower-dimensional metric, $B$-field and dilaton we obtain $2d$ vector fields $A^{(1)m'}$ and $A_{m'}^{(2)}$ (we suppress, for the moment, the external vector index) as well as a number of scalar fields contained in $g_{m'n'}$ and $B_{m'n'}$. These can be combined into fields that transform simply under the $O(d,d)$ symmetry: an $O(d,d)$ vector
\begin{equation}
\mathcal A_M=
\left(
\begin{array}{c}
	A^{(1)m'}\\
	A^{(2)}_{m'}
\end{array}
\right)
\end{equation}
and a symmetric $O(d,d)$ matrix, sometimes called the generalized metric,
\begin{equation}
\mathcal H_{MN}=
\left(
\begin{array}{cc}
	g^{m'n'} & -g^{m'k'}B_{k'n'}\\
	B_{m'k'}g^{k'n'} & g_{m'n'}-B_{m'k'}g^{k'l'}B_{l'n'}
\end{array}
\right)\,.
\end{equation}
What will be important for us is that from two $\mathcal A$s we can form precisely two $O(d,d)$ scalars, namely 
\begin{equation}
\mathcal A_M\eta^{MN}\mathcal A_N\qquad\mbox{and}\qquad\mathcal A_M\mathcal H^{MN}\mathcal A_N\,,
\end{equation}
where $\mathcal H^{MN}=\eta^{MK}\mathcal H_{KL}\eta^{LN}$ and we introduced the $O(d,d)$ invariant metric
\begin{equation}
\eta^{MN}=
\left(
\begin{array}{cc}
0 & \delta^{m'}{}_{n'}\\
\delta_{m'}{}^{n'} & 0	
\end{array}
\right)\,.
\end{equation}
In the following we will simply set the scalars that arise in the dimensional reduction to zero (this is consistent since we only work up to quadratic order in the KK vectors \cite{Wulff:2024ips}). In this case the two $O(d,d)$ scalars constructed from the KK vectors become
\begin{equation}
A_{(m}^{(1)m'}A^{(2)}_{n)m'}\qquad\mbox{and}\qquad A_m^{(1)m'}A^{(1)}_{nm'}+A_m^{(2)m'}A^{(2)}_{nm'}\,.
\end{equation}
We now define the combinations
\begin{equation}
A=-\frac12(A^{(1)}+A^{(2)})\qquad\mbox{and}\qquad\hat A=\frac12(A^{(1)}-A^{(2)})\,.
\end{equation}
In terms of these the two $O(d,d)$ invariants are
\begin{equation}
A_m\cdot A_n\qquad\mbox{and}\qquad\hat A_m\cdot\hat A_n\,,
\end{equation}
where the $\cdot$ denotes contraction of the internal index. Conversely, the inner product between the two vectors,
\begin{equation}
A_m\cdot\hat A_n\,,
\end{equation}
explicitly \emph{violates} the $O(d,d)$ symmetry. Such terms arise in the dimensional reduction, but they have to cancel out (up to total derivatives and equation of motion terms) in the reduced action, since otherwise $O(d,d)$ symmetry would be violated. This gives a powerful constraint on the form of the effective action.

The reduction of $H$, the spin connection and the torsionful Riemann tensor, which all enter the effective action, is given by
\begin{equation}
\begin{aligned}
\underline H_{a'b'c'}&=0\,,& & & & & \underline H_{abc'}&=-F_{c'ab}-\hat F_{c'ab}\,,\\
\underline H_{ab'c'}&=0\,, & & & & & \underline H_{abc}&=\hat H_{abc}\,,
\end{aligned}
\label{eq:Hred}
\end{equation}
where $\hat H_{abc}=H_{abc}+3(\hat A_{[a}\cdot \hat F_{bc]}-A_{[a}\cdot F_{bc]})$ (whose form we will not need since the contraction of internal indices respects the $O(d,d)$ symmetry),
\begin{equation}
\begin{aligned}
%
\underline\omega_{a'b'c'}&=0\,,& & & & & \underline\omega_{a'bc}&=\frac12(F_{a'bc}-\hat F_{a'bc})\,,\\
\underline\omega_{a'b'c}&=0\,,& & & & & \underline\omega_{abc'}&=\frac12(-F_{c'ab}+\hat F_{c'ab})\,,\\
\underline\omega_{ab'c'}&=0\,,& & & & & \underline\omega_{abc}&=\omega_{abc}
%
\end{aligned}
\end{equation}
and\footnote{It is understood that $H$ is everywhere replaced by $\hat H$ defined below (\ref{eq:Hred}) in these expressions. To go from the conventions of \cite{Wulff:2021fhr} to those of the present paper one sets
$$
F_{a',\mathrm{there}}=\frac{1}{\sqrt2}\hat F_{a',\mathrm{here}}\,,\qquad F^{a'}_{\mathrm{there}}=\frac{1}{\sqrt2}F^{a'}_{\mathrm{here}}\,.
$$
}
\begin{equation}
\begin{aligned}
\underline{\mathcal R}_{a'b'c'd'}&=0\,,& & & & & \underline{\mathcal R}_{abc'd'}&=-2\hat F_{c'[a}{}^e\hat F_{d'b]e}\,,& & & & & \underline{\mathcal R}_{abc'd}&=\nabla^{(+)}_d\hat F_{c'ab}\,,\\
\underline{\mathcal R}_{a'b'c'd}&=0\,,& & & & & \underline{\mathcal R}_{a'b'cd}&=-2F_{a'[c}{}^eF_{b'd]e}\,,& & & & & \underline{\mathcal R}_{a'bcd}&=-\nabla^{(-)}_bF_{a'cd}\,,\\
\underline{\mathcal R}_{ab'c'd'}&=0\,,& & & & & \underline{\mathcal R}_{a'bc'd}&=\hat F_{c'b}{}^eF_{a'ed}\,,& & & & & \underline{\mathcal R}_{abcd}&=\hat{\mathcal R}_{abcd}+\hat F_{ab}\cdot F_{cd}\,,
\end{aligned}
\label{eq:Rred}
\end{equation}
where we defined $\hat{\mathcal R}_{abcd}=\mathcal R_{abcd}+2\hat F_{a[c}\cdot\hat F_{d]b}-F_{ab}\cdot F_{cd}$, whose form we will again not need since the contraction of internal indices respects the $O(d,d)$ symmetry. In contrast, the second term in the last expression violates the $O(d,d)$ symmetry and will therefore be important for us. The reduction of the torsionful Ricci tensor and scalar can be found in \cite{Wulff:2024ips}.

\section{\texorpdfstring{$t_8$}{t8}-terms}\label{sec:t8}
The Lagrangian is known to contain two terms with the $t_8$ structure. Upon dimensional reduction they give rise to the following $O(d,d)$ violating terms quadratic in the KK field strengths
\begin{equation}
\begin{split}
\frac{1}{4!}t_8t_8\mathcal R^4 &\rightarrow\frac16t_8t_8\hat F\cdot F\mathcal R^3\,,
\\
-\frac{1}{12}t_8t_8H^2\mathcal R^3 &\rightarrow-\frac{1}{12}t_8t_8(\hat F\cdot F+F\cdot\hat F)\mathcal R^3+t_8HF\nabla F\mathcal R^2\,,
\end{split}
\label{eq:t8-red}
\end{equation}
where the last term is defined as\footnote{Adding the same terms with the sign of the $B$-field changed and $F$ and $\hat F$ exchanged enforces the symmetry of the $D$-dimensional effective action under $B\rightarrow-B$.}
\begin{equation}
\begin{split}
t_8HF\nabla F\mathcal R^2
&=
4t^{a_1\cdots a_8}\hat F^{be}\cdot\nabla^{(-)}_dF_{a_1a_2}H_{ea_3a_4}\mathcal R_{bca_5a_6}\mathcal R^{cd}{}_{a_7a_8}
\\
&\quad
+t^{a_1\cdots a_8}\hat F^{be}\cdot\nabla^{(-)}_bF_{a_1a_2}H_{ea_3a_4}\mathcal R_{cda_5a_6}\mathcal R^{cd}{}_{a_7a_8}
+(H\rightarrow-H, F\leftrightarrow\hat F)\,.
\end{split}
\end{equation}
Our strategy will be to try to rewrite these terms as terms without derivatives on $F$ and $\hat F$, but still with at most one derivative on $H$ and no derivatives on $\mathcal R$. As we will see, this requires additional terms 
in the $D$-dimensional action.

It is convenient to organize the terms according to the number of pairs of contracted antisymmetrized indices. We will refer to this loosely as the number of ``traces'' they contain. We have
\begin{equation}
t_8HF\nabla F\mathcal R^2
=
[t_8HF\nabla F\mathcal R^2]_{<2\tr}
+[t_8HF\nabla F\mathcal R^2]_{\geq2\tr}
\end{equation}
where
\begin{equation}
\begin{split}
{}[t_8H\cdot F\nabla F\mathcal R^2]_{<2\tr}
&=
32\hat F^{bk}\cdot\nabla^{(-)}_dF_{a_1a_2}\mathcal R_{bc}{}^{a_2a_3}H_{ka_3a_4}\mathcal R^{cda_4a_1}
\\
&\quad
+64\hat F^{(b}{}_k\cdot\nabla^{(-)d)}F_{a_1a_2}H^{ka_2a_3}\mathcal R^c{}_{da_3a_4}\mathcal R_{bc}{}^{a_4a_1}
\\
&\quad
+8\hat F^{bk}\cdot\nabla^{(-)}_bF_{a_1a_2}\mathcal R^{cda_2a_3}H_{ka_3a_4}\mathcal R_{cd}{}^{a_4a_1}
\\
&\quad
+16\hat F^{bk}\cdot\nabla^{(-)}_bF_{a_1a_2}H_{ka_2a_3}\mathcal R^{cd}{}_{a_3a_4}\mathcal R_{cd}{}^{a_4a_1}
+(H\rightarrow-H, F\leftrightarrow\hat F)
\end{split}
\label{eq:HFdFR2-1}
\end{equation}
and
\begin{equation}
\begin{split}
{}[t_8H\cdot F\nabla F\mathcal R^2]_{\geq2\tr}
&=
-8\hat F^{bk}\cdot\nabla^{(-)}_dF^{a_1a_2}H_{ka_1a_2}\mathcal R_{bca_3a_4}\mathcal R^{cda_3a_4}
\\
&\quad
-16\hat F^{(b}{}_k\cdot\nabla^{(-)d)}F^{a_1a_2}H^{ka_3a_4}\mathcal R_{bca_1a_2}\mathcal R^c{}_{da_3a_4}
\\
&\quad
-2\hat F^{bk}\cdot\nabla^{(-)}_bF^{a_1a_2}H_{ka_1a_2}\mathcal R^{cda_3a_4}\mathcal R_{cda_3a_4}
\\
&\quad
-4\hat F^{bk}\cdot\nabla^{(-)}_bF_{a_1a_2}H_{ka_3a_4}\mathcal R^{cda_1a_2}\mathcal R_{cd}{}^{a_3a_4}
+(H\rightarrow-H, F\leftrightarrow\hat F)\,.
\end{split}
\label{eq:HFdFR2-2}
\end{equation}

The first step is to get rid of all terms which have a contraction between $\hat F$ ($F$) without derivative and $H$. We focus first on the terms with fewer than two traces. The first term in (\ref{eq:HFdFR2-1}) we simply integrate by parts to get (dropping a total derivative term)
\begin{equation}
\begin{split}
32\hat F^{bk}\cdot\nabla^{(-)}_dF_{a_1a_2}H_{ka_3a_4}\mathcal R_{bc}{}^{a_2a_3}\mathcal R^{cda_4a_1}
&=
-16\hat F^{bk}\cdot F_{a_1a_2}H_{ka_3a_4}\nabla^{(-)}_b\mathcal R_{cd}{}^{a_1a_3}\mathcal R^{cda_4a_2}
\\
&\quad
+32\nabla^{(-)}_d\hat F^{bk}\cdot F_{a_1a_2}H_{ka_3a_4}\mathcal R_{bc}{}^{a_1a_3}\mathcal R^{cda_4a_2}
\\
&\quad
+32\hat F^{bk}\cdot F_{a_1a_2}\nabla^{(-)}_dH_{ka_3a_4}\mathcal R_{bc}{}^{a_1a_3}\mathcal R^{cda_4a_2}
\\
&\quad
+32\hat F^{bk}\cdot F_{a_1a_2}H_{ka_3a_4}\mathcal R_{bc}{}^{a_1a_3}(\nabla^{(-)}_d-2\partial_d\Phi)\mathcal R^{cda_4a_2}
\\
&\quad
+48\hat F^{bk}\cdot F_{a_1a_2}H_{ka_3a_4}\nabla^{(-)}_{[d}\mathcal R_{bc]}{}^{a_1a_3}\mathcal R^{cda_4a_2}\,.
\end{split}
\label{eq:HFdFR2t1}
\end{equation}
Using (\ref{eq:id1}) and (\ref{eq:id4}) the last two terms reduce to terms proportional to the equations of motion and terms of sixth order in the fields. Ignoring these, only the first term still has a contraction between the $F$ without derivative and $H$, but we will see that it cancels against another term.

Consider now the second term in (\ref{eq:HFdFR2-1}) which, using the $F$ Bianchi identity, becomes
\begin{equation}
\begin{split}
64\hat F^{k(b}\cdot\nabla^{(-)d)}F_{a_1a_2}H_k{}^{a_1a_3}\mathcal R^c{}_{da_3a_4}\mathcal R_{bc}{}^{a_4a_2}
&=
-32\hat F^{kb}\cdot\nabla_{a_1}F_{a_2d}H_k{}^{a_1a_3}\mathcal R^{cd}{}_{a_3a_4}\mathcal R_{bc}{}^{a_4a_2}
\\
&\quad
-32\hat F^{kd}\cdot\nabla_{a_1}F_{a_2}{}^bH_k{}^{a_1a_3}\mathcal R^c{}_{da_3a_4}\mathcal R_{bc}{}^{a_4a_2}
\\
&\quad
-64\hat F^{k(b}\cdot\nabla_{a_2}F^{d)}{}_{a_1}H_k{}^{a_1a_3}\mathcal R^c{}_{da_3a_4}\mathcal R_{bc}{}^{a_4a_2}
\\
&\quad
+64\hat F^{e(b}\cdot F_{k[a_1}H^{d)k}{}_{a_2]}H_e{}^{a_1a_3}\mathcal R^c{}_{da_3a_4}\mathcal R_{bc}{}^{a_4a_2}\,.
\end{split}
\label{eq:HFdFR2-1-2nd}
\end{equation}
The last term is of higher order in fields while the third term can be integrated by parts and the second involves a trace as the indices on $F$ are both contracted with the same $\mathcal R$. This leaves the first term, which can be integrated by parts to give
\begin{equation}
\begin{split}
-32\hat F^{kb}\cdot\nabla_{a_1}F_{a_2d}H_k{}^{a_1a_3}\mathcal R^{cd}{}_{a_3a_4}\mathcal R_{bc}{}^{a_4a_2}
&=
32\hat F^{kb}\cdot F_{a_2d}H_k{}^{a_1a_3}\mathcal R^{cd}{}_{a_3a_4}\nabla_{a_1}\mathcal R_{bc}{}^{a_4a_2}
\\
&\quad
+32\hat F^{kb}\cdot F_{a_2d}H_k{}^{a_1a_3}\nabla_{a_1}\mathcal R^{cd}{}_{a_3a_4}\mathcal R_{bc}{}^{a_4a_2}
\\
&\quad
+32\nabla_{a_1}\hat F^{kb}\cdot F_{a_2d}H_k{}^{a_1a_3}\mathcal R^{cd}{}_{a_3a_4}\mathcal R_{bc}{}^{a_4a_2}
\\
&\quad
+32\hat F^{kb}\cdot F_{a_2d}\mathcal R^{cd}{}_{a_3a_4}\mathcal R_{bc}{}^{a_4a_2}\mathbbm B^{a_3}{}_k\,.
\end{split}
\end{equation}
The last term is proportional to the equation of motion for the $B$-field (\ref{eq:eom1}), while the third does not have a contraction between $F$ and $H$ and the second gives a trace plus higher order terms. This again leaves the first term, which becomes
\begin{equation}
\begin{split}
32\hat F^{kb}\cdot F_{a_2d}H_k{}^{a_1a_3}\mathcal R^{cd}{}_{a_3a_4}\nabla_{a_1}\mathcal R_{bc}{}^{a_4a_2}
&=
-32\hat F^{kb}\cdot F_{a_2d}H_k{}^{a_1a_3}\mathcal R^{cd}{}_{a_3a_4}\nabla_b\mathcal R_{ca_1}{}^{a_4a_2}
\\
&\quad
-32\hat F^{kb}\cdot F_{a_2}{}^dH_k{}^{a_1a_3}\mathcal R^{cd}{}_{a_3a_4}\nabla_c\mathcal R_{a_1b}{}^{a_4a_2}
\\
&\quad
+96\hat F^{eb}\cdot F_{a_2}{}^dH_{ea_1}{}^{a_3}\mathcal R^{cd}{}_{a_3a_4}\nabla_{[a_1}\mathcal R_{bc]}{}^{a_4a_2}\,.
\end{split}
\end{equation}
The last term is of higher order, the second term can be integrated by parts to give terms with two traces, while the first term gives
\begin{equation}
\begin{split}
-32\hat F^{kb}\cdot F_{a_2d}H_k{}^{a_1a_3}\mathcal R^{cd}{}_{a_3a_4}\nabla_b\mathcal R_{ca_1}{}^{a_4a_2}
&=
-32\hat F^{kb}\cdot F_{a_2d}H_k{}^{a_1a_3}\mathcal R^{cd}{}_{a_3a_4}\nabla_b\mathcal R^{a_4a_2}{}_{ca_1}
\\
&\quad
-64\hat F^{kb}\cdot F_{a_2d}H_k{}^{a_1a_3}\mathcal R^{cd}{}_{a_3a_4}\nabla_b(\nabla H)^{a_4a_2}{}_{ca_1}
\\
&=
-16\hat F^{kb}\cdot F_{a_2d}H_k{}^{a_1a_3}\nabla_b(\mathcal R^{cd}{}_{a_3a_4}\mathcal R^{a_4a_2}{}_{ca_1})
\\
&\quad
-64\hat F^{kb}\cdot F_{a_2d}H_k{}^{a_1a_3}\mathcal R^{cd}{}_{a_3a_4}\nabla_b(\nabla H)^{a_4a_2}{}_{ca_1}
\\
&=
16\hat F^{kb}\cdot\nabla_bF_{a_2d}H_k{}^{a_1a_3}\mathcal R^{cd}{}_{a_3a_4}\mathcal R^{a_4a_2}{}_{ca_1}
\\
&\quad
+16\hat F^{kb}\cdot F_{a_2d}\nabla_bH_k{}^{a_1a_3}\mathcal R^{cd}{}_{a_3a_4}\mathcal R^{a_4a_2}{}_{ca_1}
\\
&\quad
-64\hat F^{kb}\cdot F_{a_2d}H_k{}^{a_1a_3}\mathcal R^{cd}{}_{a_3a_4}\nabla_b(\nabla H)^{a_4a_2}{}_{ca_1}
\\
&\quad
+16(\nabla_b-2\partial_b\Phi)\hat F^{kb}\cdot F_{a_2d}H_k{}^{a_1a_3}\mathcal R^{cd}{}_{a_3a_4}\mathcal R^{a_4a_2}{}_{ca_1}\,,
\end{split}
\end{equation}
where the first term becomes
\begin{equation}
\begin{split}
16\hat F^{kb}\cdot\nabla_bF_{a_2d}H_k{}^{a_1a_3}\mathcal R^{cd}{}_{a_3a_4}\mathcal R^{a_4a_2}{}_{ca_1}
&=
-32\hat F^{kb}\cdot\nabla_dF_{ba_2}H_k{}^{a_1a_3}\mathcal R^{cd}{}_{a_3a_4}\mathcal R^{a_4a_2}{}_{ca_1}
\\
&=
16\hat F^{kb}\cdot F_{ba_2}H_k{}^{a_1a_3}\mathcal R^{cd}{}_{a_3a_4}\nabla_{a_1}\mathcal R^{a_4a_2}{}_{cd}
\\
&\quad
+32\nabla_d\hat F^{kb}\cdot F_{ba_2}H_k{}^{a_1a_3}\mathcal R^{cd}{}_{a_3a_4}\mathcal R^{a_4a_2}{}_{ca_1}
\\
&\quad
+32\hat F^{kb}\cdot F_{ba_2}\nabla_dH_k{}^{a_1a_3}\mathcal R^{cd}{}_{a_3a_4}\mathcal R^{a_4a_2}{}_{ca_1}
\\
&\quad
+32\hat F^{kb}\cdot F_{ba_2}H_k{}^{a_1a_3}(\nabla_d-2\partial_d\Phi)\mathcal R^{cd}{}_{a_3a_4}\mathcal R^{a_4a_2}{}_{ca_1}
\\
&\quad
+48\hat F^{kb}\cdot F_{ba_2}H_k{}^{a_1a_3}\mathcal R^{cd}{}_{a_3a_4}\nabla_{[d}\mathcal R^{a_4a_2}{}_{ca_1]}
\end{split}
\end{equation}
and finally integrating the first term by parts gives
\begin{equation}
\begin{split}
16\hat F^{kb}\cdot F_{ba_2}H_k{}^{a_1a_3}\mathcal R^{cd}{}_{a_3a_4}\nabla_{a_1}\mathcal R^{a_4a_2}{}_{cd}
&=
-16\hat F^{kb}\cdot\nabla_bF_{a_1a_2}H_k{}^{a_1a_3}\mathcal R^{cd}{}_{a_3a_4}\mathcal R^{a_4a_2}{}_{cd}
\\
&\quad
-16\hat F^{kb}\cdot\nabla_{a_2}F_{ba_1}H_k{}^{a_1a_3}\mathcal R^{cd}{}_{a_3a_4}\mathcal R^{a_4a_2}{}_{cd}
\\
&\quad
-16\hat F^{kb}\cdot F_{ba_2}H_k{}^{a_1a_3}\nabla_{a_1}\mathcal R^{cd}{}_{a_3a_4}\mathcal R^{a_4a_2}{}_{cd}
\\
&\quad
+8\nabla^b\hat F_{a_1k}\cdot F_{ba_2}H^{ka_1a_3}\mathcal R^{cd}{}_{a_3a_4}\mathcal R^{a_4a_2}{}_{cd}
\\
&\quad
-16\hat F^{kb}\cdot F_{ba_2}\mathcal R^{cd}{}_{a_3a_4}\mathcal R^{a_4a_2}{}_{cd}\mathbbm B^{a_3}{}_k\,,
\end{split}
\end{equation}
where the first term cancels the fourth term from (\ref{eq:HFdFR2-1}) and the rest are fine. 

The only remaining term with $H$ contracted with both $F$s is the third term in (\ref{eq:HFdFR2-1-2nd}), which becomes
\begin{equation}
\begin{split}
-64&\hat F^{k(b}\cdot\nabla_{a_2}F^{d)a_1}H_{ka_1}{}^{a_3}\mathcal R^c{}_{da_3a_4}\mathcal R_{bc}{}^{a_4a_2}
+(H\rightarrow-H, F\leftrightarrow\hat F)
\\
&=
-64F_{k(b}\cdot\nabla_{a_2}\hat F_{d)a_1}H^{ka_1a_3}(\nabla H)_{a_3a_4c}{}^{d}\mathcal R_S^{bca_4a_2}
\\
&\quad
-64F_{k(b}\cdot\nabla_{a_2}\hat F_{d)a_1}H^{ka_1a_3}\mathcal R_{Sa_3a_4c}{}^d(\nabla H)^{a_4a_2bc}
\\
&\quad
+32\nabla_{a_2}(F_{k(b}\cdot\hat F_{d)a_1})H^{ka_1a_3}\mathcal R_c{}^d{}_{a_3a_4}\mathcal R^{bca_4a_2}
+(H\rightarrow-H, F\leftrightarrow\hat F)\,.
\label{eq:FdotHdotdF}
\end{split}
\end{equation}
The first two terms cannot be integrated by parts. In fact, the first one also cannot cancel against other terms since all other terms turn out to involve at least one trace (and so does the second term above). Instead, it must be canceled by adding additional terms of the form $H^2(\nabla H)^2\mathcal R$ to the Lagrangian. Consider the term
\begin{equation}
l_1=(H^2)^{ab}{}_{ef}(\nabla H)_{bc}{}^{fg}\mathcal R^{cd}{}_{gh}(\nabla H)_{da}{}^{he}\,,
\label{eq:l1}
\end{equation}
which involves tracing over the first pair of indices and over the second pair in each factor. Upon dimensional reduction it gives rise to the $O(d,d)$-violating terms
\begin{equation}
\begin{split}
l_1
&\rightarrow
-\hat F_{ab}\cdot F_{cd}(\nabla H)^{eacf}(\nabla H)^{hbdg}\mathcal R_{Sehgf}
-2F^{bk}\cdot\nabla_e\hat F_{cf}H_k{}^{cd}(\nabla H)_{hbdg}\mathcal R_S^{ehgf}
\\
&\quad
+\mathcal O(H^3)
+(H\rightarrow-H, F\leftrightarrow\hat F)\,,
\end{split}
\end{equation}
where we ignored terms qubic and higher in the $B$-field. The second term has the right form to cancel the unwanted term above (up to terms with more traces). We see that we need to add to the Lagrangian a term
\begin{equation}
+32l_1\,.
\end{equation}
To determine the remaining terms that need to be added to the Lagrangian it will be convenient to first work only to leading order in traces, i.e. we first consider only terms with no contracted pairs of anti-symmetrized indices.

\subsection{Missing terms: leading order in traces}
We know all the $H^2\mathcal R^3$ terms in the Lagrangian from before. We will now determine the remaining $H^2(\nabla H)^2\mathcal R$ that are needed. To simplify this procedure we will in the following drop all terms of order $H^3$ or higher, terms of sixth order or higher in the fields, equation of motion terms and terms without derivatives on $F$. The latter two will be included later. Finally, we will, in this first step, also drop terms with one or more traces (contracted pairs of anti-symmetrized indices). Using integration by parts and Bianchi identities the relevant terms appearing upon dimensional reduction can be reduced to the following set of $F\nabla FH\mathcal R^2$ terms
\begin{equation}
\begin{split}
f_1&=F_{ef}\cdot\nabla_g\hat F_{bk}H^k{}_{cd}\mathcal R^{ged}{}_h\mathcal R^{cbhf}+(H\rightarrow-H, F\leftrightarrow\hat F)\,,\\
f_2&=F_{ef}\cdot\nabla^g\hat F_{bk}H^k{}_{cd}\mathcal R_{gh}{}^{df}\mathcal R^{hbce}+(H\rightarrow-H, F\leftrightarrow\hat F)\,,\\
f_3&=F_{be}\cdot\nabla^f\hat F^{bk}H_{kcd}\mathcal R^{gceh}\mathcal R^d{}_{hfg}+(H\rightarrow-H, F\leftrightarrow\hat F)
\end{split}
\end{equation}
and $F\nabla FH\nabla H\mathcal R$ terms
\begin{equation}
\begin{split}
g_1&=F^{ef}\cdot\nabla_g\hat F^{bk}H_{kcd}\nabla^gH^d{}_{he}\mathcal R^{ch}{}_{bf}+(H\rightarrow-H, F\leftrightarrow\hat F)\,,\\
g_2&=F_{ef}\cdot\nabla_g\hat F^{bk}H_{kcd}\nabla^gH^{dhe}\mathcal R^{cf}{}_{bh}+(H\rightarrow-H, F\leftrightarrow\hat F)\,,\\
g_3&=F_{ef}\cdot\nabla_g\hat F^{bk}H_{kcd}\nabla^eH^{dgh}\mathcal R^{cf}{}_{bh}+(H\rightarrow-H, F\leftrightarrow\hat F)\,,\\
g_4&=F_{ef}\cdot\nabla^k\hat F_{bg}H_{kcd}\nabla^cH^{eb}{}_h\mathcal R^{ghdf}+(H\rightarrow-H, F\leftrightarrow\hat F)\,,\\
g_5&=F_{ef}\cdot\nabla^k\hat F_{bg}H_{kcd}\nabla_hH^{cbe}\mathcal R^{ghdf}+(H\rightarrow-H, F\leftrightarrow\hat F)\,,\\
g_6&=F_{ef}\cdot\nabla^k\hat F_{bg}H_{kcd}\nabla^eH^{bc}{}_h\mathcal R^{gfhd}+(H\rightarrow-H, F\leftrightarrow\hat F)\,,\\
g_7&=\nabla^b\hat F^{gd}\cdot\nabla_eF_{gf}H^2_{abcd}R^{e(ac)f}+(H\rightarrow-H, F\leftrightarrow\hat F)\,,\\
g_8&=F_{be}\cdot\nabla^k\hat F^{bf}H_{kcd}\nabla_gH^d{}_{hf}\mathcal R^{chge}+(H\rightarrow-H, F\leftrightarrow\hat F)\,,\\
g_9&=F_{be}\cdot\nabla^f\hat F^{bk}H_{kcd}\nabla_gH^d{}_{hf}\mathcal R^{chge}+(H\rightarrow-H, F\leftrightarrow\hat F)\,,\\
g_{10}&=F_{be}\cdot\nabla^f\hat F^{bk}H_{kcd}\nabla^gH^{dhe}\mathcal R^c{}_{hgf}+(H\rightarrow-H, F\leftrightarrow\hat F)\,.
\end{split}
\end{equation}
In terms of these the terms in the Lagrangian we have discussed so far give
\begin{equation}
-\frac{1}{12}t_8t_8H^2\mathcal R^3+32l_1
\rightarrow
32f_1%
-32f_2%
-32f_3%
-32g_1%
-32g_6%
-32g_9
+32g_{10}
+\ldots\,,
\end{equation}
where the ellipsis denotes the terms we are dropping for the moment. We know from \cite{Wulff:2021fhr} that the Lagrangian also contains the following two terms of the form $H^2\mathcal R^3$ (there are more terms of this form but they don't contribute to the terms we are considering here)
\begin{equation}
\frac23\varepsilon_5H^2\varepsilon_5\mathcal R^3
+\frac{1}{2(3!)^3}(\varepsilon_9\varepsilon_9)'[H^2]\mathcal R^3
\end{equation}
where we have defined
\begin{equation}
\varepsilon_5H^2\varepsilon_5\mathcal R^3
=
-\frac{5!}{2}H^{[a_1a_2a_3}H^{a_4a_5]}{}_b\mathcal R^{bc}{}_{a_1e}\mathcal R_{cda_2a_3}\mathcal R^{de}{}_{a_4a_5}
+(H\rightarrow-H)\,.
\end{equation}
Upon dimensional reduction these give the $O(d,d)$-violating terms
\begin{equation}
\begin{split}
\frac23\varepsilon_5H^2\varepsilon_5\mathcal R^3
&\rightarrow
32f_2%
+32g_3%
-32g_6%
+\ldots\,,
\\
\frac{1}{2(3!)^3}(\varepsilon_9\varepsilon_9)'[H^2]\mathcal R^3
&\rightarrow
-32f_1
+32f_3
-32g_1
+32g_2
+32g_3
-16g_4
-16g_5
\\
&\quad
-16g_6
+32g_9
+\ldots\,.
\end{split}
\end{equation}
While the contributions to the $f$-terms cancel, as they must, it is clear that the contributions to the $g$-terms don't cancel. We need to add additional terms to the Lagrangian in order to cancel these. It is natural to consider terms of a similar structure as those needed at first order in $H$, i.e.
\begin{equation}
\begin{split}
\frac{1}{8(3!)^2}(\varepsilon_9\varepsilon_9)'[H^2](\nabla H)^2\mathcal R
&\rightarrow
-16g_1
+8g_2
+8g_3
-8g_4
-8g_5
-8g_6
+8g_8
+8g_{10}
+\ldots\,,
\\
\varepsilon_5H^2\varepsilon_5\nabla H\mathcal R\nabla H
&\rightarrow
-12g_4%
-12g_5%
-12g_6%
+\ldots\,,
\\
\varepsilon_5H^2\varepsilon_5\mathcal R\nabla H\nabla H
&\rightarrow
+12g_4
+12g_5
-36g_6
+\ldots\,.
\end{split}
\end{equation}
It is easy to see that this is still not enough and it is clear that we need to consider how we should complete the $l_1$ term we added already (\ref{eq:l1}), which did not have a similar structure to the other terms. This term is in fact part of a term with a familiar structure namely\footnote{A similar term with $\varepsilon\varepsilon$-structure would only give, after integration by parts, terms without derivatives on $F$ and terms with more traces, both of which we are dropping for the moment.}
\begin{equation}
t_8t_8H^2(\nabla H)^2\mathcal R\,.
\end{equation}
Explicitly we have
\begin{equation}
\frac{1}{64}t_8t_8H^2(\nabla H)^2\mathcal R
%
%
=
l_1
+2l_2
+2l_3
+4l_4
+\ldots
\end{equation}
where $l_1$ was defined in (\ref{eq:l1}) and
\begin{equation}
\begin{aligned}
l_2&=(H^2)^{ab}{}_{ef}(\nabla H)_{bc}{}^{fg}(\nabla H)^{cd}{}_{gh}\mathcal R_{da}{}^{he}\,,\\
l_3&=(H^2)^{ab}{}_{ef}(\nabla H)^{cdfg}(\nabla H)_{dagh}\mathcal R_{bc}{}^{he}\,,\\
l_4&=(H^2)^{ab}{}_{ef}(\nabla H)_{bc}{}^{he}(\nabla H)^{cdfg}\mathcal R_{Sdagh}
\end{aligned}
\end{equation}
and the ellipsis denotes terms with more traces. However, it turns out that $l_4$ produces terms which are of a different structure to the ones we want to cancel, so adding the term $t_8t_8H^2(\nabla H)^2\mathcal R$ to the Lagrangian does not work. Let us nevertheless consider the terms $l_2$ and $l_3$. Upon dimensional reduction they give the $O(d,d)$-violating terms
\begin{equation}
l_2\rightarrow-\frac12g_7+\ldots\,,\qquad
l_3\rightarrow-\frac12g_3-\frac12g_7-\frac12g_8+\ldots\,.
\label{eq:l23}
\end{equation}
Remarkably, with just these two terms we can cancel the coefficients of all terms but $g_3$ and $g_6$. A contribution to these terms comes, among others, from the following simple term
\begin{equation}
l_5=H^2_{abcd}\nabla^gH^{hae}\nabla_gH_h{}^{cf}\mathcal R_S^{bd}{}_{ef}
\label{eq:l5}
\end{equation}
with
\begin{equation}
l_5\rightarrow-2g_3+2g_6-2g_{10}+\ldots\,.
\end{equation}
Finally, the extra $g_{10}$ contribution can be canceled by the term
\begin{equation}
l_6=H^2_{abcd}\nabla_eH^{agh}\nabla_fH^c{}_{gh}\mathcal R_S^{bdef}\,,
\label{eq:l6}
\end{equation}
with
\begin{equation}
l_6\rightarrow-4g_{10}+\ldots\,.
\end{equation}

We have found that all the $O(d,d)$-violating terms we are considering cancel by adding to the Lagrangian the terms
\begin{equation}
\frac{a}{8(3!)^2}(\varepsilon_9\varepsilon_9)'[H^2](\nabla H)^2\mathcal R
+b\varepsilon_5H^2\varepsilon_5\nabla H\mathcal R\nabla H
+c\varepsilon_5H^2\varepsilon_5\mathcal R\nabla H\nabla H
+dl_2
+el_3
+fl_5
+gl_6
\end{equation}
for the choice
\begin{equation}
a=-4\,,\quad
b=\frac43\,,\quad
c=0\,,\quad
d=64\,,\quad
e=-64\,,\quad
f=32\,,\quad
g=-16\,.
\end{equation}
We can cast these terms in a nicer form by defining
\begin{equation}
\begin{split}
\hat t_8M_1M_2M_3M_4=
-8\tr(M_1M_2M_3M_4)
+8\tr(M_1M_4M_2M_3)
+8\tr(M_1M_3M_4M_2)\,.
\end{split}
\end{equation}
Note the change of sign of the first term compared to $t_8$ (\ref{eq:t8}). Then we have
\begin{equation}
\frac12\hat t_8\hat t_8H^2(\nabla H)^2\mathcal R=32l_1+64l_2-64l_3\,,
\label{eq:t8hatterms}
\end{equation}
which is precisely the combination of terms we need.

Furthermore we note that the $\varepsilon_5$-terms combine to
\begin{equation}
\begin{split}
\frac23\varepsilon_5H^2\varepsilon_5\mathcal R^3
+\frac43&\varepsilon_5H^2\varepsilon_5\nabla H\mathcal R\nabla H
\\
&=
2\varepsilon_4H^2\varepsilon_4\mathcal R^2\mathcal R
-4\varepsilon_4H^2\varepsilon_4(\nabla H)^2\mathcal R
\\
&\quad
+\frac{16(4!)}{3}H_{a_1a_2a_3}H_{a_4a_5}{}^b(\nabla H)_{be}{}^{a_4c}(\nabla H)_{cd}{}^{[a_1a_2}\mathcal R_S^{a_3a_5]de}\,,
\end{split}
\end{equation}
where we may drop the last term since it contains more traces and we didn't fix those terms yet. The first and second term are defined in (\ref{eq:e4}).

We are now ready to repeat the calculations including the terms we dropped, in order to find the remaining missing terms of fifth order in the fields.

\subsection{Determining the remaining \texorpdfstring{$H^2(\nabla H)^2\mathcal R$}{H2dH2R}-terms}
We now repeat the calculation including all terms, expect terms of sixth order and higher in the fields. Including all terms of the form $\mathcal R^4$, $H^2\mathcal R^3$ as well as the terms of the form $H^2(\nabla H)^2\mathcal R$ we have found above the Lagrangian takes the form
\begin{equation}
L'+32l_5-16l_6\,,
\label{eq:Lprimeplus}
\end{equation}
where
\begin{equation}
\begin{split}
L'
&=
\frac{1}{4!}t_8t_8\mathcal R^4
-\frac{1}{2(3!)}t_8t_8H^2\mathcal R^3
+\frac12\hat t_8\hat t_8H^2(\nabla H)^2\mathcal R
+\frac{1}{4(4!)}(\varepsilon_8\varepsilon_8)'\mathcal R^4
\\
&\quad
+\frac{1}{2(3!)^3}(\varepsilon_9\varepsilon_9)'H^2\mathcal R^3
-\frac{1}{2(3!)^2}H^2(\varepsilon_6\varepsilon_6)'\mathcal R^3
+\frac{1}{2(3!)^3}(\varepsilon_9\varepsilon_9)'[H^2]\mathcal R^3
\\
&\quad
-\frac{1}{2(3!)^2}(\varepsilon_9\varepsilon_9)'[H^2](\nabla H)^2\mathcal R
+2\varepsilon_4H^2\varepsilon_4\mathcal R^2\mathcal R
-4\varepsilon_4H^2\varepsilon_4(\nabla H)^2\mathcal R\,.
\end{split}
\label{eq:Lprime}
\end{equation}
We will first compute the $O(d,d)$-violating terms quadratic in the KK field strengths which arise from dimensional reduction of $L'$. For the first three terms the calculation follows the same steps as before. The details are given in appendix \ref{app:t8-red}. The reduction of the other terms are given in appendix \ref{app:eeR4-red}--\ref{app:e4e4-red}. Using these results we have
\begin{equation}
\begin{split}
L'
&\rightarrow
\frac12t_8t_8\hat F\cdot F(\nabla H)^2\mathcal R_S
-\frac18(\varepsilon_8\varepsilon_8)'\hat F\cdot F(\nabla H)^2\mathcal R_S
+k_3^{(F^2H^2\mathcal R)}
+2k_5^{(F^2H^2\mathcal R)}
\\
&\quad
+k_4^{(F\nabla FH\nabla H\mathcal R)}
+k_4^{(F^2(\nabla H)^2\mathcal R)}
+k_4^{(\mathrm{e.o.m.})}
+(H\rightarrow-H, F\leftrightarrow\hat F)
+\mathcal O(H^3)
+\mathrm{H.O.}\,,
\end{split}
\end{equation}
where $k_3$ is given in (\ref{eq:k3H2}) and $k_5$ in (\ref{eq:k5}) and we have defined
{
\allowdisplaybreaks
\begin{align}
k_4^{(F\nabla FH\nabla H\mathcal R)}
&=
64F^{ef}\cdot\nabla^g\hat F_b{}^kH_{kcd}(\nabla H)_{he}{}^{bc}\mathcal R_{gf}{}^{dh}
+64F_{ef}\cdot\nabla_g\hat F^{bk}H_{kcd}(\nabla H)^{df}{}_{bh}\mathcal R^{ghce}
\nonumber\\
&\quad
+64F_{ef}\cdot\nabla^k\hat F_{bg}H_{kcd}(\nabla H)_h{}^{bec}\mathcal R^{dhfg}
+32F_{ef}\cdot\nabla^k\hat F_{bg}H_{kcd}(\nabla H)_h{}^{ebc}\mathcal R^{ghfd}
\nonumber\\
&\quad
-32F_{ef}\cdot\nabla^k\hat F_{bg}H_{kcd}(\nabla H)^{fdg}{}_h\mathcal R^{bche}
+32F_{ef}\cdot\nabla^k\hat F^{eb}H_{kcd}(\nabla H)_{bgh}{}^d\mathcal R^{fhcg}
\nonumber\\
&\quad
-32F_{ef}\cdot\nabla^k\hat F^{eb}H_{kcd}(\nabla H)^{gcfh}\mathcal R^d{}_{hbg}
+32F^{ef}\cdot\nabla_g\hat F^{cd}H_{kcd}(\nabla H)_{be}{}^{gh}\mathcal R_{fh}{}^{bk}
\nonumber\\
&\quad
-16F_{ef}\cdot\nabla^g\hat F^{bk}H_{kcd}\nabla_bH^d{}_{hg}\mathcal R^{efch}
+16F_{ef}\cdot\nabla^g\hat F^{bk}H_{kcd}\nabla_bH^e{}_{hg}\mathcal R^{hfcd}
\nonumber\\
&\quad
-16F_{ef}\cdot\nabla^e\hat F_{bg}H_{kcd}(\nabla H)^{bfk}{}_h\mathcal R^{ghcd}
-16F^{ef}\cdot\nabla_e\hat F_{bg}H_{kcd}(\nabla H)^{bk}{}_{fh}\mathcal R^{cdgh}
\nonumber\\
&\quad
+16F^{ef}\cdot\nabla_e\hat F_{bg}H_{kcd}(\nabla H)^{ghcd}\mathcal R_{fh}{}^{bk}
+16F_{ef}\cdot\nabla_e\hat F_{bg}H_{kcd}(\nabla H)^{ghcd}\mathcal R_h{}^{kfb}
\nonumber\\
&\quad
-16F_{ef}\cdot\nabla_g\hat F^{bk}H_{kcd}(\nabla H)_{hb}{}^{cd}\mathcal R^{ghef}
-8F_{ef}\cdot\nabla_h\hat F_{bg}H_{kcd}\nabla^kH^{ebg}\mathcal R^{hfcd}
\nonumber\\
&\quad
-16F_{ef}\cdot\nabla^k\hat F_{gh}H_{kcd}(\nabla H)_b{}^{egh}\mathcal R^{cdbf}
+16F_{ef}\cdot\nabla^k\hat F_{gh}H_{kcd}(\nabla H)_b{}^{fcd}\mathcal R^{ghbe}
\nonumber\\
&\quad
+8F_{ef}\cdot\nabla_b\hat F_{gh}H_{kcd}(\nabla H)^{ekgh}\mathcal R^{cdfb}
-8F_{ef}\cdot\nabla_b\hat F_{gh}H_{kcd}(\nabla H)^{fbcd}\mathcal R^{ghek}
\nonumber\\
&\quad
-8F_{ef}\cdot\nabla^k\hat F_{gh}H_{kcd}(\nabla H)_b{}^{cef}\mathcal R^{bdgh}
-8F_{ef}\cdot\nabla^k\hat F_{gh}H_{kcd}(\nabla H)_b{}^{dgh}\mathcal R^{efbc}
\nonumber\\
&\quad
+8F^{ef}\cdot\nabla^k\hat F_e{}^bH_{kcd}(\nabla H)_{ghbf}\mathcal R^{cdgh}
+8F_{ef}\cdot\nabla^k\hat F^e{}_bH_{kcd}(\nabla H)^{cdgh}\mathcal R^{bf}{}_{gh}
\nonumber\\
&\quad
+8F_{ef}\cdot\nabla^e\hat F_{gh}H_{kcd}(\nabla H)^{ghbk}\mathcal R_b{}^{fcd}
+8F_{ef}\cdot\nabla^e\hat F_{gh}H_{kcd}(\nabla H)^{bfcd}\mathcal R_b{}^{kgh}
\nonumber\\
&\quad
+16F_{ef}\cdot\nabla^e\hat F_{gh}H_{kcd}(\nabla H)^{kbgh}\mathcal R^{cdf}{}_b
-16F_{ef}\cdot\nabla^e\hat F_{gh}H_{kcd}(\nabla H)^{fbcd}\mathcal R^{ghk}{}_b
\nonumber\\
&\quad
+8F_{ef}\cdot\nabla_b\hat F^{cd}H_{kcd}(\nabla H)^{bk}{}_{gh}\mathcal R^{efgh}
-8F_{ef}\cdot\nabla_b\hat F_{gh}H_{kcd}(\nabla H)^{bkef}\mathcal R^{cdgh}
\nonumber\\
&\quad
-8F_{ef}\cdot\nabla_b\hat F_{gh}H_{kcd}(\nabla H)^{cdgh}\mathcal R^{efbk}
-4F_{ef}\cdot\nabla_b\hat F_{gh}H_{kcd}(\nabla H)^{bkgh}\mathcal R^{cdef}
\nonumber\\
&\quad
+4F_{ef}\cdot\nabla_b\hat F_{gh}H_{kcd}(\nabla H)^{efcd}\mathcal R^{ghbk}
\end{align}
}
and
{
\allowdisplaybreaks
\begin{align}
k_4^{(F^2(\nabla H)^2\mathcal R)}
&=
-32\hat F_{ab}\cdot F_{ef}\nabla_cH^{bfg}(\nabla H)^{hacd}\mathcal R_S^e{}_{hgd}
-32\hat F_{ab}\cdot F_{ef}\nabla_cH^{bfg}(\nabla H)^e{}_{hgd}\mathcal R^{hacd}
\nonumber\\
&\quad
-32\hat F_{ab}\cdot F_{ef}\nabla_cH^{agh}(\nabla H)_g{}^{bed}\mathcal R^{cf}{}_{dh}
-32\hat F_{ab}\cdot F_{ef}\nabla_cH^{agh}(\nabla H)^{cf}{}_{dh}\mathcal R_{Sg}{}^{bed}
\nonumber\\
&\quad
+32\hat F^{ab}\cdot F_{ef}\nabla_cH_{agh}(\nabla H)_{bd}{}^{fg}\mathcal R^{cdhe}
+32\hat F^{ab}\cdot F_{ef}\nabla_cH_{agh}(\nabla H)^{cdhe}\mathcal R_{Sbd}{}^{fg}
\nonumber\\
&\quad
+32\hat F^{ae}\cdot F_{eb}\nabla^fH_{acd}(\nabla H)^{bghc}\mathcal R_{fhg}{}^d
+32\hat F^{ae}\cdot F_{eb}\nabla^fH_{acd}(\nabla H)_{fhg}{}^d\mathcal R_S^{bghc}
\nonumber\\
&\quad
-16\hat F^{ab}\cdot F^{ef}(\nabla H)_{abcd}(\nabla H)_{eh}{}^{gc}\mathcal R_{fg}{}^{hd}
-16\hat F^{ab}\cdot F^{ef}(\nabla H)_{abcd}(\nabla H)_{fg}{}^{hd}\mathcal R_{Seh}{}^{gc}
\nonumber\\
&\quad
-16\hat F^{ab}\cdot F_{ef}\nabla_cH_{agh}(\nabla H)^{degh}\mathcal R_{bd}{}^{fc}
-16\hat F^{ab}\cdot F_{ef}\nabla_cH_{agh}(\nabla H)_{bd}{}^{fc}\mathcal R_S^{degh}
\nonumber\\
&\quad
-16\hat F^{ab}\cdot F_{ef}\nabla_gH_a{}^{cd}(\nabla H)_{bd}{}^{gh}\mathcal R^{ef}{}_{ch}
-16\hat F^{ab}\cdot F_{ef}\nabla_gH_a{}^{cd}(\nabla H)^{ef}{}_{ch}\mathcal R_{Sbd}{}^{gh}
\nonumber\\
&\quad
+16\hat F_{ab}\cdot F_{ef}\nabla_hH_{kcd}(\nabla H)^{aecd}\mathcal R^{fhbk}
+16\hat F_{ab}\cdot F_{ef}\nabla_hH_{kcd}(\nabla H)^{fhbk}\mathcal R_S^{aecd}
\nonumber\\
&\quad
-16\hat F_{ab}\cdot F_{ef}\nabla^aH_{kcd}(\nabla H)_h{}^{ecd}\mathcal R^{bfkh}
-16\hat F_{ab}\cdot F_{ef}\nabla^aH_{kcd}(\nabla H)^{bfkh}\mathcal R_{Sh}{}^{ecd}
\nonumber\\
&\quad
+8\hat F^{ab}\cdot F_{ef}\nabla_aH_{kcd}(\nabla H)^{hecd}\mathcal R_{bh}{}^{fk}
+8\hat F^{ab}\cdot F_{ef}\nabla_aH_{kcd}(\nabla H)_{bh}{}^{fk}\mathcal R_S^{hecd}
\nonumber\\
&\quad
+8\hat F^{ab}\cdot F_{ef}\nabla_aH_{kcd}(\nabla H)^{cdhe}\mathcal R^{fk}{}_{bh}
+8\hat F^{ab}\cdot F_{ef}\nabla_aH_{kcd}(\nabla H)^{fk}{}_{bh}\mathcal R_S^{cdhe}
\nonumber\\
&\quad
+8\hat F_{ab}\cdot F^{ac}\nabla^eH^b{}_{cd}(\nabla H)_{efgh}\mathcal R^{ghfd}
+8\hat F_{ab}\cdot F^{ac}\nabla^eH^b{}_{cd}(\nabla H)^{ghfd}\mathcal R_{Sefgh}
\nonumber\\
&\quad
+8\hat F_{ab}\cdot F^a{}_e\nabla_gH_{kcd}(\nabla H)^{hkbe}\mathcal R^{cdg}{}_h
+8\hat F_{ab}\cdot F^a{}_e\nabla_gH_{kcd}(\nabla H)^{cdg}{}_h\mathcal R_S^{hkbe}
\nonumber\\
&\quad
-8\hat F^{ab}\cdot F_{ag}\nabla_hH_{kcd}(\nabla H)^{ghke}\mathcal R_{be}{}^{cd}
-8\hat F^{ab}\cdot F_{ag}\nabla_hH_{kcd}(\nabla H)_{be}{}^{cd}\mathcal R_S^{ghke}
\nonumber\\
&\quad
-8\hat F^{ab}\cdot F_{ag}\nabla_hH_{kcd}(\nabla H)^{kegh}\mathcal R^{cd}{}_{be}
-8\hat F^{ab}\cdot F_{ag}\nabla_hH_{kcd}(\nabla H)^{cd}{}_{be}\mathcal R_S^{kegh}
\nonumber\\
&\quad
-16\hat F_{ab}\cdot F^{ef}\nabla_aH^k{}_{cd}(\nabla H)^{bhcd}\mathcal R_{khef}
-16\hat F_{ab}\cdot F^{ef}\nabla_aH^k{}_{cd}(\nabla H)_{efkh}\mathcal R^{bhcd}
\nonumber\\
&\quad
-4\hat F_{ab}\cdot F_{ef}\nabla_hH_{kcd}(\nabla H)^{bkef}\mathcal R^{cdha}
-4\hat F_{ab}\cdot F_{ef}\nabla_hH_{kcd}(\nabla H)^{cdha}\mathcal R_S^{bkef}
\nonumber\\
&\quad
-12\hat F_{ab}\cdot F_{ef}\nabla_hH_{kcd}(\nabla H)^{efbk}\mathcal R^{hacd}
-12\hat F_{ab}\cdot F_{ef}\nabla_hH_{kcd}(\nabla H)^{hacd}\mathcal R_S^{efbk}
\nonumber\\
&\quad
-8\hat F^{ab}\cdot F_{ef}\nabla_gH_{acd}(\nabla H)_{ghef}\mathcal R_{bh}{}^{cd}
-8\hat F^{ab}\cdot F_{ef}\nabla_gH_{acd}(\nabla H)_{bh}{}^{cd}\mathcal R_S^{ghef}
\nonumber\\
&\quad
+8\hat F^{ab}\cdot F_{ef}(\nabla H)_{abcd}(\nabla H)^{fc}{}_{gh}\mathcal R^{degh}
+8\hat F^{ab}\cdot F_{ef}(\nabla H)_{abcd}(\nabla H)^{degh}\mathcal R_S^{fc}{}_{gh}
\nonumber\\
&\quad
+8\hat F^{ab}\cdot F_{ae}\nabla_fH_{bcd}(\nabla H)^{efgh}\mathcal R_{gh}{}^{cd}
+8\hat F^{ab}\cdot F_{ae}\nabla_fH_{bcd}(\nabla H)_{gh}{}^{cd}\mathcal R_S^{efgh}
\nonumber\\
&\quad
-4\hat F_{ab}\cdot F_{ef}(\nabla H)_{cdgh}(\nabla H)^{abgh}\mathcal R^{efcd}
-4\hat F_{ab}\cdot F_{ef}(\nabla H)_{cdgh}(\nabla H)^{efcd}\mathcal R_S^{abgh}\,,
\end{align}
}
while the terms proportional to the lowest order equations of motion, (\ref{eq:eom1}) and (\ref{eq:eom2}), are
\begin{equation}
\begin{split}
k_4^{(\mathrm{e.o.m.})}
&=
k_3^{(\mathrm{e.o.m.})}
-\frac14k_5^{(\mathrm{e.o.m.})}
+\frac{1}{96}k_6^{(\mathrm{e.o.m.})}
+\frac{1}{2(3!)^3}k_7^{(\mathrm{e.o.m.})}
\\
&\quad
+32F_{ef}\cdot\hat F_{bg}\mathcal R_h{}^{ebc}\mathcal R^{fgdh}\mathbbm B_{cd}
-16F_{ef}\cdot\hat F_{bg}\mathcal R_h{}^{bec}\mathcal R^{fhgd}\mathbbm B_{cd}
\\
&\quad
+16F^{ef}\cdot\hat F_{eb}\mathcal R_{fhg}{}^c\mathcal R^{bgdh}\mathbbm B_{cd}
-4F_{ef}\cdot\hat F^e{}_b\mathcal R^{bf}{}_{gh}\mathcal R^{ghcd}\mathbbm B_{cd}
\\
&\quad
-8F_{ef}\cdot\hat F_{bg}\mathcal R^{efb}{}_h\mathcal R^{ghcd}\mathbbm B_{cd}
+4F_{ef}\cdot\hat F_{gh}\mathcal R^{efc}{}_b\mathcal R^{ghdb}\mathbbm B_{cd}
\\
&\quad
+8F_{ef}\cdot\hat{\mathbbm A}^bH_{kcd}\mathcal R_{hb}{}^{ek}\mathcal R^{fhcd}%
-16F_{ef}\cdot\hat{\mathbbm A}_gH_{kcd}\mathcal R^{fgk}{}_h\mathcal R^{hecd}%
\\
&\quad
+8\mathbbm A^f\cdot\hat F_{bg}H_{kcd}\mathcal R^{hbcd}\mathcal R_{fh}{}^{gk}%
-8\mathbbm A^b\cdot\hat F_{ef}H_{kcd}\mathcal R^{efkh}\mathcal R_{hb}{}^{cd}%
\\
&\quad
+8F_{ef}\cdot\hat{\mathbbm A}^gH_{kcd}\mathcal R^{efkh}\mathcal R_{gh}{}^{cd}
-32F_{ef}\cdot\hat F_{bg}H_{kcd}\mathcal R^{fgcd}\nabla^{[b}\mathbbm G^{k]e}
\\
&\quad
+32F_{ef}\cdot\hat F^{eb}H_{kcd}\mathcal R_{bg}{}^{cd}\nabla^{[g}\mathbbm G^{k]f}
-16F_{ef}\cdot\hat F^{be}H_{kcd}\mathcal R^{hkf}{}_b\nabla^c\mathbbm G^d{}_h
\\
&\quad
+32F_{ef}\cdot\hat F_{gh}H_{kcd}\mathcal R^{efkg}\nabla^c\mathbbm G^{dh}
-8F_{ef}\cdot\hat F_{gh}H_{kcd}\mathcal R^{efcd}\nabla^g\mathbbm G^{hk}
+\mathcal O(H^2)\,,
\end{split}
\label{eq:k4eom}
\end{equation}
with $k_3$, $k_5$, $k_6$ and $k_7$ given in (\ref{eq:k3eom}), (\ref{eq:k5eom}), (\ref{eq:k6eom}) and (\ref{eq:k7eom}) respectively. In the above calculation we used the fact that
\begin{equation}
\begin{split}
t_8t_8\hat F\cdot F&\nabla H\mathcal R_S^2
-\frac14(\varepsilon_8\varepsilon_8)'\hat F\cdot F\nabla H\mathcal R_S^2
\\
&=
128\hat F_{ab}\cdot F_{ef}\nabla_cH^{bfg}\mathcal R_S^e{}_{hgd}\mathcal R_S^{hacd}
+128\hat F_{ab}\cdot F_{ef}\nabla_cH^a{}_{gh}\mathcal R_S^{bgf}{}_d\mathcal R_S^{cedh}
\\
&\quad
-128\hat F_{ab}\cdot F_{ef}\nabla^cH^a{}_{gh}\mathcal R_S^{bdfg}\mathcal R_{Scd}{}^{he}
+64\hat F_{ab}\cdot F^{ef}\nabla^cH^{agh}\mathcal R_{Sdegh}\mathcal R_S^{bd}{}_{fc}
\\
&\quad
+64\hat F^{ab}\cdot F^{ef}(\nabla H)_{abcd}\mathcal R_{Seh}{}^{gc}\mathcal R_{Sfg}{}^{hd}
-32\hat F_{ab}\cdot F^{ef}(\nabla H)_{cdef}\mathcal R_S^{da}{}_{gh}\mathcal R_S^{bcgh}
\\
&\quad
-64\hat F_{ab}\cdot F_{ef}\nabla_cH_{dgh}\mathcal R_S^{aegh}\mathcal R_S^{bcfd}
-128\hat F_{ab}\cdot F_{ef}\nabla^aH_{cgh}\mathcal R_{Sd}{}^{egh}\mathcal R_S^{bdfc}
\\
&\quad
+64\hat F_{ab}\cdot F_{ef}\nabla^aH_{cgh}\mathcal R_S^{degh}\mathcal R_S^{bcf}{}_d
+64\hat F_{ab}\cdot F_{ef}\nabla_gH_h{}^{ac}\mathcal R_S^{dghb}\mathcal R_S^{ef}{}_{cd}
\\
&\quad
+64\hat F_{ab}\cdot F_{ef}\nabla_dH_{cgh}\mathcal R_S^{dagh}\mathcal R_S^{bcef}
+32\hat F_{ab}\cdot F_{ef}\nabla^cH^{bgh}\mathcal R_S^{da}{}_{gh}\mathcal R_{Scd}{}^{ef}
\\
&\quad
+16\hat F_{ab}\cdot F_{ef}(\nabla H)_{cdgh}\mathcal R_S^{abgh}\mathcal R_S^{cdef}
-128\hat F^{ke}\cdot F_{eb}\nabla^fH_{kcd}\mathcal R_S^{bghc}\mathcal R_{Sfhg}{}^d
\\
&\quad
+64\hat F^{ab}\cdot F_{ag}\nabla_hH_{kcd}\mathcal R_S^{kegh}\mathcal R_{Sbe}{}^{cd}%
-32\hat F^a{}_b\cdot F_{ae}\nabla^gH_{kcd}\mathcal R_S^{behk}\mathcal R_{Sgh}{}^{cd}%
\\
&\quad
-32\hat F^{ab}\cdot F_a{}^c\nabla^dH_{cbe}\mathcal R_S^{efgh}\mathcal R_{Sdfgh}%
-32\hat F^{ak}\cdot F_{ae}\nabla_fH_{kcd}\mathcal R_S^{efgh}\mathcal R_S^{cd}{}_{gh}
\\
&\quad
%
+\mathcal O(H^3)
+\mathrm{H.O.}\,.
\end{split}
\end{equation}

All terms linear in $H$ have now canceled out but we are left with many terms quadratic in $H$. The last step is to find the terms we need to add to the Lagrangian for these to cancel. So far we did this only for the terms without additional traces. 

In the next step we use the fact that
\begin{equation}
\begin{split}
\frac12t_8t_8&\hat F\cdot F(\nabla H)^2\mathcal R_S
+\frac12\hat t_8\hat t_8\hat F\cdot F(\nabla H)^2\mathcal R_S
\\
&=
128\hat F_{ef}\cdot F^{ab}(\nabla H)^{fg}{}_{bc}(\nabla H)_{gh}{}^{cd}\mathcal R_S^{he}{}_{da}
+64\hat F_{ef}\cdot F^{ab}(\nabla H)^{he}{}_{bc}(\nabla H)^{fgcd}\mathcal R_{Sghda}
\\
&\quad
+64\hat F_{ef}\cdot F^{ab}(\nabla H)^{fg}{}_{da}(\nabla H)_{ghbc}\mathcal R_S^{hecd}
+64\hat F_{ef}\cdot F^{ab}(\nabla H)^{he}{}_{da}(\nabla H)^{fg}{}_{bc}\mathcal R_{Sgh}{}^{cd}
\\
&\quad
-16\hat F_{ef}\cdot F^{ab}(\nabla H)^{fg}{}_{ab}(\nabla H)_{ghcd}\mathcal R_S^{hecd}
-16\hat F_{ef}\cdot F^{ab}(\nabla H)^{he}{}_{ab}(\nabla H)^{fgcd}\mathcal R_{Sghcd}
\\
&\quad
-16\hat F_{ef}\cdot F^{ab}(\nabla H)_{ghab}(\nabla H)^{hecd}\mathcal R_S^{fg}{}_{cd}
-16\hat F_{ef}\cdot F^{ab}(\nabla H)^{fgcd}(\nabla H)_{ghcd}\mathcal R_S^{he}{}_{ab}
\\
&\quad
-8\hat F_{ef}\cdot F^{ab}(\nabla H)^{hecd}(\nabla H)^{fg}{}_{cd}\mathcal R_{Sghab}
-16\hat F_{ef}\cdot F^{ab}(\nabla H)^{ef}{}_{bc}(\nabla H)^{ghcd}\mathcal R_{Sghda}
\\
&\quad
-16\hat F_{ef}\cdot F^{ab}(\nabla H)^{ef}{}_{da}(\nabla H)_{ghbc}\mathcal R_S^{ghcd}
-16\hat F_{ef}\cdot F^{ab}(\nabla H)^{efcd}(\nabla H)_{ghda}\mathcal R_S^{gh}{}_{bc}
\\
&\quad
-16\hat F_{ef}\cdot F^{ab}(\nabla H)_{ghbc}(\nabla H)^{ghcd}\mathcal R_S^{ef}{}_{da}
-8\hat F_{ef}\cdot F^{ab}(\nabla H)_{ghda}(\nabla H)^{gh}{}_{bc}\mathcal R_S^{efcd}
\\
&\quad
+4\hat F_{ef}\cdot F^{ab}(\nabla H)_{ghab}(\nabla H)^{ef}{}_{cd}\mathcal R_S^{ghcd}
+4\hat F_{ef}\cdot F^{ab}(\nabla H)_{ghab}(\nabla H)^{ghcd}\mathcal R_S^{ef}{}_{cd}
\\
&\quad
+4\hat F_{ef}\cdot F^{ab}(\nabla H)^{efcd}(\nabla H)_{ghcd}\mathcal R_S^{gh}{}_{ab}
+2\hat F_{ef}\cdot F^{ab}(\nabla H)_{ghcd}(\nabla H)^{ghcd}\mathcal R_S^{ef}{}_{ab}
\end{split}
\end{equation}
and that
\begin{equation}
\begin{split}
-\frac18&(\varepsilon_8\varepsilon_8)'\hat F\cdot F(\nabla H)^2\mathcal R_S
\\
&=
32\hat F_{ef}\cdot F^{ab}(\nabla H)^{fg}{}_{bc}(\nabla H)^{he}{}_{da}\mathcal R_S^{cd}{}_{gh}
+64\hat F^{ef}\cdot F_{ab}(\nabla H)^{bc}{}_{dh}(\nabla H)^{gh}{}_{ce}\mathcal R_S^{da}{}_{fg}
\\
&\quad
+128\hat F_{ef}\cdot F^{ab}(\nabla H)^{fd}{}_{bg}(\nabla H)^{gh}{}_{ca}\mathcal R_S^{ec}{}_{dh}
-64\hat F^{ae}\cdot F_{ef}(\nabla H)^{fg}{}_{bc}(\nabla H)^{hb}{}_{da}\mathcal R_S^{cd}{}_{gh}
\\
&\quad
+64\hat F_{ef}\cdot F^{ab}(\nabla H)^{fg}{}_{bc}(\nabla H)^{cd}{}_{gh}\mathcal R_S^{he}{}_{da}
-64\hat F^{ae}\cdot F_{ef}(\nabla H)^{cd}{}_{gh}(\nabla H)^{hb}{}_{da}\mathcal R_S^{fg}{}_{bc}
\\
&\quad
-64\hat F^{ae}\cdot F_{ef}(\nabla H)^{fg}{}_{bc}(\nabla H)^{cd}{}_{gh}\mathcal R_S^{hb}{}_{da}
-32\hat F^{ab}\cdot F_{ef}(\nabla H)^{cg}{}_{ah}(\nabla H)^{dh}{}_{bg}\mathcal R_S^{ef}{}_{cd}
\\
&\quad
-64\hat F^{ab}\cdot F_{ef}(\nabla H)^{ef}{}_{cd}(\nabla H)^{cg}{}_{ah}\mathcal R_S^{dh}{}_{bg}
-32\hat F^{ab}\cdot F_{ef}(\nabla H)^{ec}{}_{gh}(\nabla H)^{gh}{}_{bd}\mathcal R_S^{df}{}_{ac}
\\
&\quad
+64\hat F_{ab}\cdot F^{ef}(\nabla H)^{ac}{}_{ed}(\nabla H)^{gh}{}_{cf}\mathcal R_S^{bd}{}_{gh}
+64\hat F_{ae}\cdot F^{ef}(\nabla H)^{ba}{}_{df}(\nabla H)^{cd}{}_{gh}\mathcal R_S^{gh}{}_{bc}
\\
&\quad
+32\hat F^{ae}\cdot F_{ef}(\nabla H)^{cb}{}_{gh}(\nabla H)^{gh}{}_{dc}\mathcal R_S^{df}{}_{ba}
-32\hat F^{ab}\cdot F_{ef}(\nabla H)^{cd}{}_{gh}(\nabla H)^{gh}{}_{da}\mathcal R_S^{ef}{}_{bc}
\\
&\quad
-32\hat F^{ab}\cdot F_{ef}(\nabla H)^{ef}{}_{bc}(\nabla H)^{gh}{}_{da}\mathcal R_S^{cd}{}_{gh}
-32\hat F^{ab}\cdot F_{ef}(\nabla H)^{ef}{}_{bc}(\nabla H)^{cd}{}_{gh}\mathcal R_S^{gh}{}_{da}
\\
&\quad
-32\hat F^{ae}\cdot F_{ef}(\nabla H)^{fb}{}_{cd}(\nabla H)^{cd}{}_{gh}\mathcal R_S^{gh}{}_{ba}
-16\hat F^{ae}\cdot F_{ef}(\nabla H)^{fb}{}_{cd}(\nabla H)^{gh}{}_{ba}\mathcal R_S^{cd}{}_{gh}
\\
&\quad
-16\hat F^{ab}\cdot F_{ab}(\nabla H)^{gh}{}_{ce}(\nabla H)^{cd}{}_{hf}\mathcal R^{ef}{}_{dg}
+2\hat F^{ab}\cdot F_{ef}(\nabla H)^{cd}{}_{gh}(\nabla H)^{gh}{}_{cd}\mathcal R_S^{ef}{}_{ab}
\\
&\quad
+8\hat F^{ab}\cdot F_{ef}(\nabla H)^{ef}{}_{cd}(\nabla H)^{cd}{}_{gh}\mathcal R_S^{gh}{}_{ab}
+4\hat F^{ab}\cdot F_{ef}(\nabla H)^{ef}{}_{cd}(\nabla H)^{gh}{}_{ab}\mathcal R_S^{cd}{}_{gh}
\\
&\quad
+4\hat F^{ab}\cdot F_{ab}(\nabla H)^{cd}{}_{ef}(\nabla H)^{ef}{}_{gh}\mathcal R^{gh}{}_{cd}\,.
\end{split}
\end{equation}
We need also the following nontrivial relation, which is proven by first integrating half of the term by parts,
\begin{equation}
\begin{split}
-64\hat F_{ab}\cdot F_{ef}\nabla^cH^{bfg}\nabla^dH^a{}_{ch}\mathcal R_{Sdg}{}^{he}
&=
-32\hat F_{ab}\cdot F_{ef}\nabla^cH^{bfg}\nabla^dH^a{}_{ch}\mathcal R_{Sdg}{}^{he}
\\
&\quad
-16\hat F_{ab}\cdot F_{ef}\nabla^cH^{bfg}\nabla^dH^{aeh}\mathcal R_{Schgd}
\\
&\quad
-32\hat F_{ab}\cdot F^e{}_cH^{bfg}\nabla_f\nabla^hH^{acd}\mathcal R_{Sehgd}
\\
&\quad
-32\hat F_{ab}\cdot F^{ec}H^{bfg}\nabla_f\nabla^aH_c{}^{hd}\mathcal R_{Sehgd}
\\
&\quad
-32\hat F_{ab}\cdot F^{ec}H^{bfg}\nabla_f\nabla_cH^{adh}\mathcal R_{Sehgd}
+\ldots
\\
&=
-16F_{ab}\cdot F_{ef}\nabla_cH^{aeg}\nabla_dH^{bfh}\mathcal R_S^{cd}{}_{gh}
\\
&\quad
-32F_{ab}\cdot F_{ef}\nabla_cH^{dga}\nabla^cH_d{}^{he}\mathcal R_S^{bf}{}_{gh}
+\ldots\,,
\end{split}
\end{equation}
where the ellipsis denotes (many) terms with additional traces, equation of motion terms and higher order terms in the number of fields. First we must take into account the last two terms in (\ref{eq:Lprimeplus}) which we did not yet include
\begin{equation}
32l_5-16l_6\,.
\end{equation}
Their reductions are given in (\ref{eq:l5red}) and (\ref{eq:l6red}). After a straightforward but long calculation one finds that only one more term with less than two traces is needed, namely the term
\begin{equation}
-32l_7
\end{equation}
where
\begin{equation}
l_7=H_{kcd}H_{aef}\nabla^gH^{hac}\nabla_gH_{hb}{}^d\mathcal R_S^{efkb}\,.
\end{equation}
The reduction of this term is given in (\ref{eq:l7red}). As for the terms with at least two traces one finds that many of these are needed. A relatively minimal set is given by the following 29 terms
\begin{equation}
\begin{aligned}
l_8&=H^2_{abcd}(\nabla H)^{acgh}(\nabla H)_{efgh}\mathcal R_S^{bdef} &&&
l_{23}&=H_{abe}H_{kcd}\nabla^eH^{kgh}(\nabla H)^c{}_{fgh}\mathcal R_S^{fdab}
\\
l_9&=H^2_{abcd}(\nabla H)^{abgh}(\nabla H)_{efgh}\mathcal R_S^{efcd} &&&
l_{24}&=\nabla^eH^2_{fc}H^{kcd}\nabla_kH^{fgh}\mathcal R_{Sdegh}
\\
l_{10}&=H^2_{abcd}(\nabla H)^{begh}(\nabla H)_{ef}{}^{cd}\mathcal R_S^{fa}{}_{gh} &&&
l_{25}&=\nabla^cH^2_{ef}H_{kcd}\nabla^eH^d{}_{gh}\mathcal R_S^{fkgh}
\\
l_{11}&=H^2_{abcd}\nabla^eH^{agh}\nabla_eH_{fgh}\mathcal R_S^{fbcd}&&&
l_{26}&=H^2_{ef}(\nabla H)^{ekcd}(\nabla H)_k{}^{fgh}\mathcal R_{Scdgh}
\\
l_{12}&=\nabla^e(H^2_{abcd}H^{agh})\nabla_eH_{fgh}\mathcal R_S^{fbcd}&&&
l_{27}&=H^2_{ef}\nabla_kH^{eab}(\nabla H)_{abgh}\mathcal R_S^{kfgh}
\\
l_{13}&=H_{abf}H_{kcd}\nabla^eH^{abk}\nabla_eH^{fgh}\mathcal R_S^{cd}{}_{gh}&&&
l_{28}&=H^2_{ef}\nabla^eH_{kcd}\nabla^fH^{kgh}\mathcal R_S^{cd}{}_{gh}
\\
l_{14}&=\nabla^kH^2_{ef}\nabla_k(H^e{}_{cd}H^{fgh})\mathcal R^{cd}{}_{gh}&&&
l_{29}&=\nabla_eH^2_{fk}H^{kcd}(\nabla H)_{cdgh}\mathcal R_S^{ghef}
\\
l_{15}&=\nabla^kH^{abe}\nabla_kH_{abf}H_{ecd}H^{fgh}\mathcal R^{cd}{}_{gh}&&&
l_{30}&=\nabla_kH^2_{ef}H^{kcd}\nabla_gH_{hcd}\mathcal R_S^{eghf}
\\
l_{16}&=H^2_{abcd}\nabla_eH^{cgh}\nabla_fH^d{}_{gh}\mathcal R_S^{abef}&&&
l_{31}&=\nabla^eH^2_{abcd}\nabla_eH^2\mathcal R_S^{abcd}
\\
l_{17}&=H_{abh}H_{kcd}\nabla_eH^{abg}\nabla_fH_g{}^{cd}\mathcal R_S^{efkh}&&&
l_{32}&=H^2_{cd}\nabla_aH_{efg}\nabla_bH^{efg}\mathcal R_S^{acdb}
\\
l_{18}&=H_{ab}{}^cH_{def}(\nabla H)_{ckgh}(\nabla H)^{efgh}\mathcal R_S^{kdab}&&&
l_{33}&=H^2\nabla_cH_{agh}\nabla_dH_b{}^{gh}\mathcal R_S^{abcd}
\\
l_{19}&=H^{abc}H_{def}(\nabla H)_{ck}{}^{gd}(\nabla H)^{hk}{}_{ab}\mathcal R_S^{ef}{}_{gh}&&&
l_{34}&=H^2\nabla_bH_{kcd}\nabla^bH^{kgh}\mathcal R_S^{cd}{}_{gh}
\\
l_{20}&=H^{abc}H_{def}(\nabla H)^{dk}{}_{ab}(\nabla H)_{ckgh}\mathcal R_S^{efgh}&&&
l_{35}&=\nabla^hH_{efg}\nabla^kH^{efg}H_{kcd}H_{abh}\mathcal R_S^{cdab}
\\
l_{21}&=H^{abc}H_{def}(\nabla H)^{ef}{}_{ab}(\nabla H)_{ckgh}\mathcal R_S^{dkgh}&&&
l_{36}&=H^2(\nabla H)_{abcd}(\nabla H)^{abgh}\mathcal R_S^{cd}{}_{gh}
\\
l_{22}&=H_{abc}H^{def}(\nabla H)_{dk}{}^{ab}(\nabla H)_{efgh}\mathcal R_S^{ckgh}&&&
\end{aligned}
\label{eq:l8to36}
\end{equation}
and after a long calculation one finds that the remaining $O(d,d)$-violating terms cancel if we add to the Lagrangian the combination
\begin{equation}
\begin{split}
L''
&=
32l_8
-8l_9
+32l_{10}
-8l_{11}
-8l_{12}
-8l_{13}
+2l_{14}
-4l_{15}
-4l_{16}
-8l_{17}
+16l_{18}
+64l_{19}
\\
&\quad
-16l_{20}
-16l_{21}
+8l_{22}
+16l_{23}
-8l_{24}
+8l_{25}
+16l_{26}
+8l_{27}
+4l_{28}
-4l_{29}
-4l_{30}
\\
&\quad
-\frac13l_{31}
-\frac43l_{32}
-\frac23l_{33}
-\frac23l_{34}
+\frac43l_{35}
+\frac43l_{36}\,.
\end{split}
\end{equation}
To summarize we have found that up to fifth order in fields the NSNS sector Lagrangian should take the form
\begin{equation}
L'+32l_5-16l_6-32l_7+L''\,,
\end{equation}
with $L'$ given in (\ref{eq:Lprime}). We can simplify this result, at least a little bit, as follows. First we define $\tilde t_8$ as in (\ref{eq:t8tilde}), i.e.
\begin{equation}
\begin{split}
\tilde t_8M_1M_2M_3M_4
&=
\hat t_8M_1M_2M_3M_4
+2\tr(M_1M_2)\tr(M_3M_4)
-2\tr(M_1M_3)\tr(M_4M_2)
\\
&\quad
+2\tr(M_1M_4)\tr(M_2M_3)\,.
\end{split}
\end{equation}
With this definition we have
\begin{equation}
\begin{split}
\tilde t_8\tilde t_8H^2(\nabla H)^2\mathcal R
&=
\hat t_8\hat t_8H^2(\nabla H)^2\mathcal R
-64H^2_{abcd}(\nabla H)^{becd}(\nabla H)_{ef}{}^{gh}\mathcal R_S^{fa}{}_{gh}
\\
&\quad
+64H^2_{abcd}(\nabla H)^{begh}(\nabla H)_{ef}{}^{cd}\mathcal R_S^{fa}{}_{gh}
+32H^2_{abcd}(\nabla H)^{fagh}(\nabla H)_{begh}\mathcal R_S^{efcd}
\\
&\quad
-8H^2_{abcd}(\nabla H)^{abgh}(\nabla H)_{ef}{}^{cd}\mathcal R_S^{ef}{}_{gh}
+4H^2_{abcd}(\nabla H)^{efgh}(\nabla H)_{efgh}\mathcal R_S^{abcd}\,.
%
\end{split}
\end{equation}
Furthermore, noting that
\begin{equation}
\nabla^2H_{abc}
=
-3R^{de}{}_{[ab}H_{c]de}
+\mbox{e.o.m. terms}
+\mathrm{H.O.}
\end{equation}
and that we may write
\begin{equation}
\begin{split}
l_{24}
&=
-H^2_{fc}\nabla^eH^{kcd}\nabla_kH^{fgh}\mathcal R_{Sdegh}
-\frac12H^2_{fk}\nabla_eH^{kcd}\nabla^eH^{fgh}\mathcal R_{Scdgh}
\\
&\quad
-\frac12\nabla_eH^2_{fk}H^{kcd}\nabla^eH^{fgh}\mathcal R_{Scdgh}
+\nabla_kH^2_{fc}H^{kcd}\nabla^eH^{fgh}\mathcal R_{Sdegh}
\\
&\quad
+\mbox{e.o.m. terms}
+\mathrm{H.O.}\,,
\end{split}
\end{equation}
where we can remove the equation of motion terms by field redefinitions, one finds with a little bit of work that the Lagrangian up to fifth order in fields can be written as in (\ref{eq:L}).

\vspace{4cm}

\section*{Acknowledgments}
This work is supported by the grant ``Dualities and higher derivatives'' (GA23-06498S) from the Czech Science Foundation (GA\v CR).

\newpage

\appendix

\section{Useful identities}
The torsionful Riemann tensor satisfies the following useful identities
\begin{equation}
\nabla^{(+)}_{[d}\mathcal R^{ef}{}_{ef]}=-H_{[de}{}^g\mathcal R^{ef}{}_{f]g}\,,
\qquad
\nabla^{(-)[d}\mathcal R^{ef]}{}_{ef}=H^{[de}{}_g\mathcal R^{f]g}{}_{ef}\,,
\label{eq:id1}
\end{equation}
\begin{equation}
\begin{split}
\nabla^{(-)[b}\nabla^{(+)}_e\hat F^{cd]}&=-\mathcal R^{[bc}{}_{eg}\hat F^{d]g}+H^{[bc}{}_g\nabla^{(+)}_e\hat F^{d]g}\,,
\\
\nabla^{(+)}_{[b}\nabla^{(-)e}F_{cd]}&=-\mathcal R^{eg}{}_{[bc}F_{d]g}-H_{[bc}{}^g\nabla^{(-)e}F_{d]g}\,,
\end{split}
\label{eq:id2}
\end{equation}
and
\begin{equation}
(\nabla^{(-)}_c-2\partial_c\Phi)\mathcal R^{bc}{}_{ef}+H^b{}_{cd}\mathcal R^{cd}{}_{ef}
=
-2\nabla^{(+)}_{[e}(\mathcal R^b{}_{f]}+2\nabla^{(-)b}\partial_{f]}\Phi)
+H_{cef}(\mathcal R^{bc}+2\nabla^{(-)b}\partial^c\Phi)\,,
\label{eq:id4}
\end{equation}
where the RHS is proportional to the lowest order equations of motion since
\begin{equation}
\mathcal R_{ab}+2\nabla^{(-)}_a\partial_b\Phi=\mathbbm G_{ab}+\frac12\mathbbm B_{ab}\,.
\end{equation}
The lowest order equations of motion are $\mathbbm G_{ab}=0$ and $\mathbbm B_{ab}=0$ with
\begin{equation}
\mathbbm B_{ab}=(\nabla^c-2\partial^c\Phi)H_{abc}\,,\qquad
\mathbbm G_{ab}=R_{ab}+2\nabla_a\partial_b\Phi-\frac14H^2_{ab}
\label{eq:eom1}
\end{equation}
and in the reduced theory we also have the equations of motion of the KK vectors $\mathbbm A_a=0$ and $\hat{\mathbbm A}_a=0$ with
\begin{equation}
\mathbbm A^b_{a'}=(\nabla^{(-)}_a-2\partial_a\Phi)F^{ab}_{a'}\,,\qquad \hat{\mathbbm A}^b_{a'}=(\nabla^{(+)}_a-2\partial_a\Phi)\hat F^{ab}_{a'}\,.
\label{eq:eom2}
\end{equation}
We did not introduce a notation for the dilaton equation of motion since it does not enter our calculations.

\section{Dimensional reduction and simplification of various terms}\label{app:dim-red}
In this appendix we collect the calculation of the $O(d,d)$-violating terms quadratic in the KK vectors from various terms that we need, using (\ref{eq:Hred}) and (\ref{eq:Rred}). We also simplify them as needed for our calculations.

\subsection{\texorpdfstring{$t_8$}{t8} and \texorpdfstring{$\hat t_8$}{t8hat} terms}\label{app:t8-red}
Using (\ref{eq:t8hatterms}), (\ref{eq:l1}) and (\ref{eq:l23}) the reduction of the $\hat t_8$ term gives
\begin{equation}
\begin{split}
\frac12\hat t_8\hat t_8H^2(\nabla H)^2\mathcal R
&\rightarrow
\frac12\hat t_8\hat t_8\hat F\cdot F(\nabla H)^2\mathcal R_S
+\hat t_8FH\nabla F\nabla H\mathcal R
+\hat t_8H^2(\nabla F)^2\mathcal R
\\
&\quad
+\frac14\hat t_8\hat t_8H^2(\nabla H)^2\hat F\cdot F
-\frac14\hat t_8\hat t_8H^2\nabla H\hat F\cdot F\mathcal R_S
\\
&\quad
-\frac14\hat t_8\hat t_8H^2\hat F\cdot F\nabla H\mathcal R_S
+(H\rightarrow-H, F\leftrightarrow\hat F)\,,
\end{split}
\end{equation}
where the last three terms are of sixth order in fields and will not be needed here, while the second term is defined as
\begin{equation}
\begin{split}
\hat t_8FH\nabla F\nabla H\mathcal R
&=
64\hat F^{bk}\cdot\nabla^{(-)}_dF^{a_1a_2}H_{ka_2a_3}(\nabla H)_{bc}{}^{a_3a_4}\mathcal R^{cd}{}_{a_4a_1}
\\
&\quad
+64\hat F^{bk}\cdot\nabla^{(-)}_dF^{a_1a_2}H_{ka_2a_3}(\nabla H)^{cd}{}_{a_4a_1}\mathcal R_{Sbc}{}^{a_3a_4}
\\
&\quad
+64\hat F^{bk}\cdot\nabla^{(-)}_dF^{a_1a_2}H_{ka_2a_3}(\nabla H)_{bca_4a_1}(\nabla H)^{cda_3a_4}
\\
&\quad
-64\hat F^b{}_k\cdot\nabla^{(-)}_dF^{a_1a_2}H^{ka_3a_4}(\nabla H)^{cd}{}_{a_2a_3}\mathcal R_{Sbca_4a_1}\,,
\end{split}
\label{eq:HFdFR3-1}
\end{equation}
and the third as
\begin{equation}
\begin{split}
\hat t_8H^2(\nabla F)^2\mathcal R
&=
-32\nabla^{(+)f}\hat F^{bc}\cdot\nabla^{(-)}_hF_{cd}H^2_{abef}\mathcal R_S^{heda}
+32\nabla^{(+)}_h\hat F^{bc}\cdot\nabla^{(-)f}F_{cd}H^2_{abef}\mathcal R_S^{heda}
\\
&\quad
-32\nabla^{(+)}_d\hat F^{fg}\cdot\nabla^{(-)b}F^{he}H^2_{abef}(\nabla H)^{da}{}_{gh}\,.
\end{split}
\label{eq:HFdFR3-2}
\end{equation}
Rather than simplify these terms by themselves we will first combine them with the $t_8$-terms in (\ref{eq:t8-red}) and then simplify the result. Firstly, (\ref{eq:HFdFR3-1}) combines with (\ref{eq:HFdFR2-1}) to give
\begin{equation}
\begin{split}
&32\hat F^{bk}\cdot\nabla^{(-)}_dF_{a_1a_2}H_{ka_3a_4}\mathcal R^{cda_2a_3}\mathcal R_{bc}{}^{a_4a_1}
-64\hat F^{bk}\cdot\nabla^{(-)}_dF_{a_1a_2}H_{ka_3a_4}(\nabla H)^{cda_2a_3}\mathcal R_{Sbc}{}^{a_4a_1}
\\
&\quad
+32\hat F^{bk}\cdot\nabla^{(-)}_dF^{a_1a_2}H_{ka_2a_3}\mathcal R^{cda_3a_4}\mathcal R_{bca_4a_1}
+32\hat F^{bk}\cdot\nabla^{(-)}_dF^{a_1a_2}H_{ka_2a_3}\mathcal R^{a_3a_4}{}_{bc}\mathcal R_{a_4a_1}{}^{cd}
\\
&\quad
+64\hat F^{bk}\cdot\nabla^{(-)}_dF^{a_1a_2}H_{ka_2a_3}(\nabla H)^{cda_3a_4}(\nabla H)_{bca_4a_1}
\\
&\quad
+64\hat F^{bk}\cdot\nabla^{(-)}_dF^{a_1a_2}H_{ka_2a_3}(\nabla H)^{a_3a_4}{}_{bc}(\nabla H)^{cd}{}_{a_4a_1}
+8\hat F^{be}\cdot\nabla^{(-)}_bF_{a_1a_2}\mathcal R^{cda_2a_3}H_{ea_3a_4}\mathcal R_{cd}{}^{a_4a_1}
\\
&\quad
+16\hat F_{be}\cdot\nabla^{(-)b}F_{a_1a_2}H^{ea_2a_3}\mathcal R^{cd}{}_{a_3a_4}\mathcal R_{cd}{}^{a_4a_1}
+(H\rightarrow-H, F\leftrightarrow\hat F)\,.
\end{split}
\end{equation}
Apart from the second, fifth and sixth terms this is the same as (\ref{eq:HFdFR2-1}), except that the order of indices on $\mathcal R$ is swapped in one term. The first term was dealt with in (\ref{eq:HFdFR2t1}). Dropping higher order terms it becomes
\begin{equation}
\begin{split}
&
32\nabla_d\hat F^{bk}\cdot F_{a_1a_2}H_{ka_3a_4}\mathcal R_{bc}{}^{a_1a_3}\mathcal R^{cda_4a_2}
+8\hat F^{bk}\cdot\nabla_bF_{a_1a_2}H_{ka_3a_4}\mathcal R_{cd}{}^{a_1a_3}\mathcal R^{cda_4a_2}
\\
&\quad
+32\hat F^{bk}\cdot F_{a_1a_2}\nabla_dH_{ka_3a_4}\mathcal R_{bc}{}^{a_1a_3}\mathcal R^{cda_4a_2}
+8\hat F^{bk}\cdot F_{a_1a_2}\nabla_bH_{ka_3a_4}\mathcal R_{cd}{}^{a_1a_3}\mathcal R^{cda_4a_2}
\\
&\quad
+32\hat F^{bk}\cdot F_{a_1a_2}H_{ka_3a_4}\mathcal R_{bc}{}^{a_1a_3}(\nabla_d-2\partial_d\Phi)\mathcal R^{cda_4a_2}
\\
&\quad
+8\hat{\mathbbm A}^k\cdot F_{a_1a_2}H_{ka_3a_4}\mathcal R_{cd}{}^{a_1a_3}\mathcal R^{cda_4a_2}
+\mathrm{H.O.}\,.
\end{split}
\end{equation}
Similarly, the second term is integrated by parts to give
\begin{equation}
\begin{split}
&
64\nabla_d\hat F^{bk}\cdot F_{a_1a_2}H_{ka_3a_4}(\nabla H)^{cda_2a_3}\mathcal R_{Sbc}{}^{a_4a_1}
+32\hat F^{bk}\cdot\nabla_bF_{a_1a_2}H_{ka_3a_4}(\nabla H)_{cd}{}^{a_2a_3}\mathcal R_S^{cda_4a_1}
\\
&\quad
+32\hat F^{bk}\cdot F_{a_1a_2}H_{ka_3a_4}\nabla_b(\nabla H)_{cd}{}^{a_2a_3}\mathcal R_S^{cda_4a_1}
+64\hat F^{bk}\cdot F_{a_1a_2}\nabla_dH_{ka_3a_4}(\nabla H)^{cda_2a_3}\mathcal R_{Sbc}{}^{a_4a_1}
\\
&\quad
+32\hat F^{bk}\cdot F_{a_1a_2}(\nabla H)_{bka_3a_4}(\nabla H)_{cd}{}^{a_2a_3}\mathcal R_S^{cda_4a_1}
\\
&\quad
+64\hat F^{bk}\cdot F_{a_1a_2}H_{ka_3a_4}(\nabla_d-2\partial_d\Phi)(\nabla H)^{cda_2a_3}\mathcal R_{Sbc}{}^{a_4a_1}
\\
&\quad
+32(\nabla_b-2\partial_b\Phi)\hat F^{bk}\cdot F_{a_1a_2}H_{ka_3a_4}(\nabla H)_{cd}{}^{a_2a_3}\mathcal R_S^{cda_4a_1}\,.
\end{split}
\end{equation}
Finally the third term gives, first using the Bianchi identity and then integrating one term by parts,
\begin{equation}
\begin{split}
&32\hat F^{bk}\cdot\nabla^{a_2}F_{a_1d}H_k{}^{a_2a_3}\mathcal R^{cd}{}_{a_3a_4}\mathcal R_{bc}{}^{a_4a_1}
-32F^{bk}\cdot\nabla^{a_1}\hat F^{a_2d}H_{ka_2a_3}\mathcal R_{bc}{}^{a_3a_4}\mathcal R^c{}_{da_4a_1}
\\
&\quad
+16\hat F^{bk}\cdot F^{da_2}H_{ka_2a_3}\nabla^{a_3}\mathcal R^c{}_{da_4a_1}\mathcal R_{bc}{}^{a_4a_1}
-32\hat F^{bk}\cdot F^{da_2}\nabla^{a_1}H_{ka_2a_3}\mathcal R^c{}_d{}^{a_3a_4}\mathcal R_{bca_4a_1}
\\
&\quad
-32\hat F^{bk}\cdot F^{da_2}H_{ka_2a_3}\mathcal R^c{}_d{}^{a_3a_4}(\nabla^{a_1}-2\partial^{a_1}\Phi)\mathcal R_{bca_4a_1}
+\mathrm{H.O.}\,,
\end{split}
\end{equation}
where the third term can be further simplified by integrating half of it by parts. Putting this together and simplifying we find that (\ref{eq:HFdFR3-1}) and (\ref{eq:HFdFR2-1}) reduce to
\begin{equation}
k_1^{(F\nabla\hat FH\mathcal R^2)}
+k_1^{(F\hat F\nabla H\mathcal R^2)}
+k_1^{(F^2H^2\mathcal R)}
+k_1^{(F\nabla\hat FH(\nabla H)^2)}
+k_1^{(\mathrm{e.o.m.})}
+(H\rightarrow-H, F\leftrightarrow\hat F)
+\mathrm{H.O.}\,,
\end{equation}
where
\begin{equation}
\begin{split}
k_1^{(F\nabla\hat FH\mathcal R^2)}
&=
32\hat F^{bk}\cdot\nabla^{a_2}F^{a_1}{}_dH_{ka_2a_3}\mathcal R^{cda_3a_4}\mathcal R_{bca_4a_1}
-32F_{a_1a_2}\cdot\nabla_d\hat F^{bk}H_{ka_3a_4}\mathcal R^{cda_2a_3}\mathcal R_{bc}{}^{a_4a_1}
\\
&\quad
+16\hat F^{bk}\cdot\nabla^{a_2}F_{ef}H_{ka_2a_3}\mathcal R_{a_4}{}^{cef}\mathcal R^{a_3a_4}{}_{bc}
+16\hat F^{bk}\cdot\nabla_bF^{a_1a_2}H_{ka_2a_3}\mathcal R^{cda_3a_4}\mathcal R_{cda_4a_1}
\\
&\quad
+8\hat F^{bk}\cdot\nabla_dF^{a_2a_3}H_{ka_2a_3}\mathcal R_{bca_4a_1}\mathcal R^{a_4a_1cd}\,,
\end{split}
\end{equation}
\begin{equation}
\begin{split}
k_1^{(F\hat F\nabla H\mathcal R^2)}
&=
-32\hat F^{bk}\cdot F_{a_1a_2}\nabla_dH_{ka_3a_4}\mathcal R_{bc}{}^{a_2a_3}\mathcal R^{cda_4a_1}
-32\hat F^{bk}\cdot F_d{}^{a_2}\nabla^{a_1}H_{ka_2a_3}\mathcal R^{cda_3a_4}\mathcal R_{bca_4a_1}
\\
&\quad
-8\hat F^{bk}\cdot F_{a_1a_2}\nabla_bH_{ka_3a_4}\mathcal R^{a_2a_3}{}_{cd}\mathcal R^{a_4a_1cd}\,,
\end{split}
\end{equation}
\begin{equation}
\begin{split}
k_1^{(F^2H^2\mathcal R)}
&=
32\hat F^{bk}\cdot F_{a_1a_2}H_{ka_3a_4}\nabla_b(\nabla H)_{cd}{}^{a_2a_3}\mathcal R_S^{cda_4a_1}
\\
&\quad
+32\hat F^{bk}\cdot F_d{}^{a_2}H_{ka_2a_3}\nabla^{a_3}(\nabla H)^{a_4a_1cd}\mathcal R_{a_4a_1bc}
\\
&\quad
+64F_{a_1a_2}\cdot\nabla_d\hat F^{bk}H_{ka_3a_4}(\nabla H)^{cda_2a_3}\mathcal R_{Sbc}{}^{a_4a_1}
\\
&\quad
+16\hat F^{bk}\cdot\nabla^{a_2}F_{ef}H_{ka_2a_3}\nabla_{a_4}H^{cef}\mathcal R^{a_3a_4}{}_{bc}
\\
&\quad
+64\hat F^{bk}\cdot F_{a_1a_2}\nabla_dH_{ka_3a_4}(\nabla H)^{cda_4a_1}\mathcal R_{Sbc}{}^{a_2a_3}
\\
&\quad
+32\hat F^{bk}\cdot\nabla_bF_{a_1a_2}H_{ka_3a_4}(\nabla H)_{cd}{}^{a_2a_3}\mathcal R_S^{cda_4a_1}
\\
&\quad
+64\hat F^{bk}\cdot F_{a_1a_2}(\nabla H)_{bka_3a_4}(\nabla H)_{cd}{}^{a_2a_3}\mathcal R_S^{cda_4a_1}\,,
\end{split}
\end{equation}
\begin{equation}
\begin{split}
k_1^{(F\nabla\hat FH(\nabla H)^2)}
&=
64\hat F^{bk}\cdot\nabla_dF^{a_1a_2}H_{ka_2a_3}(\nabla H)^{cda_3a_4}(\nabla H)_{bca_4a_1}
\\
&\quad
+64\hat F^{bk}\cdot\nabla_dF^{a_1a_2}H_{ka_2a_3}(\nabla H)^{a_3a_4}{}_{bc}(\nabla H)^{cda_4a_1}
\\
&\quad
+32\hat F^{bk}\cdot\nabla_dF^{a_2a_3}H_{ka_2a_3}(\nabla H)_{bca_4a_1}(\nabla H)^{cda_4a_1}
\end{split}
\end{equation}
and
\begin{equation}
\begin{split}
k_1^{(\mathrm{e.o.m.})}
&=
16\hat F^{bk}\cdot F_d{}^{a_2}\mathbbm B_{ka_2}(\nabla H)_{bca_4a_1}\mathcal R^{cda_4a_1}
-8\hat F^{bk}\cdot F_d{}^{a_2}\mathbbm B_{ka_2}\mathcal R^{cda_4a_1}\mathcal R_{bca_4a_1}
\\
&\quad
+32\hat F^{bk}\cdot F_{a_1a_2}H_{ka_3a_4}\mathcal R_{bc}{}^{a_1a_3}(\nabla_d-2\partial_d\Phi)\mathcal R^{cda_4a_2}
\\
&\quad
+8\hat{\mathbbm A}^k\cdot F_{a_1a_2}H_{ka_3a_4}\mathcal R_{cd}{}^{a_1a_3}\mathcal R^{cda_4a_2}
\\
&\quad
+64\hat F^{bk}\cdot F_{a_1a_2}H_{ka_3a_4}(\nabla_d-2\partial_d\Phi)(\nabla H)^{cda_2a_3}\mathcal R_{Sbc}{}^{a_4a_1}
\\
&\quad
+32\hat{\mathbbm A}^k\cdot F_{a_1a_2}H_{ka_3a_4}(\nabla H)_{cd}{}^{a_2a_3}\mathcal R_S^{cda_4a_1}
\\
&\quad
-32\hat F^{bk}\cdot F_d{}^{a_2}H_{ka_2a_3}\mathcal R^{cda_3a_4}(\nabla^{a_1}-2\partial^{a_1}\Phi)\mathcal R_{bca_4a_1}\,.
\end{split}
\end{equation}

Looking at the $F\nabla\hat FH\mathcal R^2$ terms we see that the first term is the only remaining term with a contraction of $F$ and $H$ and no trace. Integrating it by parts gives
\begin{equation}
\begin{split}
&
-32\hat F^{bk}\cdot F^{a_1}{}_dH_{ka_2a_3}\mathcal R^{cda_3a_4}\nabla_{a_2}\mathcal R_{bca_4a_1}
+16F^{a_1}{}_d\cdot\nabla^b\hat F^{a_2a_3}H_{ka_2a_3}\mathcal R^{cdka_4}\mathcal R_{bca_4a_1}
\\
&
+16\hat F^{bk}\cdot F^{a_1}{}_dH_{ka_2a_3}\nabla^{a_4}\mathcal R^{cda_2a_3}\mathcal R_{bca_4a_1}
-32\hat F^{bk}\cdot F^{a_1}{}_d\mathbbm B_{a_3k}\mathcal R^{cda_3a_4}\mathcal R_{bca_4a_1}
+\mathrm{H.O.}\,,
\end{split}
\end{equation}
where the first term can be further rewritten, using the Bianchi identity and integrating by parts, as
\begin{equation}
\begin{split}
&
-16\hat F^{bk}\cdot F^{a_1d}H_{ka_2a_3}\nabla_b(\mathcal R_{cd}{}^{a_3a_4}\mathcal R_{a_4a_1}{}^{a_2c})
+64\hat F^{bk}\cdot F^{a_1d}H_{ka_2a_3}\nabla_b(\nabla H)^{a_2c}{}_{a_4a_1}\mathcal R_{cd}{}^{a_3a_4}
\\
&
+16\hat F_b{}^k\cdot\nabla^{a_1}F_{cd}H_{ka_2a_3}\mathcal R^{cda_3a_4}\mathcal R^{ba_2}{}_{a_4a_1}
+32\nabla^c\hat F_b{}^k\cdot F^{a_1d}H_{ka_2a_3}\mathcal R_{cd}{}^{a_3a_4}\mathcal R^{ba_2}{}_{a_4a_1}
\\
&
+32\hat F_b{}^k\cdot F^{a_1d}\nabla^cH_{ka_2a_3}\mathcal R_{cd}{}^{a_3a_4}\mathcal R^{ba_2}{}_{a_4a_1}
+32\hat F_b{}^k\cdot F^{a_1d}H_{ka_2a_3}(\nabla^c-2\partial^c\Phi)\mathcal R_{cd}{}^{a_3a_4}\mathcal R^{ba_2}{}_{a_4a_1}\,,
\end{split}
\end{equation}
where the first term can be further integrated by parts resulting in one term with the unwanted structure, which, using the Bianchi identity and another integration by parts, leads to terms with additional traces. After these steps we obtain
\begin{equation}
k_2^{(F\nabla\hat FH\mathcal R^2)}
+k_2^{(F\hat F\nabla H\mathcal R^2)}
+k_2^{(F^2H^2\mathcal R)}
+k_2^{(F\nabla\hat FH(\nabla H)^2)}
+k_2^{(\mathrm{e.o.m.})}
+(H\rightarrow-H, F\leftrightarrow\hat F)
+\mathrm{H.O.}\,,
\end{equation}
where
\begin{equation}
\begin{split}
k_2^{(F\nabla\hat FH\mathcal R^2)}
&=
32F^{ef}\cdot\nabla^g\hat F_b{}^kH_{kcd}\mathcal R_{gf}{}^{dh}\mathcal R^{bc}{}_{he}
+32F_{ef}\cdot\nabla_g\hat F^{bk}H_{kcd}\mathcal R^{ghce}\mathcal R_{bhdf}
\\
&\quad
+32F_{ef}\cdot\nabla^b\hat F^{ek}H_{kcd}\mathcal R^{fhgc}\mathcal R_{bgh}{}^d
+16F_{ef}\cdot\nabla^g\hat F^{cd}H_{kcd}\mathcal R^{fhbk}\mathcal R_{ghb}{}^e
\\
&\quad
+16\hat F_b{}^k\cdot\nabla^gF_{ef}H_{kcd}\mathcal R^{efch}\mathcal R^{bd}{}^{gh}
-16\hat F^{bk}\cdot\nabla^fF_b{}^cH_{kcd}\mathcal R_{efgh}\mathcal R^{ghed}
\\
&\quad
+16F_{ef}\cdot\nabla_g\hat F^{bk}H_{kcd}\mathcal R_{bh}{}^{gf}\mathcal R^{hecd}
-8\hat F^{bk}\cdot\nabla_gF^{ef}H_{kcd}\mathcal R_{bhef}\mathcal R^{ghcd}
\\
&\quad
+8\hat F^{bk}\cdot\nabla_fF^{cd}H_{kcd}\mathcal R_{begh}\mathcal R^{ghef}
-8\hat F_b{}^k\cdot\nabla^cF_{ef}H_{kcd}\mathcal R^{ghef}\mathcal R^{bd}{}_{gh}
\\
&\quad
+8F_{ef}\cdot\nabla_b\hat F^{ek}H_{kcd}\mathcal R^{bf}{}_{gh}\mathcal R^{ghcd}
+8F^{ef}\cdot\nabla_e\hat F^{cd}H_{kcd}\mathcal R_{bfgh}\mathcal R^{ghbk}
\\
&\quad
+4\hat F^{bk}\cdot\nabla_bF^{ef}H_{kcd}\mathcal R_{efgh}\mathcal R^{ghcd}\,,
\end{split}
\end{equation}
\begin{equation}
\begin{split}
k_2^{(F\hat F\nabla H\mathcal R^2)}
&=
32\hat F_{ab}\cdot F_{ef}\nabla_cH^{bfg}\mathcal R^e{}_{hgd}\mathcal R^{hacd}
+32\hat F_{ab}\cdot F^{ef}\nabla^cH^a{}_{gh}\mathcal R^{bg}{}_{de}\mathcal R_{cf}{}^{dh}
\\
&\quad
-32\hat F_{ab}\cdot F^{ef}\nabla_cH^{agh}\mathcal R^b{}_{dfg}\mathcal R^{cd}{}_{he}
-32\hat F_{ke}\cdot F^{eb}\nabla_fH^k{}_{cd}\mathcal R_{bh}{}^{gc}\mathcal R^{fg}{}^{hd}
\\
&\quad
+16\hat F_{ab}\cdot F^{ef}\nabla^cH^{agh}\mathcal R_{degh}\mathcal R^{bd}{}_{fc}
+16\hat F_{ab}\cdot F_{ef}(\nabla H)^{abcd}\mathcal R^{eh}{}_{gc}\mathcal R^{fg}{}_{hd}
\\
&\quad
-8\hat F^{ab}\cdot F_{ef}(\nabla H)_{abcd}\mathcal R^{fcgh}\mathcal R^{de}{}_{gh}
+8\hat F^{kb}\cdot F_{be}\nabla_fH_{kcd}\mathcal R^{efgh}\mathcal R_{gh}{}^{cd}\,,
\end{split}
\end{equation}
\begin{equation}
\begin{split}
k_2^{(F^2H^2\mathcal R)}
&=
k_1^{(F^2H^2\mathcal R)}
+64\hat F^{bk}\cdot F_{ef}H_{kcd}\nabla_b(\nabla H)^{cghe}\mathcal R_{fgh}{}^d
\\
&\quad
+32\hat F^{bk}\cdot\nabla_bF^{ec}H_{kcd}(\nabla H)_{fegh}\mathcal R^{ghdf}
+8\hat F^{bk}\cdot\nabla^cF_{ef}H_{kcd}\nabla^dH_{bgh}\mathcal R^{ghef}\,,
\end{split}
\end{equation}
\begin{equation}
k_2^{(F\nabla\hat FH(\nabla H)^2)}=k_1^{(F\nabla\hat FH(\nabla H)^2)}\,,
\end{equation}
and
\begin{equation}
\begin{split}
k_2^{(\mathrm{e.o.m.})}
&=
k_1^{(\mathrm{e.o.m.})}
+32\hat F_b{}^k\cdot F^{ef}H_{kcd}\mathcal R^{bc}{}_{he}(\nabla^g-2\partial^g\Phi)\mathcal R_{gf}{}^{dh}
\\
&\quad
-32\hat F^{bk}\cdot F_{ef}\mathcal R^{gf}{}_{ch}\mathcal R_{bg}{}^{he}\mathbbm B_{ck}
+16\hat{\mathbbm A}^k\cdot F^{ef}H_{kcd}\mathcal R_{gf}{}^{dh}\mathcal R_{he}{}^{cg}
\\
&\quad
-32\hat F_{bk}\cdot F^{be}H^{kcd}\mathcal R_{efdg}(\nabla_h-2\partial_h\Phi)\mathcal R^{ghf}{}_c
+16\hat F^{bk}\cdot F_{be}\mathcal R^{efgh}\mathcal R_{ghf}{}^c\mathbbm B_{ck}
\\
&\quad
-16\hat F^{bk}\cdot F_{ef}H_{kcd}\mathcal R^{gfcd}(\nabla_h-2\partial_h\Phi)\mathcal R_{bg}{}^{he}
\\
&\quad
-8\hat F^{bk}\cdot F_{be}H_{kcd}(\nabla_f-2\partial_f\Phi)\mathcal R^{efgh}\mathcal R_{gh}{}^{cd}\,.
\end{split}
\end{equation}
The fifth and sixth terms are the only terms left with the unwanted structure and less than 2 traces. The former can be integrated by parts and for the latter we integrate half of it by parts. 

Finally we add in the terms with two or more traces in (\ref{eq:HFdFR2-2}) and we get
\begin{equation}
\begin{split}
\frac{1}{4!}t_8t_8\mathcal R^4 
-\frac{1}{12}t_8t_8H^2\mathcal R^3
+\frac12\hat t_8\hat t_8H^2(\nabla H)^2\mathcal R
&\rightarrow
k_3^{(F\nabla\hat FH\mathcal R^2)}
+k_3^{(F\hat F\nabla H\mathcal R^2)}
+k_3^{(F^2H^2\mathcal R)}
\\
&\quad
+k_3^{(F\nabla\hat FH(\nabla H)^2)}
+k_3^{(\mathrm{e.o.m.})}
\\
&\quad
+(H\rightarrow-H, F\leftrightarrow\hat F)
+\mathrm{H.O.}\,,
\end{split}
\end{equation}
where
\begin{equation}
\begin{split}
k_3^{(F\nabla\hat FH\mathcal R^2)}
&=
32F^{ef}\cdot\nabla^g\hat F_b{}^kH_{kcd}\mathcal R_{eg}{}^{dh}\mathcal R^{bc}{}_{hf}
+32F_{ef}\cdot\nabla_g\hat F^{bk}H_{kcd}\mathcal R^{ghce}\mathcal R_{bh}{}^{df}
\\
&\quad
+32F^{ef}\cdot\nabla^b\hat F_e{}^kH_{kcd}\mathcal R_{fh}{}^{gc}\mathcal R_{bg}{}^{hd}
+16F_{ef}\cdot\nabla^g\hat F^{cd}H_{kcd}\mathcal R^{hebk}\mathcal R_{ghb}{}^f
\\
&\quad
-16F_{ef}\cdot\nabla^g\hat F_b{}^kH_{kcd}\mathcal R^{efch}\mathcal R^{bd}{}_{gh}
+16F_{ef}\cdot\nabla_g\hat F^{bk}H_{kcd}\mathcal R_{bh}{}^{eg}\mathcal R^{hfcd}
\\
&\quad
+8F^{ef}\cdot\nabla_e\hat F^{cd}H_{kcd}\mathcal R_{bfgh}\mathcal R^{ghbk}
+8F^{ef}\cdot\nabla^h\hat F^{bk}H_{kcd}\mathcal R_{ghef}\mathcal R^{bgcd}
\\
&\quad
+8F_{ef}\cdot\nabla_b\hat F^{ek}H_{kcd}\mathcal R^{bf}{}_{gh}\mathcal R^{ghcd}
-4F^{ef}\cdot\nabla_b\hat F^{cd}H_{kcd}\mathcal R_{efgh}\mathcal R^{bkgh}\,,
\end{split}
\end{equation}
\begin{equation}
\begin{split}
k_3^{(F\hat F\nabla H\mathcal R^2)}
&=
-\frac12t_8t_8\hat F\cdot F\nabla H\mathcal R^2
+32\hat F_{ab}\cdot F_{ef}\nabla_cH^{bfg}\mathcal R^{eh}{}_{gd}\mathcal R_h{}^{acd}
\\
&\quad
+32\hat F_{ab}\cdot F_{ef}\nabla_cH^a{}_{gh}\mathcal R^{bgde}\mathcal R^{fch}{}_d
-32\hat F_{ab}\cdot F_{ef}\nabla^cH_{agh}\mathcal R^{bdfg}\mathcal R_{cd}{}^{he}
\\
&\quad
-32\hat F^{ae}\cdot F_{eb}\nabla^fH_{acd}\mathcal R^{bhgc}\mathcal R_{fgh}{}^d
+16\hat F_{ab}\cdot F_{ef}\nabla_cH^{agh}\mathcal R^e{}_{dgh}\mathcal R^{bdcf}
\\
&\quad
+16\hat F_{ab}\cdot F_{ef}(\nabla H)^{abcd}\mathcal R^{eh}{}_{gc}\mathcal R^{fg}{}_{hd}
+16\hat F_{ab}\cdot F_{ef}\nabla^gH^a{}_{cd}\mathcal R^{efch}\mathcal R^{bd}{}_{gh}
\\
&\quad
-8\hat F_{ab}\cdot F^{bc}\nabla^eH^a{}_{cd}\mathcal R_{efgh}\mathcal R^{ghdf}
+8\hat F_{ab}\cdot F^{ef}\nabla^gH^a{}_{cd}\mathcal R_{ghef}\mathcal R^{bhcd}
\\
&\quad
-8\hat F^{ab}\cdot F_{ef}(\nabla H)_{abcd}\mathcal R^{ce}{}_{gh}\mathcal R^{dfgh}
+8\hat F^{ab}\cdot F_{be}\nabla_fH_{acd}\mathcal R^{ef}{}_{gh}\mathcal R^{ghcd}\,,
\end{split}
\end{equation}
\begin{equation}
\begin{split}
k_3^{(F^2H^2\mathcal R)}
&=
\frac12\hat t_8\hat t_8\hat F\cdot F(\nabla H)^2\mathcal R_S
+64F_{ef}\cdot\nabla_g\hat F^{bk}H_{kcd}(\nabla H)^{ghec}\mathcal R_{Sbh}{}^{df}
\\
&\quad
+64\hat F^{bk}\cdot F^{ef}H_{kcd}\nabla_b(\nabla H)_{he}{}^{cg}\mathcal R_{fg}{}^{dh}
+32\hat F^{bk}\cdot F_{ef}H_{kcd}\nabla_b(\nabla H)^{ec}{}_{gh}\mathcal R_S^{ghdf}
\\
&\quad
+32\hat F^{bk}\cdot F^c{}_fH_{kcd}\nabla^d(\nabla H)^{efgh}\mathcal R_{ghbe}
-16\hat F^{bk}\cdot\nabla^cF^{ef}H_{kcd}\nabla^gH_{hef}\mathcal R^{hd}{}_{bg}
\\
&\quad
-32\nabla^f\hat F^{bc}\cdot\nabla_hF_{cd}H^2_{abef}\mathcal R^{heda}
+32\nabla_h\hat F^{bc}\cdot\nabla^fF_{cd}H^2_{abef}\mathcal R_S^{dahe}
\\
&\quad
+32\hat F^{bk}\cdot\nabla_bF^{ce}H_{kcd}(\nabla H)_{efgh}\mathcal R^{ghdf}
+32\hat F^{bk}\cdot\nabla_bF_{ef}H_{kcd}(\nabla H)^{cf}{}_{gh}\mathcal R_S^{ghde}
\\
&\quad
-16\hat F^{bk}\cdot\nabla_eF^{cd}H_{kcd}(\nabla H)^{efgh}\mathcal R_{bfgh}
+16\hat F^{bk}\cdot\nabla^eF_b{}^cH_{kcd}(\nabla H)_{efgh}\mathcal R^{ghdf}
\\
&\quad
-16\hat F^{bk}\cdot\nabla^eF_b{}^cH_{kcd}(\nabla H)^{dfgh}\mathcal R_{ghef}
+8\hat F^{bk}\cdot\nabla^cF_{ef}H_{kcd}\nabla^dH_{bgh}\mathcal R^{ghef}
\\
&\quad
+8\hat F^{bk}\cdot F^{ef}H_{kcd}\nabla_b(\nabla H)_{efgh}\mathcal R^{cdgh}
-8\hat F^{bk}\cdot F^{ef}H_{kcd}\nabla_b(\nabla H)^{cdgh}\mathcal R_{ghef}
\\
&\quad
-64\hat F^{bk}\cdot F_{ef}\nabla_hH_{kcd}(\nabla H)^{ghec}\mathcal R_{Sbg}{}^{df}
-64\hat F^{bk}\cdot F_{ef}(\nabla H)_{bkcd}(\nabla H)^{ce}{}_{gh}\mathcal R_S^{dfgh}
\\
&\quad
+8\hat F^{bk}\cdot F^{ef}(\nabla H)_{bkcd}(\nabla H)_{efgh}\mathcal R^{cdgh}\,,
\end{split}
\end{equation}
\begin{equation}
\begin{split}
k_3^{(F\nabla\hat FH(\nabla H)^2)}
&=
k_1^{(F\nabla\hat FH(\nabla H)^2)}
-\frac16t_8t_8\hat F\cdot F(\nabla H)^3
-32\nabla^f\hat F^{bc}\cdot\nabla_hF_{cd}H^2_{abef}(\nabla H)^{heda}
\\
&\quad
-32\nabla_d\hat F^{fg}\cdot\nabla^bF^{he}H^2_{abef}(\nabla H)^{da}{}_{gh}
-16\hat F^{bk}\cdot\nabla_bF^{ef}H_{kcd}(\nabla H)_{efgh}(\nabla H)^{cdgh}
\\
&\quad
+16\hat F^{bk}\cdot\nabla^cF_{ef}H_{kcd}(\nabla H)^{ghef}\nabla^dH_{bgh}
+4\hat F^{bk}\cdot\nabla_bF^{cd}H_{kcd}(\nabla H)_{efgh}(\nabla H)^{ghef}
\end{split}
\end{equation}
and
\begin{equation}
\begin{split}
k_3^{(\mathrm{e.o.m.})}
&=
k_2^{(\mathrm{e.o.m.})}
-2\hat F^{bk}\cdot F_b{}^c\mathcal R_{efgh}\mathcal R^{ghef}\mathbbm B_{kc}
+8\hat F^{bk}\cdot F_{ef}H_{kcd}(\nabla_h-2\partial_h\Phi)\mathcal R^{ghef}\mathcal R_{bg}{}^{cd}
\\
&\quad
+8\hat F^{bk}\cdot F_{ef}\mathcal R^{efgh}\mathcal R_{bdgh}\mathbbm B^d{}_k
+8\hat F^{bk}\cdot F_b{}^cH_{kcd}(\nabla^f-2\partial^f\Phi)\mathcal R_{efgh}\mathcal R^{ghed}
\\
&\quad
-16\hat F^{bk}\cdot F_{ef}H_{kcd}\mathcal R^{efc}{}_h(\nabla_g-2\partial_g\Phi)\mathcal R_b{}^{dgh}
+8\hat{\mathbbm A}^k\cdot F^{ef}H_{kcd}(\nabla H)_{efgh}\mathcal R^{cdgh}
\\
&=
-8\hat F_{ab}\cdot F^{ef}\mathcal R^{acgh}\mathcal R_{cegh}\mathbbm B^b{}_f
-32\hat F_{ab}\cdot F_e{}^f\mathcal R_{cfgh}\mathcal R^{acge}\mathbbm B^{bh}
\\
&\quad
+16\hat F_{ab}\cdot F^{bc}\mathcal R_{cdgh}\mathcal R^{ghd}{}_e\mathbbm B^{ae}
+2\hat F_{ab}\cdot F^{bc}\mathcal R_{efgh}\mathcal R^{ghef}\mathbbm B^a{}_c
\\
&\quad
+16\hat{\mathbbm A}^k\cdot F^{ef}H_{kcd}\mathcal R_{gf}{}^{dh}\mathcal R_{he}{}^{cg}
+8\hat F_{ab}\cdot F_{ef}\mathcal R^{efgh}\mathcal R^b{}_{dgh}\mathbbm B^{ad}
\\
&\quad
-8\hat{\mathbbm A}^k\cdot F_{ef}H_{kcd}\mathcal R_{gh}{}^{ce}\mathcal R^{ghdf}
-64\hat F^{bk}\cdot F_{ef}H_{kcd}\mathcal R_{bg}{}^{ec}\nabla^{[d}\mathbbm G^{f]g}
\\
&\quad
+64\hat F^{bk}\cdot F^{ef}H_{kcd}\mathcal R^{bc}{}_{ge}\nabla^{[d}\mathbbm G^{g]}{}_f
+64\hat F^{bk}\cdot F^{ec}H_{kcd}\mathcal R_e{}^{ghd}\nabla_{[b}\mathbbm G_{g]h}
\\
&\quad
-64\hat F^{bk}\cdot F_{be}H_{kcd}\mathcal R_f{}^{edg}\nabla^{[f}\mathbbm G^{c]}{}_g
-32\hat F^{bk}\cdot F_{ef}H_{kcd}\mathcal R^{gfcd}\nabla_{[b}\mathbbm G_{g]}{}^e
\\
&\quad
+16\hat F^{bk}\cdot F_{be}H_{kcd}\mathcal R^{ghcd}\nabla_g\mathbbm G_h{}^e
-16\hat F^{bk}\cdot F^{ef}H_{kcd}\mathcal R_{bg}{}^{cd}\nabla_e\mathbbm G_f{}^g
\\
&\quad
-16\hat F^{bk}\cdot F_b{}^cH_{kcd}\mathcal R^{ghed}\nabla_g\mathbbm G_{he}
-32\hat F_b{}^k\cdot F_{ef}H_{kcd}\mathcal R^{efch}\nabla^{[b}\mathbbm G^{d]}{}_h
\\
&\quad
+\mathcal O(H^2)
+\mathrm{H.O.}\,.
\end{split}
\label{eq:k3eom}
\end{equation}

To further simplify the $F^2H^2\mathcal R$ terms we note that the third term in $k_3^{(F^2H^2\mathcal R)}$ can be rewritten as
\begin{equation}
\begin{split}
64&\hat F^{bk}\cdot F^{ef}H_{kcd}\nabla_b(\nabla H)_{he}{}^{cg}\mathcal R_{fg}{}^{dh}
=
-32F_{ef}\cdot\nabla_g\hat F^{bk}H_{kcd}\nabla_bH^{che}\mathcal R^{fgd}{}_h%
\\
&\quad
+16F_{ef}\cdot\nabla_g\hat F^{bk}H_{kcd}\nabla_bH^{cgh}\mathcal R^{efd}{}_h%
-16\hat F^{bk}\cdot\nabla_bF_{ef}H_{kcd}\nabla^cH^{egh}\mathcal R^{df}{}_{gh}%
\\
&\quad
+16F_{ef}\cdot\nabla^e\hat F_b{}^kH_{kcd}\nabla^cH^{fgh}\mathcal R^{bd}{}_{gh}%
-16\hat F^{bk}\cdot\nabla^cF^{ef}H_{kcd}\nabla^dH_e{}^{gh}\mathcal R_{bfgh}%
\\
&\quad
+16\hat F_b{}^k\cdot F^{ef}H_{kcd}\nabla^c(\nabla H)_{efgh}\mathcal R^{bdgh}%
+8F_{ef}\cdot\nabla_b\hat F^{cd}H_{kcd}\nabla^kH^{egh}\mathcal R^{bf}{}_{gh}%
\\
&\quad
+8\hat F^{bk}\cdot F_{ef}H_{kcd}\nabla_bH^{cgh}\nabla^d\mathcal R^{ef}{}_{gh}%
-32\hat F^{bk}\cdot F_{ef}\nabla_gH_{kcd}\nabla_bH^{che}\mathcal R^{fgd}{}_h%
\\
&\quad
+16\hat F^{bk}\cdot F_{ef}\nabla_gH_{kcd}\nabla_bH^{cgh}\mathcal R^{efd}{}_h%
-16\hat F^{bk}\cdot F_{ef}\nabla_bH_{kcd}\nabla^cH^{egh}\mathcal R^{df}{}_{gh}%
\\
&\quad
+16\hat F_b{}^k\cdot F^{ef}\nabla_eH_{kcd}\nabla^cH_{fgh}\mathcal R^{bdgh}%
-32\hat F^{bk}\cdot F_{ef}H_{kcd}\nabla_bH^{che}(\nabla_g-2\partial_g\Phi)\mathcal R^{fgd}{}_h
\\
&\quad
-16\hat F^{bk}\cdot F_{ef}H_{kcd}\mathcal R^{efdh}\nabla_b\mathbbm B^c{}_h
+16\hat{\mathbbm A}^k\cdot F_{ef}H_{kcd}\nabla^cH^{egh}\mathcal R^{fd}{}_{gh}
\\
&\quad
+16\hat F_b{}^k\cdot\mathbbm A^eH_{kcd}\nabla^cH_{egh}\mathcal R^{bdgh}
-16\hat F_{bk}\cdot F_{ef}\nabla_cH^{egh}\mathcal R^{fb}{}_{gh}\mathbbm B^{kc}
+\mathcal O(H^3)
+\mathrm{H.O.}\,.
\end{split}
\end{equation}
The fourth term in $k_3^{(F^2H^2\mathcal R)}$ can be rewritten
\begin{equation}
\begin{split}
32&\hat F^{bk}\cdot F_{ef}H_{kcd}\nabla_b(\nabla H)^{ec}{}_{gh}\mathcal R_S^{ghdf}
=
-16F_{ef}\cdot\nabla^e\hat F^{bk}H_{kcd}\nabla_bH^c{}_{gh}\mathcal R_S^{ghdf}%
\\
&\quad
+16\hat F^{bk}\cdot\nabla_bF_{ef}H_{kcd}\nabla^cH^e{}_{gh}\mathcal R_S^{ghdf}%
-16\hat F^{bk}\cdot\nabla^dF^{ef}H_{kcd}\nabla^cH_{egh}\mathcal R_S^{gh}{}_{bf}%
\\
&\quad
-16F^{ef}\cdot\nabla_e\hat F_b{}^kH_{kcd}\nabla^cH_{fgh}\mathcal R_S^{ghbd}%
-8\hat F^{bk}\cdot F^{ef}H_{kcd}\nabla_bH^{cgh}\nabla^d\mathcal R_{Sghef}%
\\
&\quad
-8F^{ef}\cdot\nabla^b\hat F^{cd}H_{kcd}\nabla^kH_{egh}\mathcal R_S^{gh}{}_{bf}%
-16\hat F_b{}^k\cdot F^{ef}H_{kcd}\nabla^c(\nabla H)_{efgh}\mathcal R_S^{ghbd}%
\\
&\quad
-16\hat F^{bk}\cdot F_{ef}\nabla^eH_{kcd}\nabla_bH^{cgh}\mathcal R_S^{df}{}_{gh}%
-16\hat F_b{}^k\cdot F^{ef}\nabla_eH_{kcd}\nabla^cH_{fgh}\mathcal R_S^{ghbd}%
\\
&\quad
+16\hat F^{bk}\cdot F_{ef}(\nabla H)_{bkcd}\nabla^cH^e{}_{gh}\mathcal R_S^{ghdf}%
-16\hat F^{bk}\cdot\mathbbm A_fH_{kcd}\nabla_bH^c{}_{gh}\mathcal R_S^{ghdf}
\\
&\quad
+16\hat{\mathbbm A}^k\cdot F_{ef}H_{kcd}\nabla^cH^e{}_{gh}\mathcal R_S^{ghdf}
-16\hat F^{bk}\cdot F^{ef}\nabla^cH_{egh}\mathcal R_S^{gh}{}_{bf}\mathbbm B_{kc}
\\
&\quad
+16\hat F^{bk}\cdot\mathbbm A^eH_{kcd}\nabla^cH_{egh}\mathcal R_S^{ghd}{}_b
+\mathcal O(H^3)
+\mathrm{H.O.}\,.
\end{split}
\end{equation}
The fifth term in $k_3^{(F^2H^2\mathcal R)}$ can be rewritten
\begin{equation}
\begin{split}
32&\hat F^{bk}\cdot F^c{}_fH_{kcd}\nabla^d(\nabla H)^{efgh}\mathcal R_{ghbe}
=
16\hat F^{bk}\cdot\nabla^cF^{ef}H_{kcd}\nabla^dH_{egh}\mathcal R^{gh}{}_{bf}%
\\
&\quad
-8F^{bk}\cdot\nabla^c\hat F_{ef}H_{kcd}\nabla^dH_{bgh}\mathcal R^{ghef}%
-16\hat F^{bk}\cdot F^{cf}\nabla^eH_{kcd}\nabla^dH_{fgh}\mathcal R^{gh}{}_{be}%
\\
&\quad
+8\hat F^{bk}\cdot F^{ce}\nabla_fH_{kcd}\nabla^dH^{fgh}\mathcal R_{Sghbe}%
-16\hat F^{bk}\cdot F^{cf}H_{kcd}\nabla^dH_{fgh}(\nabla^e-2\partial^e\Phi)\mathcal R^{gh}{}_{be}
\\
&\quad
+8\hat F^{bk}\cdot F^{ce}H_{kcd}\mathcal R_S^{gh}{}_{be}\nabla^d\mathbbm B_{gh}
+\mathcal O(H^3)
+\mathrm{H.O.}\,.
\end{split}
\end{equation}
Finally the seventh and eighth term in $k_3^{(F^2H^2\mathcal R)}$ can be rewritten
\begin{equation}
\begin{split}
-32&\nabla^f\hat F^{bc}\cdot\nabla_hF_{cd}H^2_{abef}\mathcal R^{heda}
+32\nabla_h\hat F^{bc}\cdot\nabla^fF_{cd}H^2_{abef}\mathcal R_S^{dahe}
\\
&=
32F_{ef}\cdot\nabla^k\hat F^{be}H_{kcd}\nabla_hH_{ab}{}^c\mathcal R_S^{fahd}%
+32F_{ef}\cdot\nabla_b\hat F^{ke}H_{kcd}(\nabla H)^{cb}{}_{gh}\mathcal R_S^{fdgh}%
\\
&\quad
+32F^{ef}\cdot\nabla_e\hat F_{bg}H_{kcd}(\nabla H)^{abcd}\mathcal R_S^{gk}{}_{fa}%
-16F_{ef}\cdot\nabla^e\hat F_{bg}H_{kcd}(\nabla H)^{abcd}\mathcal R_S^{akfg}%
\\
&\quad
+16F_{ef}\cdot\nabla_g\hat F^{be}H_{kcd}(\nabla H)_{ab}{}^{cd}\mathcal R_S^{akfg}%
+16F_{ef}\cdot\nabla^h\hat F^{be}H^2_{abcd}\nabla_h\mathcal R^{cdfa}%
\\
&\quad
+8F^{ef}\cdot\hat F_{eg}H_{kcd}(\nabla H)^{cd}{}_{ab}\nabla_f\mathcal R_S^{abgk}%
+16F_{ef}\cdot\hat F_g{}^e\nabla^bH_{kcd}(\nabla H)_{ab}{}^{cd}\mathcal R_S^{afgk}
\\
&\quad
-32\nabla_h\hat F^{bc}\cdot\nabla^fF_{cd}H^2_{abef}(\nabla_h-2\partial_h\Phi)\mathcal R_S^{dahe}
+32\hat F^{bc}\cdot\nabla_hF_{cd}H_{abk}\mathcal R^{heda}\mathbbm B^k{}_e
\\
&\quad
+32\hat F^{be}\cdot F_{ef}\nabla^hH_{abc}\mathcal R_S^{fa}{}_{hd}\mathbbm B^{cd}
+32\hat F^{ke}\cdot F_{ef}H_{kcd}(\nabla_b-2\partial_b\Phi)(\nabla H)^{cbgh}\mathcal R_S^{fd}{}_{gh}
\\
&\quad
-32\hat F^{be}\cdot F_{ef}H_{kcd}(\nabla H)^{cd}{}_{ab}(\nabla_h-2\partial_h\Phi)\mathcal R_S^{hkfa}
\\
&\quad
+16F^{ef}\cdot\hat F_{eg}H_{kcd}(\nabla_b-2\partial_b\Phi)(\nabla H)^{cdab}\mathcal R_{Saf}{}^{gk}
\\
&\quad
+16\hat F^{be}\cdot F_{ef}H^2_{abcd}(\nabla_h-2\partial_h\Phi)\nabla^h\mathcal R^{cdfa}
+\mathcal O(H^3)
+\mathrm{H.O.}\,.
\end{split}
\end{equation}

Using these results we get
\begin{equation}
\begin{split}
k_3^{(F^2H^2\mathcal R)}
&=
\frac12\hat t_8\hat t_8\hat F\cdot F(\nabla H)^2\mathcal R_S
+64F_{ef}\cdot\nabla_g\hat F^{bk}H_{kcd}(\nabla H)^{ghec}\mathcal R_{Sbh}{}^{df}
\\
&\quad
-32F_{ef}\cdot\nabla_g\hat F^{bk}H_{kcd}\nabla_bH^{che}\mathcal R^{fgd}{}_h
+32F_{ef}\cdot\nabla^k\hat F^{be}H_{kcd}\nabla_hH_{ab}{}^c\mathcal R_S^{fahd}
\\
&\quad
+32\hat F^{bk}\cdot\nabla_bF^{ce}H_{kcd}(\nabla H)_{efgh}\mathcal R^{ghdf}
+32\hat F^{bk}\cdot\nabla_bF_{ef}H_{kcd}(\nabla H)^{cf}{}_{gh}\mathcal R_S^{ghde}
\\
&\quad
+16\hat F^{bk}\cdot\nabla^eF_b{}^cH_{kcd}(\nabla H)_{efgh}\mathcal R^{ghdf}
-16\hat F^{bk}\cdot\nabla^eF_b{}^cH_{kcd}(\nabla H)^{dfgh}\mathcal R_{ghef}
\\
&\quad
+16\hat F^{bk}\cdot\nabla^cF^{ef}H_{kcd}\nabla^dH_{egh}\mathcal R_S^{gh}{}_{bf}
-16\hat F^{bk}\cdot\nabla^cF^{ef}H_{kcd}\nabla^gH_{hef}\mathcal R^{hd}{}_{bg}
\\
&\quad
+16F_{ef}\cdot\nabla^g\hat F^{bk}H_{kcd}\nabla_bH^c{}_{gh}\mathcal R^{efdh}
-16F_{ef}\cdot\nabla^e\hat F^{bk}H_{kcd}\nabla_bH^{cgh}\mathcal R_S^{df}{}_{gh}
\\
&\quad
+32F^{ef}\cdot\nabla_e\hat F_{bg}H_{kcd}(\nabla H)^{abcd}\mathcal R_S^{gk}{}_{fa}
+16F_{ef}\cdot\nabla^b\hat F_g{}^eH_{kcd}(\nabla H)_{ab}{}^{cd}\mathcal R_S^{akfg}
\\
&\quad
+32F_{ef}\cdot\nabla_b\hat F^{ke}H_{kcd}(\nabla H)^{cbgh}\mathcal R_S^{fd}{}_{gh}
+16F_{ef}\cdot\nabla_h\hat F^{be}H^2_{abcd}\nabla^h\mathcal R^{cdfa}%
\\
&\quad
-16\hat F^{bk}\cdot\nabla_eF^{cd}H_{kcd}(\nabla H)^{efgh}\mathcal R_{bfgh}
+8F^{ef}\cdot\hat F_{eg}H_{kcd}(\nabla H)^{cd}{}_{ab}\nabla_f\mathcal R_S^{abgk}
\\
&\quad
+8\hat F^{bk}\cdot F_{ef}H_{kcd}\nabla_b(\nabla H)^{ef}{}_{gh}\mathcal R^{cdgh}
-8\hat F^{bk}\cdot F_{ef}H_{kcd}\nabla_b(\nabla H)^{cd}{}_{gh}\mathcal R^{ghef}
\\
&\quad
-64\hat F^{bk}\cdot F_{ef}\nabla_hH_{kcd}(\nabla H)^{ghec}\mathcal R_{Sbg}{}^{df}
-32\hat F^{bk}\cdot F_{ef}\nabla_gH_{kcd}\nabla_bH^{che}\mathcal R^{fgd}{}_h
\\
&\quad
-64\hat F^{bk}\cdot F_{ef}(\nabla H)_{bkcd}(\nabla H)^{ce}{}_{gh}\mathcal R_S^{dfgh}
-16\hat F^{bk}\cdot F_{ef}\nabla^eH_{kcd}\nabla_bH^{cgh}\mathcal R_S^{df}{}_{gh}
\\
&\quad
-16\hat F^{bk}\cdot F^{cf}\nabla^eH_{kcd}\nabla^dH_{fgh}\mathcal R^{gh}{}_{be}
+8\hat F^{bk}\cdot F^{ce}\nabla^fH_{kcd}\nabla^dH_{fgh}\mathcal R_S^{gh}{}_{be}
\\
&\quad
+16F^{ef}\cdot\hat F_{ge}\nabla_bH_{kcd}(\nabla H)^{abcd}\mathcal R_{Saf}{}^{gk}
+8\hat F^{bk}\cdot F^{ef}(\nabla H)_{bkcd}(\nabla H)_{efgh}\mathcal R^{cdgh}
\\
&\quad
+16\hat F^{bk}\cdot F_{ef}\nabla^gH_{kcd}\nabla_bH^c{}_{gh}\mathcal R^{efdh}
%
+\mathcal O(H^3)
+\mathrm{H.O.}
+\mbox{e.o.m. terms of }\mathcal O(H^2)\,.
\end{split}
\label{eq:k3H2}
\end{equation}

\subsection{\texorpdfstring{$(\varepsilon_8\varepsilon_8)'\mathcal R^4$}{e8e8R4}}\label{app:eeR4-red}
The reduction of this term, defined in (\ref{eq:e8e8R4}), gives rise to the following $O(d,d)$-violating terms quadratic in the KK vectors
\begin{equation}
(\varepsilon_8\varepsilon_8)'\mathcal R^4
\rightarrow
2(\varepsilon_8\varepsilon_8)'[\hat F\cdot F]\mathcal R^3
-24(\varepsilon_7\varepsilon_7)'\nabla F\cdot\nabla\hat F\mathcal R^2
+(H\rightarrow-H, F\leftrightarrow\hat F)
\,.
\label{eq:eeR4-red1}
\end{equation}
The square brackets in the first term means that no terms with contraction between $\hat F$ and $F$ are included. It can be written
\begin{equation}
(\varepsilon_8\varepsilon_8)'[\hat F\cdot F]\mathcal R^3
=
(\varepsilon_8\varepsilon_8)'\hat F\cdot F\mathcal R^3
+4(7!)\hat F^{ab}\cdot F_{a[b}(\mathcal R^{cd}{}_{cd}\mathcal R^{ef}{}_{ef}\mathcal R^{gh}{}_{gh]})'
+2(\varepsilon_6\varepsilon_6)'\hat F\cdot F\mathcal R^3\,,
\end{equation}
where both indices on $F$ and $\hat F$ are understood to be contracted in the last term. The second term in (\ref{eq:eeR4-red1}) is defined as
\begin{equation}
(\varepsilon_7\varepsilon_7)'\nabla F\cdot\nabla\hat F\mathcal R^2
=
-7!(\nabla^{(-)b}F_{[bc}\cdot\nabla^{(+)}_d\hat F^{cd}\mathcal R^{ef}{}_{ef}\mathcal R^{gh}{}_{gh]})'\,.
\end{equation}
We wish to integrate by parts to remove the derivatives from $F$ and $\hat F$. To do this we have to take care about the removal of self-contractions. We first write this term as
\begin{equation}
-7!\nabla^{(-)b}F_{[bc}(\cdot\nabla^{(+)}_d\hat F^{cd}\mathcal R^{ef}{}_{ef}\mathcal R^{gh}{}_{gh]})'
+2(6!)\nabla^{(-)b}F_{b[c}(\cdot\nabla^{(+)}_d\hat F^{cd}\mathcal R^{ef}{}_{ef}\mathcal R^{gh}{}_{gh]})'\,.
\end{equation}
The first term can now be integrated by parts yielding
\begin{equation}
\begin{split}
&(\varepsilon_7\varepsilon_7)'\nabla F\cdot\nabla\hat F\mathcal R^2
\\
&\quad =
-7!(F_{[bc}\cdot\hat F^{dk}\mathcal R^{bc}{}_{d|k|}\mathcal R^{ef}{}_{ef}\mathcal R^{gh}{}_{gh]})'
+6!(F_{[cd}\cdot\hat F^{bk}\mathcal R^{cd}{}_{|bk|}\mathcal R^{ef}{}_{ef}\mathcal R^{gh}{}_{gh]})'
\\
&\quad\quad
+7!(F_{[bc}\cdot\nabla^{(+)}_d\hat F^{dk}H^{bc}{}_{|k|}\mathcal R^{ef}{}_{ef}\mathcal R^{gh}{}_{gh]})'
+2(6!)(F_{[cd}\cdot\nabla^{(+)b}\hat F^{kc}H^d{}_{|bk|}\mathcal R^{ef}{}_{ef}\mathcal R^{gh}{}_{gh]})'
\\
&\quad\quad
+2(7!)(F_{[bc}\cdot\nabla^{(+)}_d\hat F^{cd}H^{be}{}_{|k|}\mathcal R^{fk}{}_{ef}\mathcal R^{gh}{}_{gh]})'
-8(6!)(F_{[bc}\cdot\nabla^{(+)}_d\hat F^{cd}H^b{}_{|ek|}\mathcal R^{fke}{}_f\mathcal R^{gh}{}_{gh]})'
\\
&\quad\quad
-4(5!)(F_{[fc}\cdot\nabla^{(+)}_d\hat F^{cd}H_{|bek|}\mathcal R^{fkeb}\mathcal R^{gh}{}_{gh]})'
-6!(F_{[dc}\cdot(\nabla^{(-)b}-2\partial^b\Phi)\nabla^{(+)}_{|b|}\hat F^{cd}\mathcal R^{ef}{}_{ef}\mathcal R^{gh}{}_{gh]})'
\\
&\quad\quad
-4(6!)(F_{[ec}\cdot\nabla^{(+)}_d\hat F^{cd}(\nabla^{(-)b}-2\partial^b\Phi)\mathcal R^{ef}{}_{|b|f}\mathcal R^{gh}{}_{gh]})'
+2(6!)(\mathbbm A_{[c}\cdot\nabla^{(+)}_d\hat F^{cd}\mathcal R^{ef}{}_{ef}\mathcal R^{gh}{}_{gh]})'
\end{split}
\end{equation}
where we used the identities (\ref{eq:id1}) and (\ref{eq:id2}). Note the need for subtractions of terms with additional contractions. Next we note that the first term can be rewritten
\begin{equation}
\frac16(\varepsilon_8\varepsilon_8)'\hat F\cdot F\mathcal R^3
-\frac{7!}{3}(\hat F^{bk}\cdot F_{k[b}\mathcal R^{cd}{}_{cd}\mathcal R^{ef}{}_{ef}\mathcal R^{gh}{}_{gh]})'\,,
\end{equation}
while the third and fifth combine to give
\begin{equation}
\frac{8!}{2}(F_{[bc}\cdot\nabla^{(+)}_d\hat F^{bc}H^{dk}{}_k\mathcal R^{ef}{}_{ef}\mathcal R^{gh}{}_{gh]})'\,.
\end{equation}
The eighth term becomes
\begin{equation}
\begin{split}
&
2(6!)(F_{[cd}\cdot\nabla^c\hat{\mathbbm A}^d\mathcal R^{ef}{}_{ef}\mathcal R^{gh}{}_{gh]})'
-6!(F_{[cd}\cdot\hat F^{bk}\mathcal R^{cd}{}_{|bk|}\mathcal R^{ef}{}_{ef}\mathcal R^{gh}{}_{gh]})'
\\
&
-2(6!)(F_{[cd}\cdot\nabla^{(+)b}\hat F^{kc}H^d{}_{|bk|}\mathcal R^{ef}{}_{ef}\mathcal R^{gh}{}_{gh]})'
-2(6!)(F_{[cd}\cdot\hat F^{bc}(\mathcal R^d{}_{|b|}+2\nabla^{(-)d}\partial_{|b|}\Phi)\mathcal R^{ef}{}_{ef}\mathcal R^{gh}{}_{gh]})'
\end{split}
\end{equation}
and finally the ninth term becomes
\begin{equation}
\begin{split}
&
4(6!)(F_{[ec}\cdot\nabla^{(+)}_d\hat F^{cd}[(\nabla^{(+)b}-2\partial^b\Phi)\mathcal R^{ef}{}_{f|b|}-H^{bk}{}_f\mathcal R^{ef}{}_{|bk|}]\mathcal R^{gh}{}_{gh]})'
\\
&
-8(6!)(F_{[ec}\cdot\nabla^{(+)}_d\hat F^{cd}H^{ebk}\mathcal R^f{}_{|bk|f}\mathcal R^{gh}{}_{gh]})'
+\frac{4(5!)}{3}(F_{[fc}\cdot\nabla^{(+)}_d\hat F^{cd}\nabla^fH^2\mathcal R^{gh}{}_{gh]})'\,,
\end{split}
\end{equation}
where we noted that $H^{bek}\mathcal R^f{}_{bek}=-\frac13H^{bek}\nabla^{(-)f}H_{bek}=-\frac16\nabla^fH^2$. In the first term the expression in square brackets (which is treated as a single object as far as the removal of self-contractions) is proportional to the equations of motion, see (\ref{eq:id4}).

Putting this together we see that most of the terms with extra contractions cancel out and we are left with
\begin{equation}
\begin{split}
&(\varepsilon_7\varepsilon_7)'\nabla F\cdot\nabla\hat F\mathcal R^2
\\
&\quad =
\frac16(\varepsilon_8\varepsilon_8)'\hat F\cdot F\mathcal R^3
-\frac{7!}{3}(\hat F^{bk}\cdot F_{k[b}\mathcal R^{cd}{}_{cd}\mathcal R^{ef}{}_{ef}\mathcal R^{gh}{}_{gh]})'
\\
&\quad\quad
+\frac{8!}{2}(F_{[bc}\cdot\nabla^{(+)}_d\hat F^{bc}H^{dk}{}_k\mathcal R^{ef}{}_{ef}\mathcal R^{gh}{}_{gh]})'
+\frac{2(5!)}{3}(F_{[fc}\cdot\nabla^{(+)}_d\hat F^{cd}\nabla^fH^2\mathcal R^{gh}{}_{gh]})'
+k_5^{(\mathrm{e.o.m.})}\,,
\end{split}
\end{equation}
where we have collected the terms proportional to the equations of motion in the last term
\begin{equation}
\begin{split}
k_5^{(\mathrm{e.o.m.})}
&=
2(6!)(F_{[cd}\cdot\nabla^c\hat{\mathbbm A}^d\mathcal R^{ef}{}_{ef}\mathcal R^{gh}{}_{gh]})'
+2(6!)(\mathbbm A_{[c}\cdot\nabla^{(+)}_d\hat F^{cd}\mathcal R^{ef}{}_{ef}\mathcal R^{gh}{}_{gh]})'
\\
&\quad
-2(6!)(F_{[cd}\cdot\hat F^{bc}(\mathcal R^d{}_{|b|}+2\nabla^{(-)d}\partial_{|b|}\Phi)\mathcal R^{ef}{}_{ef}\mathcal R^{gh}{}_{gh]})'
\\
&\quad
+4(6!)(F_{[ec}\cdot\nabla^{(+)}_d\hat F^{cd}[(\nabla^{(+)b}-2\partial^b\Phi)\mathcal R^{ef}{}_{f|b|}-H^{bk}{}_f\mathcal R^{ef}{}_{|bk|}]\mathcal R^{gh}{}_{gh]})'
\,.
\end{split}
\label{eq:k5eom}
\end{equation}

Putting this together we have shown that
\begin{equation}
\begin{split}
(\varepsilon_8\varepsilon_8)'\mathcal R^4
&\rightarrow
-2(\varepsilon_8\varepsilon_8)'\hat F\cdot F\mathcal R^3
+4(\varepsilon_6\varepsilon_6)'\hat F\cdot F\mathcal R^3
-12(8!)(F_{[bc}\cdot\nabla^{(+)}_d\hat F^{bc}H^{dk}{}_k\mathcal R^{ef}{}_{ef}\mathcal R^{gh}{}_{gh]})'
\\
&\quad
-16(5!)(F_{[cd}\cdot\nabla^{(+)}_f\hat F^{cd}\nabla^fH^2\mathcal R^{gh}{}_{gh]})'
-24k_5^{(\mathrm{e.o.m.})}
+(H\rightarrow-H, F\leftrightarrow\hat F)
\,.
\end{split}
\end{equation}
Since we will not need terms of sixth order or higher in fields we can easily integrate the third term by parts to get
\begin{equation}
-12(8!)(F_{[bc}\cdot\nabla^{(+)}_d\hat F^{bc}H^{dk}{}_k\mathcal R^{ef}{}_{ef}\mathcal R^{gh}{}_{gh]})'
=
12(\varepsilon_8\varepsilon_8)'\hat F\cdot F(\nabla H)\mathcal R^2
+k_6^{(\mathrm{e.o.m.})}
+\mathrm{H.O.}
\end{equation}
with
\begin{equation}
\begin{split}
k_6^{(\mathrm{e.o.m.})}
&=
-48(7!)(F_{[bc}\cdot\hat F^{bc}H^{dk}{}_k[(\nabla^{(-)}_{|e|}-2\partial_{|e|}\Phi)\mathcal R^{ef}{}_{df}-H^f{}_{|el|}\mathcal R^{el}{}_{df}]\mathcal R^{gh}{}_{gh]})'
\\
&\quad
+24(7!)(F_{[bc}\cdot\hat F^{bc}\mathbbm B^k{}_k\mathcal R^{ef}{}_{ef}\mathcal R^{gh}{}_{gh]})'
+24(7!)(F_{[cd}\cdot\hat{\mathbbm A}^cH^{dk}{}_k\mathcal R^{ef}{}_{ef}\mathcal R^{gh}{}_{gh]})'\,.
\end{split}
\label{eq:k6eom}
\end{equation}
So that finally we get
\begin{equation}
\begin{split}
(\varepsilon_8\varepsilon_8)'\mathcal R^4
&\rightarrow
-2(\varepsilon_8\varepsilon_8)'\hat F\cdot F\mathcal R^3
+12(\varepsilon_8\varepsilon_8)'\hat F\cdot F(\nabla H)\mathcal R^2
+4(\varepsilon_6\varepsilon_6)'\hat F\cdot F\mathcal R^3
\\
&\quad
-16(5!)(F_{[cd}\cdot\nabla_f\hat F^{cd}\nabla^fH^2\mathcal R^{gh}{}_{gh]})'
-24k_5^{(\mathrm{e.o.m.})}
+k_6^{(\mathrm{e.o.m.})}
\\
&\quad
+\mathrm{H.O.}
+(H\rightarrow-H, F\leftrightarrow\hat F)\,.
\end{split}
\label{eq:eeR4-red}
\end{equation}

\subsection{\texorpdfstring{$H^2(\varepsilon_6\varepsilon_6)'\mathcal R^3$}{e6eH26R3}}
Dimensional reduction of this term, defined in (\ref{eq:H2e6e6R3}), gives rise to the following $O(d,d)$-violating terms quadratic in the KK vectors
\begin{equation}
\begin{split}
H^2(\varepsilon_6\varepsilon_6)'\mathcal R^3
&\rightarrow
3(\varepsilon_6\varepsilon_6)'\hat F\cdot F\mathcal R^3
+12(5!)H^2(\nabla^{(-)b}F_{[cd}\cdot\nabla^{(+)}_b\hat F^{cd}\mathcal R^{ef}{}_{ef]})'
\\
&\quad
+\frac32H^2(\varepsilon_6\varepsilon_6)'[\hat F\cdot F]\mathcal R^2
+(H\rightarrow-H, F\leftrightarrow\hat F)\,.
\end{split}
\label{eq:H2eeR3-red}
\end{equation}
This can be used to cancel the third and fourth term from (\ref{eq:eeR4-red}). Note that the last term will not be needed as it is of sixth order in fields.

\subsection{\texorpdfstring{$(\varepsilon_9\varepsilon_9)'H^2\mathcal R^3$}{e9e9H2R3}}
Dimensional reduction of this term, defined in (\ref{eq:e9e9H2R3}), gives rise to the following $O(d,d)$-violating terms quadratic in the KK vectors
\begin{equation}
\begin{split}
(\varepsilon_9\varepsilon_9)'H^2\mathcal R^3
&\rightarrow
\frac92(\varepsilon_8\varepsilon_8)'(\hat F\cdot F+F\cdot\hat F)\mathcal R^3
-18(8!)F^{bc}\cdot(\nabla^{(+)}_{[e}\hat F^{de}H_{dbc}\mathcal R^{fg}{}_{fg}\mathcal R^{hk}{}_{hk]})'
\\
&\quad
-12(8!)H^{abc}H_{[abc}(\nabla^{(-)e}F_{fe}\cdot\nabla^{(+)}_g\hat F^{fg}\mathcal R^{hk}{}_{hk]})'
+\frac32(\varepsilon_9\varepsilon_9)'H^2[\hat F\cdot F]\mathcal R^2
\\
&\quad
+(H\rightarrow-H, F\leftrightarrow\hat F)\,.
\end{split}
\label{eq:e9e9H2R3-red1}
\end{equation}
The last term is of higher order so we will not need it here. The second term can be integrated by parts to give
\begin{equation}
\begin{split}
&
-18(8!)F^{bc}\cdot(\nabla^{(+)}_{[e}\hat F^{de}H_{dbc}\mathcal R^{fg}{}_{fg}\mathcal R^{hk}{}_{hk]})'
\\
&\quad=
-36(8!)F^{bc}\cdot\nabla^e\hat F^d{}_{[e}(H_{dbc}\mathcal R^{fg}{}_{fg}\mathcal R^{hk}{}_{hk]})'
+36(7!)F^{bc}\cdot\nabla_e\hat F^{de}(H_{[dbc}\mathcal R^{fg}{}_{fg}\mathcal R^{hk}{}_{hk]})'
+\mathrm{H.O.}
\\
&\quad=
36(8!)F^{bc}\cdot\hat F^d{}_{[e}(\nabla^eH_{dbc}\mathcal R^{fg}{}_{fg}\mathcal R^{hk}{}_{hk]})'
+k_7^{(\mathrm{e.o.m.})}
+\mathrm{H.O.}\,,
\end{split}
\end{equation}
with
\begin{equation}
\begin{split}
k_7^{(\mathrm{e.o.m.})}
&=
-36(7!)F^{bc}\cdot\hat{\mathbbm A}^dH_{[dbc}(\mathcal R^{fg}{}_{fg}\mathcal R^{hk}{}_{hk]})'%
-108(7!)F^{bc}\cdot\hat F^d{}_{[d}\mathbbm B_{bc}(\mathcal R^{fg}{}_{fg}\mathcal R^{hk}{}_{hk]})'
\\
&\quad
-144(7!)F^{bc}\cdot\hat F^d{}_{[f}(H_{dbc}(\nabla^e-2\partial^e\Phi)\mathcal R^{fg}{}_{|e|g}\mathcal R^{hk}{}_{hk]})'\,.%
\end{split}
\label{eq:k7eom}
\end{equation}
The first term can be further rewritten
\begin{equation}
\begin{split}
&36(8!)F^{bc}\cdot\hat F^d{}_{[e}(\nabla^eH_{dbc}\mathcal R^{fg}{}_{fg}\mathcal R^{hk}{}_{hk]})'
\\
&\quad =
12(8!)F^{bc}\cdot\hat F^d{}_{[e}(\nabla^eH_{dbc}\mathcal R^{fg}{}_{fg}\mathcal R^{hk}{}_{hk]})'
-48(8!)F^{bc}\cdot\hat F^d{}_{[e}((\nabla H)^{eg}{}_{bc}\nabla^{(-)f}H_{dfg}\mathcal R^{hk}{}_{hk]})'
\\
&\quad\quad
+36(\varepsilon_8\varepsilon_8)'F\cdot\hat F\nabla H\mathcal R^2
+72(8!)F^b{}_{[d}\cdot\hat F^d{}_e((\nabla H)^{ec}{}_{bc}\mathcal R^{fg}{}_{fg}\mathcal R^{hk}{}_{hk]})'\,.
\end{split}
\end{equation}
The first term has the same form as the LHS.  The second term is a total derivative up to terms of higher order and equations of motion terms of order $H^2$, as can be seen by integrating by parts twice. The last term gives, due to the symmetry in exchanging $F$ and $\hat F$,
\begin{equation}
\begin{split}
-144&(8!)F^b{}_{[d}\cdot\hat F^d{}_e((\nabla H)^{ec}{}_{bc}(\nabla H)^{fg}{}_{fg}\mathcal R_S^{hk}{}_{hk]})'
\\
&=
72(\varepsilon_8\varepsilon_8)'F\cdot\hat F(\nabla H)^2\mathcal R_S
-96(8!)F^{bc}\cdot\hat F^d{}_{[e}(\nabla^eH_{dbc}(\nabla H)^{fg}{}_{fg}\mathcal R_S^{hk}{}_{hk]})'
+\mathrm{H.O.}\,.
\end{split}
\end{equation}

Finally we need to deal with the third term in (\ref{eq:e9e9H2R3-red1}). Integrating by parts twice it becomes
\begin{equation}
-12(8!)\hat F^{fg}\cdot F_{[fe}(\nabla_gH^{abc}\nabla^eH_{abc}\mathcal R^{hk}{}_{hk]})'\,,
\end{equation}
again plus terms of higher order and equations of motion terms of order $H^2$, which we drop. But this term can also be rewritten, by using the fact that anti-symmetrizing in one more index gives zero since the indices are contracted,
\begin{equation}
\begin{split}
&
-4(8!)\hat F^{fg}\cdot F_{[fe}(\nabla_gH^{abc}\nabla^eH_{abc}\mathcal R^{hk}{}_{hk]})'
+36(\varepsilon_8\varepsilon_8)'\hat F\cdot F(\nabla H)^2\mathcal R
\\
&
-24(8!)\hat F^{fg}\cdot F_{[fe}(\nabla_gH_h{}^{bc}\nabla^eH_{abc}\mathcal R^{hka}{}_{k]})'
-24(8!)\hat F^{fg}\cdot F^a{}_{[e}(\nabla_gH_f{}^{bc}\nabla^eH_{abc}\mathcal R^{hk}{}_{hk]})'\,.
\end{split}
\end{equation}
The first term is again of the same form, while the third and fourth vanish up to terms we are dropping. Putting these results together we have
\begin{equation}
\begin{split}
(\varepsilon_9\varepsilon_9)'H^2\mathcal R^3
&\rightarrow
\frac92(\varepsilon_8\varepsilon_8)'(\hat F\cdot F+F\cdot\hat F)\mathcal R^3
-54(\varepsilon_8\varepsilon_8)'\hat F\cdot F\nabla H\mathcal R_S^2
+54(\varepsilon_8\varepsilon_8)'\hat F\cdot F(\nabla H)^2\mathcal R_S
\\
&\quad
+k_7^{(\mathrm{e.o.m.})}
+(H\rightarrow-H, F\leftrightarrow\hat F)
+\mathrm{H.O.}
+\mathcal O(H^3)
+\mbox{e.o.m. terms of }\mathcal O(H^2)\,.
\end{split}
\label{eq:e9e9H2R3-red}
\end{equation}

\subsection{\texorpdfstring{$(\varepsilon_9\varepsilon_9)'[H^2]\mathcal R^3$ and $(\varepsilon_9\varepsilon_9)'[H^2](\nabla H)^2\mathcal R$}{e9e9[H2]R3 and e9e9H2dH2R}}
Dimensional reduction of these terms, defined in (\ref{eq:e9e9HHR3}), give rise to the following $O(d,d)$-violating terms quadratic in the KK vectors
\begin{equation}
\begin{split}
\frac{1}{288}&(\varepsilon_9\varepsilon_9)'[H^2]\mathcal R^3
-\frac{1}{48}(\varepsilon_9\varepsilon_9)'[H^2](\nabla H)^2\mathcal R
\\
&\rightarrow
-\frac{5!6!}{96}H_{a_4a_5a_6}F^{a_2a_3}\cdot\left(\nabla^{(+)}_{[a_2}\hat F^{[a_4a_5}\mathcal R_{a_3a_7}{}^{a_6a_7}\mathcal R_{a_8a_9]}{}^{a_8a_9]}\right)'
\\
&\quad
+\frac{5!6!}{48}H_{a_4a_5a_6}F^{a_2a_3}\cdot\left(\nabla^{(+)}_{[a_2}\hat F^{[a_4a_5}(\nabla H)^{a_6a_7}{}_{a_3a_7}(\nabla H)^{a_8a_9]}{}_{a_8a_9]}\right)'
\\
&\quad
-\frac{5!5!}{48}H^{a_1a_2a_3}H_{a_4a_5a_6}\left(\nabla^{(-)[a_4}F_{[a_1a_2}\cdot\nabla^{(+)}_{a_3}\hat F^{a_5a_6}(\nabla H)^{a_7a_8]}{}_{a_7a_8]}\right)'
\\
&\quad
+\frac{1}{192}(\varepsilon_9\varepsilon_9)'[H^2][\hat F\cdot F]\mathcal R^2
+\frac{1}{48}(\varepsilon_9\varepsilon_9)'[H^2][\hat F\cdot F](\nabla H)\mathcal R_S
\\
&\quad
-\frac{1}{96}(\varepsilon_9\varepsilon_9)'[H^2][\hat F\cdot F](\nabla H)^2
+(H\rightarrow-H, F\leftrightarrow\hat F)\,.
\end{split}
\end{equation}
Only the first term will be needed here, since the others are of order $H^3$ or of order six and higher in fields. We therefore find
\begin{equation}
\begin{split}
\frac{1}{288}&(\varepsilon_9\varepsilon_9)'[H^2]\mathcal R^3
-\frac{1}{48}(\varepsilon_9\varepsilon_9)'[H^2](\nabla H)^2\mathcal R
\\
&\rightarrow
-48F^{ef}\cdot\nabla_g\hat F_{bk}H_{kcd}\mathcal R^{ehcg}\mathcal R^{bfdh}%
+48F^{ef}\cdot\nabla_e\hat F_{bk}H_{kcd}\mathcal R^{fhgc}\mathcal R^{bgdh}%
\\
&\quad
-24F^{ef}\cdot\nabla_b\hat F_{cd}H_{kcd}\mathcal R^{ehgb}\mathcal R^{gfkh}%
+24F^{ef}\cdot\nabla_g\hat F_{bk}H_{kcd}\mathcal R^{ebhg}\mathcal R^{hfcd}%
\\
&\quad
+24F^{ef}\cdot\nabla_g\hat F_{bk}H_{kcd}\mathcal R^{efhc}\mathcal R^{bhdg}%
-24F^{ef}\cdot\nabla_b\hat F_{gh}H_{kcd}\mathcal R^{egkb}\mathcal R^{hfcd}%
\\
&\quad
+24F^{ef}\cdot\nabla_e\hat F_{bg}H_{kcd}\mathcal R^{fbhk}\mathcal R^{ghcd}%
-12F^{ef}\cdot\nabla_g\hat F_{bk}H_{kcd}\mathcal R^{efgh}\mathcal R^{bhcd}%
\\
&\quad
-12F^{ef}\cdot\nabla_e\hat F_{cd}H_{kcd}\mathcal R^{fbgh}\mathcal R^{ghkb}%
+12F^{ef}\cdot\nabla_e\hat F_{bk}H_{kcd}\mathcal R^{fbgh}\mathcal R^{ghcd}%
\\
&\quad
-12F^{ef}\cdot\nabla_e\hat F_{gh}H_{kcd}\mathcal R^{fbcd}\mathcal R^{ghkb}%
-6F^{ef}\cdot\nabla_b\hat F_{cd}H_{kcd}\mathcal R^{efgh}\mathcal R^{ghkb}%
\\
&\quad
-6F^{ef}\cdot\nabla_b\hat F_{gh}H_{kcd}\mathcal R^{efkb}\mathcal R^{ghcd}%
-6F^{ef}\cdot\nabla_b\hat F_{gh}H_{kcd}\mathcal R^{efcd}\mathcal R^{ghkb}%
\\
&\quad
+(H\rightarrow-H, F\leftrightarrow\hat F)
+\mathcal O(H^3)
+\mathrm{H.O.}
\end{split}
\label{eq:e9e9HHbR3-red}
\end{equation}

\subsection{\texorpdfstring{$\varepsilon_4H^2\varepsilon_4\mathcal R^2\mathcal R$ and $\varepsilon_4H^2\varepsilon_4(\nabla H)^2\mathcal R$}{e4H2e4R2R and e4H2e4dH2R}}\label{app:e4e4-red}
From the definition in (\ref{eq:e4}) one finds the following $O(d,d)$-violating terms quadratic in the KK vectors upon dimensional reduction
\begin{equation}
\begin{split}
\varepsilon_4H^2\varepsilon_4\mathcal R^2\mathcal R
-2\varepsilon_4H^2\varepsilon_4&(\nabla H)^2\mathcal R
\\
&\rightarrow
-4!F_{a_1a_2}\cdot\nabla^{(+)}_b\hat F^{de}H_{ea_3a_4}\mathcal R^{bc[a_1a_2}\mathcal R^{a_3a_4]}{}_{cd}
\\
&\quad
-4(3!)F_{ea_1}\cdot\nabla^{(+)[a_1}\hat F_{cd}H_{ka_2a_3}\mathcal R^{|bc|a_2a_3]}\mathcal R_b{}^{edk}
\\
&\quad
-4(3!)F_{ea_1}\cdot\nabla^{(+)[a_1}\hat F_{cd}H_{ka_2a_3}\mathcal R^{|bc|a_2a_3]}\mathcal R_b{}^{kde}
\\
&\quad
-4!\nabla^{(-)e}F^{bf}\cdot\nabla^{(+)c}\hat F^{[a_1a_2}H_{fa_1a_2}H_{ea_3a_4}\mathcal R_S^{a_3a_4]}{}_{bc}
\\
&\quad
+2(4!)\hat F_{a_1a_2}\cdot\nabla^{(-)}_bF^{de}H_{ea_3a_4}(\nabla H)^{[a_1a_2}{}_{cd}(\nabla H)^{a_3a_4]bc}
\\
&\quad
-8(3!)\hat F_{a_2f}\cdot\nabla^{(-)[a_2}F_{cd}H_{ka_3a_4}(\nabla H)^{a_3a_4]bc}(\nabla H)_b{}^{fdk}
\\
&\quad
-8(3!)\hat F_{a_2f}\cdot\nabla^{(-)[a_2}F_{cd}H_{ka_3a_4}(\nabla H)^{a_3a_4]bc}(\nabla H)^{dfk}{}_b
\\
&\quad
+\frac{4!}{2}\nabla^{(-)}_dF^{[a_1a_2}\cdot\nabla^{(+)}_b\hat F^{a_3a_4]}H_{fa_1a_2}H_{ea_3a_4}(\nabla H)^{bfde}
\\
&\quad
-\frac{4!}{2}H_{fa_1a_2}H_{ea_3a_4}\hat F_b{}^f\cdot F^{de}\mathcal R_{cd}{}^{[a_1a_2}\mathcal R^{a_3a_4]bc}
\\
&\quad
-4!H_{fa_1a_2}H_{ea_3a_4}\hat F_b{}^f\cdot F^{de}(\nabla H)^{[a_1a_2}{}_{cd}(\nabla H)^{a_3a_4]bc}
\\
&\quad
-4!H_{fa_1a_2}H_{ea_3a_4}\hat F^c{}_d\cdot F^{[a_1a_2}\mathcal R_{bc}{}^{a_3a_4]}\mathcal R^{bfde}
\\
&\quad
-2(4!)H_{fa_1a_2}H_{ea_3a_4}\hat F_{cd}\cdot F^{[a_1a_2}(\nabla H)^{a_3a_4]bc}(\nabla H)_b{}^{fde}
\\
&\quad
+(H\rightarrow-H, F\leftrightarrow\hat F)\,,
\end{split}
\end{equation}
where the first three terms give
\begin{equation}
\begin{split}
-4!F_{a_1a_2}\cdot\nabla^{(+)}_b\hat F^{de}H_{ea_3a_4}\mathcal R^{bc[a_1a_2}\mathcal R^{a_3a_4]}{}_{cd}
&=
-16F_{ef}\cdot\nabla^{(+)}_g\hat F^{bk}H_{kcd}\mathcal R^{ghce}\mathcal R^{df}{}_{bh}%
\\
&\quad
-4F_{ef}\cdot\nabla^{(+)}_g\hat F^{bk}H_{kcd}\mathcal R^{ghcd}\mathcal R^{ef}{}_{hb}%
\\
&\quad
-4F_{ef}\cdot\nabla^{(+)}_g\hat F^{bk}H_{kcd}\mathcal R^{cd}{}_{hb}\mathcal R^{ghef}\,,%
\end{split}
\end{equation}
\begin{equation}
\begin{split}
-4(3!)F_{ea_1}\cdot\nabla^{(+)[a_1}\hat F_{cd}H_{ka_2a_3}\mathcal R^{|bc|a_2a_3]}\mathcal R_b{}^{edk}
&=
16F_{ef}\cdot\nabla^{(+)k}\hat F_{bg}H_{kcd}\mathcal R_h{}^{bec}\mathcal R^{fhgd}%
\\
&\quad
+8F^{ef}\cdot\nabla^{(+)}_e\hat F_{bg}H_{kcd}\mathcal R_{hf}{}^{gk}\mathcal R^{hbcd}%
\end{split}
\end{equation}
and
\begin{equation}
\begin{split}
-4(3!)F_{ea_1}\cdot\nabla^{(+)[a_1}\hat F_{cd}H_{ka_2a_3}\mathcal R^{|bc|a_2a_3]}\mathcal R_b{}^{kde}
&=
16F_{ef}\cdot\nabla^{(+)k}\hat F_{bg}H_{kcd}\mathcal R_h{}^{bec}\mathcal R^{hdfg}%
\\
&\quad
+8F^{ef}\cdot\nabla^{(+)}_e\hat F^{bg}H_{kcd}\mathcal R^{kh}{}_{gf}\mathcal R_{bh}{}^{cd}\,,%
\end{split}
\end{equation}
while the fourth term gives
\begin{equation}
-4!\nabla^{(-)e}F^{bf}\cdot\nabla^{(+)c}\hat F^{[a_1a_2}H_{fa_1a_2}H_{ea_3a_4}\mathcal R_S^{a_3a_4]}{}_{bc}
=
k_5^{(F^2H^2\mathcal R)}
+\mathrm{H.O.}
+\mbox{e.o.m. terms of }\mathcal O(H^2)\,,
%
\end{equation}
with
\begin{equation}
\begin{split}
k_5^{(F^2H^2\mathcal R)}
&=
-16F^{ef}\cdot\nabla^g\hat F^{bk}H_{kcd}\nabla^cH_{hbe}\mathcal R_S^{dh}{}_{fg}
-8F^{bk}\cdot\nabla^c\hat F^{ef}H_{kcd}\nabla^gH_{hef}\mathcal R_S^{hd}{}_{bg}
\\
&\quad
-8F^{ef}\cdot\nabla^g\hat F^{cd}H_{kcd}\nabla^hH_{gbe}\mathcal R_S^{bk}{}_{fh}
+8\hat F^{bk}\cdot\nabla^cF^{ef}H_{kcd}\nabla^dH_{egh}\mathcal R_S^{gh}{}_{bf}
\\
&\quad
+4F_{ef}\cdot\nabla^c\hat F^{gh}H_{kcd}\nabla^kH_{bgh}\mathcal R_S^{bdef}
-4F^{bk}\cdot\nabla^g\nabla^h\hat F^{cd}H_{hef}H_{kcd}\mathcal R_S^{ef}{}_{bg}
\\
&\quad
-4\nabla_bF^{cd}\cdot\nabla_g\hat F^{kh}H_{hef}H_{kcd}\mathcal R_S^{efbg}
+4\nabla^bF^{hk}\cdot\nabla^g\hat F^{ef}H_{bef}H_{kcd}\mathcal R_S^{cd}{}_{gh}
\\
&\quad
+4\nabla^bF^{hk}\cdot\nabla^g\hat F^{cd}H_{bef}H_{kcd}\mathcal R_S^{ef}{}_{gh}
-2F_{gh}\cdot\nabla^k\hat F^{ef}H_{kcd}H_{bef}\nabla^b\mathcal R_S^{cdgh}
\\
&\quad
+4F^{bk}\cdot\hat F^{cg}\nabla^hH_{kcd}\nabla^dH_{hef}\mathcal R_S^{ef}{}_{bg}\,.
\end{split}
\label{eq:k5}
\end{equation}

Finally, putting all this together we have
\begin{equation}
\begin{split}
\varepsilon_4H^2&\varepsilon_4\mathcal R^2\mathcal R
-2\varepsilon_4H^2\varepsilon_4(\nabla H)^2\mathcal R
\\
&\rightarrow
-16F_{ef}\cdot\nabla_g\hat F^{bk}H_{kcd}\mathcal R^{ghce}\mathcal R^{df}{}_{bh}%
+16F_{ef}\cdot\nabla^k\hat F_{bg}H_{kcd}\mathcal R_h{}^{bec}\mathcal R^{fhgd}%
\\
&\quad
+16F_{ef}\cdot\nabla^k\hat F_{bg}H_{kcd}\mathcal R_h{}^{bec}\mathcal R^{hdfg}%
+8F^{ef}\cdot\nabla_e\hat F_{bg}H_{kcd}\mathcal R_{hf}{}^{gk}\mathcal R^{hbcd}%
\\
&\quad
+8F^{ef}\cdot\nabla_e\hat F^{bg}H_{kcd}\mathcal R^{kh}{}_{gf}\mathcal R_{bh}{}^{cd}%
-4F^{ef}\cdot\nabla_g\hat F^{bk}H_{kcd}\mathcal R^{ghcd}\mathcal R^{ef}{}_{hb}%
\\
&\quad
-4F_{ef}\cdot\nabla_g\hat F^{bk}H_{kcd}\mathcal R^{cd}{}_{hb}\mathcal R^{ghef}%
+k_5^{(F^2H^2\mathcal R)}
\\
&\quad
+(H\rightarrow-H, F\leftrightarrow\hat F)
+\mathcal O(H^3)
+\mathrm{H.O.}
+\mbox{e.o.m. terms of }\mathcal O(H^2)\,.
\end{split}
\end{equation}

\subsection{\texorpdfstring{$l_5$}{l5} term}\label{app:l5}
The term
\begin{equation}
l_5=H^2_{abcd}\nabla^gH^{hae}\nabla_gH_h{}^{cf}\mathcal R_S^{bd}{}_{ef}
\end{equation}
gives rise to the following $O(d,d)$-violating terms quadratic in the KK vectors upon dimensional reduction
\begin{equation}
\begin{split}
l_5
&\rightarrow
F_{ab}\cdot\hat F_{cd}\nabla^gH^{hae}\nabla_gH_h{}^{cf}\mathcal R_S^{bd}{}_{ef}
+2F_{bk}\cdot\nabla_g\hat F_{eh}H^k{}_{cd}\nabla^gH^{hcf}\mathcal R_S^{bdef}
\\
&\quad
+\nabla^gF^{ae}\cdot\nabla_g\hat F^{cf}H^2_{abcd}\mathcal R_S^{bd}{}_{ef}
+(H\rightarrow-H, F\leftrightarrow\hat F)
+\mathcal O(H^3)\,.
\end{split}
\label{eq:l5red1}
\end{equation}
The second term can be further rewritten, by using the Bianchi identity and integrating by parts, as
\begin{equation}
\begin{split}
2&F_{bk}\cdot\nabla_g\hat F_{eh}H^k{}_{cd}\nabla^gH^{hcf}\mathcal R_S^{bdef}
\\
&=
-2F^{ef}\cdot\nabla_g\hat F_b{}^kH_{kcd}\nabla_eH^{cgh}\mathcal R_S^{bd}{}_{fh}
-2F_b{}^k\cdot\nabla^g\hat F_{ef}H_{kcd}(\nabla H)^{chef}\mathcal R_S^{bd}{}_{gh}
\\
&\quad
-F^{ef}\cdot\nabla_e\hat F_b{}^kH_{kcd}\nabla_fH^{cgh}\mathcal R_S^{bd}{}_{gh}
-2\hat F_b{}^k\cdot F^{ec}\nabla_cH^{fgh}(\nabla H)_{ghdk}\mathcal R_S^{bd}{}_{ef}
\\
&\quad
-\hat F_b{}^k\cdot F^{ef}\nabla_eH_{kcd}\nabla_fH^{cgh}\mathcal R_S^{bd}{}_{gh}
+\mathrm{H.O.}
+\mbox{e.o.m. terms of }\mathcal O(H^2)\,,
%
%
\end{split}
\end{equation}
where the first term can be further rewritten as
\begin{equation}
\begin{split}
-2&F^{ef}\cdot\nabla_g\hat F_b{}^kH_{kcd}\nabla_eH^{cgh}\mathcal R_S^{bd}{}_{fh}
\\
&=
-6F_{ef}\cdot\nabla^g\hat F_b{}^kH_{kcd}\nabla^eH^c{}_{gh}\mathcal R_S^{b[dfh]}
+2F_{ef}\cdot\nabla^g\hat F_b{}^kH_{kcd}\nabla^eH^c{}_{gh}\mathcal R_S^{bfhd}
\\
&\quad
+2F_{ef}\cdot\nabla^g\hat F_b{}^kH_{kcd}\nabla^eH^c{}_{gh}\mathcal R_S^{bhdf}
\\
&=
2F_{ef}\cdot\nabla^g\hat F_b{}^kH_{kcd}\nabla^eH^c{}_{gh}\mathcal R_S^{bhdf}
+2F_{ef}\cdot\nabla^k\hat F_b{}^gH_{kcd}\nabla^eH^c{}_{gh}\mathcal R_S^{bfhd}
\\
&\quad
+2F_{ef}\cdot\nabla_b\hat F^{gk}H_{kcd}\nabla^eH^c{}_{gh}\mathcal R_S^{bfhd}
-6F_{ef}\cdot\nabla^g\hat F_b{}^kH_{kcd}\nabla^eH^c{}_{gh}\mathcal R_S^{b[dfh]}
\\
&=
2F_{ef}\cdot\nabla^g\hat F_b{}^kH_{kcd}\nabla^eH^c_{gh}\mathcal R_S^{bhdf}
+2F_{ef}\cdot\nabla^k\hat F_b{}^gH_{kcd}\nabla^eH^c{}_{gh}\mathcal R_S^{bfhd}
\\
&\quad
+2F_{ef}\cdot\nabla^e\hat F^{bk}H_{kcd}\nabla_gH^c{}_{bh}\mathcal R_S^{gfhd}%
+F_{ef}\cdot\nabla_g\hat F^{bk}H_{kcd}\nabla^gH^c{}_{bh}\mathcal R_S^{efhd}%
\\
&\quad
+2F_{ef}\cdot\hat F^{bk}H_{kcd}\nabla_gH^c{}_{bh}\nabla^g\mathcal R_S^{efhd}%
+\hat F^{bk}\cdot F_{ef}\nabla_gH_{kcd}\nabla^gH^c{}_{bh}\mathcal R_S^{efhd}%
\\
&\quad
+\mathrm{H.O.}
+\mbox{e.o.m. terms of }\mathcal O(H^2)\,.
%
%
\end{split}
\end{equation}
Finally, the third term in (\ref{eq:l5red1}) can be rewritten as
\begin{equation}
\begin{split}
\nabla^gF^{ae}&\cdot\nabla_g\hat F^{cf}H^2_{abcd}\mathcal R_S^{bd}{}_{ef}
\\
&=
\nabla^eF^a{}_g\cdot\nabla^f\hat F^{cg}H^2_{abcd}\mathcal R_S^{bd}{}_{ef}
-\nabla_aF^e{}_g\cdot\nabla^c\hat F^{fg}H^2_{abcd}\mathcal R_S^{bd}{}_{ef}
\\
&\quad
+\nabla_aF^{ge}\cdot\nabla_g\hat F^{cf}H^2_{abcd}\mathcal R_S^{bd}{}_{ef}
+\nabla_gF^{ae}\cdot\nabla^c\hat F^{gf}H^2_{abcd}\mathcal R_S^{bd}{}_{ef}
\\
&=
\nabla^eF^a{}_g\cdot\nabla^f\hat F^{cg}H^2_{abcd}\mathcal R_S^{bd}{}_{ef}
-\nabla_aF^e{}_g\cdot\nabla^c\hat F^{fg}H^2_{abcd}\mathcal R_S^{bd}{}_{ef}
\\
&\quad
+\hat F^{ef}\cdot\nabla_eF^{ac}H_{kcd}\nabla^gH_{abk}\mathcal R_S^{bd}{}_{gf}
+F^{ef}\cdot\nabla_e\hat F^{ac}H_{kcd}\nabla_gH_{abk}\mathcal R_S^{bd}{}_{gf}
\\
&\quad
-\hat F^{ef}\cdot\nabla_eF^{ag}H_{kcd}(\nabla H)_{abcd}\mathcal R_S^{bk}{}_{gf}
-F^{ef}\cdot\nabla_e\hat F^{ag}H_{kcd}(\nabla H)_{abcd}\mathcal R_S^{bk}{}_{gf}
\\
&\quad
+\frac12\hat F_{ef}\cdot\nabla^eF^a{}_g\nabla^fH^2_{abcd}\mathcal R_S^{cdgb}
+\frac12F_{ef}\cdot\nabla^e\hat F^a{}_g\nabla^fH^2_{abcd}\mathcal R_S^{cdgb}
\\
&\quad
+\frac12\hat F_{ef}\cdot\nabla^eF^{ag}\nabla_gH^2_{abcd}\mathcal R_S^{cdbf}
+\frac12F_{ef}\cdot\nabla^e\hat F^{ag}\nabla_gH^2_{abcd}\mathcal R_S^{cdbf}
\\
&\quad
+\frac12\nabla_g\hat F_{ef}\cdot\nabla^eF^{ag}H^2_{abcd}\mathcal R_S^{cdbf}
+\frac12\nabla_gF_{ef}\cdot\nabla^e\hat F^{ag}H^2_{abcd}\mathcal R_S^{cdbf}
\\
&\quad
+\frac14\nabla^g\hat F^{ef}\cdot\nabla_gF^{ac}H^2_{abcd}\mathcal R_S^{bd}{}_{ef}
+\frac14\nabla_gF^{ef}\cdot\nabla^g\hat F^{ac}H^2_{abcd}\mathcal R_S^{bd}{}_{ef}
\\
&\quad
+\mathrm{H.O.}
+\mbox{e.o.m. terms of }\mathcal O(H^2)\,.
%
%
\end{split}
\end{equation}
Putting this together we have
\begin{equation}
\begin{split}
l_5
&\rightarrow
2F_{ef}\cdot\nabla^g\hat F_b{}^kH_{kcd}\nabla^eH^c{}_{gh}\mathcal R_S^{bhdf}
+2F_{ef}\cdot\nabla^k\hat F_b{}^gH_{kcd}\nabla^eH^c{}_{gh}\mathcal R_S^{bfhd}
\\
&\quad
+2F_{ef}\cdot\nabla^e\hat F^{bk}H_{kcd}\nabla_gH^c{}_{bh}\mathcal R_S^{gfhd}
+2\hat F^{ef}\cdot\nabla_eF^{ac}H_{kcd}\nabla^gH_{abk}\mathcal R_S^{bd}{}_{gf}
\\
&\quad
+\nabla^eF^a{}_g\cdot\nabla^f\hat F^{cg}H^2_{abcd}\mathcal R_S^{bd}{}_{ef}
-\nabla_aF^e{}_g\cdot\nabla^c\hat F^{fg}H^2_{abcd}\mathcal R_S^{bd}{}_{ef}
\\
&\quad
+F_{ef}\cdot\nabla_g\hat F^{bk}H_{kcd}\nabla^gH^c{}_{bh}\mathcal R_S^{efhd}%
-2F_b{}^k\cdot\nabla^g\hat F_{ef}H_{kcd}(\nabla H)^{chef}\mathcal R_S^{bd}{}_{gh}
\\
&\quad
-F^{ef}\cdot\nabla_e\hat F_b{}^kH_{kcd}\nabla_fH^{cgh}\mathcal R_S^{bd}{}_{gh}
-2\hat F^{ef}\cdot\nabla_eF^{ag}H_{kcd}(\nabla H)_{abcd}\mathcal R_S^{bk}{}_{gf}
\\
&\quad
+2F_{ef}\cdot\hat F^{bk}H_{kcd}\nabla_gH^c{}_{bh}\nabla^g\mathcal R_S^{efhd}%
+\hat F_{ef}\cdot\nabla^eF^a{}_g\nabla^fH^2_{abcd}\mathcal R_S^{cdgb}
\\
&\quad
+\hat F_{ef}\cdot\nabla^eF^{ag}\nabla_gH^2_{abcd}\mathcal R_S^{cdbf}
+\nabla_g\hat F_{ef}\cdot\nabla^eF^{ag}H^2_{abcd}\mathcal R_S^{cdbf}
\\
&\quad
+\frac12\nabla^g\hat F^{ef}\cdot\nabla_gF^{ac}H^2_{abcd}\mathcal R_S^{bd}{}_{ef}
+F_{ab}\cdot\hat F_{cd}\nabla^gH^{hae}\nabla_gH_h{}^{cf}\mathcal R_S^{bd}{}_{ef}
\\
&\quad
+\hat F^{bk}\cdot F_{ef}\nabla_gH_{kcd}\nabla^gH^c{}_{bh}\mathcal R_S^{efhd}
-2\hat F_b{}^k\cdot F^{ec}\nabla_cH^{fgh}(\nabla H)_{kdgh}\mathcal R_S^{bd}{}_{ef}
\\
&\quad
-\hat F_b{}^k\cdot F^{ef}\nabla_eH_{kcd}\nabla_fH^{cgh}\mathcal R_S^{bd}{}_{gh}
+(H\rightarrow-H, F\leftrightarrow\hat F)
+\mathcal O(H^3)
\\
&\quad
+\mathrm{H.O.}
+\mbox{e.o.m. terms of }\mathcal O(H^2)\,.
\end{split}
\label{eq:l5red}
\end{equation}

\subsection{\texorpdfstring{$l_6$}{l6} term}\label{app:l6}
The term
\begin{equation}
l_6=H^2_{abcd}\nabla_eH^{agh}\nabla_fH^c{}_{gh}\mathcal R_S^{bdef}
\end{equation}
gives rise to the following $O(d,d)$-violating terms quadratic in the KK vectors upon dimensional reduction
\begin{equation}
\begin{split}
l_6
&\rightarrow
F_{ab}\cdot\hat F_{cd}\nabla_eH^{agh}\nabla_fH^c{}_{gh}\mathcal R_S^{bdef}
+2\nabla_e\hat F^a{}_g\cdot\nabla_fF^{cg}H^2_{abcd}\mathcal R_S^{bdef}
\\
&\quad
+2\hat F_b{}^k\cdot\nabla_gF_{ef}H_{kcd}\nabla_hH^{cef}\mathcal R_S^{bdgh}
%
+(H\rightarrow-H,\, F\leftrightarrow\hat F)
+\mathcal O(H^3)\,.
\end{split}
\label{eq:l6red}
\end{equation}

\subsection{\texorpdfstring{$l_7$}{l7} term}
The term
\begin{equation}
l_7=H_{kcd}H_{aef}\nabla^gH^{hac}\nabla_gH_{hb}{}^d\mathcal R_S^{efkb}
\end{equation}
gives rise to the following $O(d,d)$-violating terms quadratic in the KK vectors upon dimensional reduction
\begin{equation}
\begin{split}
l_7
&\rightarrow
\hat F_{ef}\cdot\nabla^gF^{bk}H_{kcd}\nabla_gH_{hb}{}^c\mathcal R_S^{efdh}
-\hat F_{ef}\cdot\nabla^gF^{bk}H_{kcd}\nabla_gH_{hb}{}^f\mathcal R_S^{hecd}
\\
&\quad
+\hat F_{ef}\cdot\nabla_hF_{bg}H_{kcd}\nabla^hH^{kbe}\mathcal R_S^{fgcd}
+\nabla_g\hat F^{ac}\cdot\nabla^gF_b{}^dH_{kcd}H_{aef}\mathcal R_S^{efkb}
%
\\
&\quad
+(H\rightarrow-H,\, F\leftrightarrow\hat F)
+\mathcal O(H^3)\,.
\end{split}
\end{equation}
The second term can be rewritten as
\begin{equation}
\begin{split}
-\hat F_{ef}&\cdot\nabla^gF^{bk}H_{kcd}\nabla_gH_{hb}{}^f\mathcal R_S^{hecd}
\\
&=
-\hat F_{ef}\cdot\nabla^kF^{gh}H_{kcd}(\nabla H)^f{}_{bgh}\mathcal R_S^{becd}
-\frac12F^{bk}\cdot\nabla_g\hat F_{ef}H_{kcd}\nabla_bH_h{}^{ef}\mathcal R_S^{ghcd}
\\
&\quad
+\frac12\hat F_{ef}\cdot F^{bk}H_{kcd}\nabla_bH_{gh}{}^f\nabla^e\mathcal R_S^{ghcd}
+\hat F_{ef}\cdot F^{bk}\nabla_gH_{kcd}\nabla_bH_h{}^{gf}\mathcal R_S^{hecd}
\\
&\quad
+\mathrm{H.O.}
+\mbox{e.o.m. terms of }\mathcal O(H^2)\,,
%
%
\end{split}
\end{equation}
while the third can be rewritten as
\begin{equation}
\begin{split}
\frac12&\nabla^h(\hat F_{ef}\cdot F_{bg})H_{kcd}\nabla_hH^{kbe}\mathcal R_S^{fgcd}
\\
&=
\frac12\hat F_{ef}\cdot\nabla^eF_{bg}H_{kcd}\nabla_hH^{kbf}\mathcal R_S^{ghcd}
+\frac12F_{ef}\cdot\nabla^e\hat F_{bg}H_{kcd}\nabla_hH^{kbf}\mathcal R_S^{ghcd}
\\
&\quad
-\frac12\hat F_{ef}\cdot F_{bg}H_{kcd}\nabla_h(\nabla H)^{kbef}\mathcal R_S^{ghcd}
-\frac12F_{ef}\cdot\hat F_{bg}H_{kcd}\nabla_h(\nabla H)^{kbef}\mathcal R_S^{ghcd}
\\
&\quad
-\frac12\hat F_{ef}\cdot F_{bg}\nabla^fH_{kcd}\nabla_hH^{kbe}\mathcal R_S^{ghcd}
-\frac12F_{ef}\cdot\hat F_{bg}\nabla^fH_{kcd}\nabla_hH^{kbe}\mathcal R_S^{ghcd}
\\
&\quad
-\frac12\hat F_{ef}\cdot F_{bg}\nabla^hH_{kcd}\nabla_hH^{kbe}\mathcal R_S^{fgcd}
+\mathrm{H.O.}
+\mbox{e.o.m. terms of }\mathcal O(H^2)\,.
%
%
\end{split}
\end{equation}
Putting this together we have
\begin{equation}
\begin{split}
l_7
&\rightarrow
\hat F_{ef}\cdot\nabla^gF^{bk}H_{kcd}\nabla_gH_{hb}{}^c\mathcal R_S^{efdh}
-\hat F_{ef}\cdot\nabla^kF^{gh}H_{kcd}(\nabla H)^f{}_{bgh}\mathcal R_S^{becd}
\\
&\quad
+\hat F_{ef}\cdot\nabla^eF_{bg}H_{kcd}\nabla_hH^{kbf}\mathcal R_S^{ghcd}
-\hat F_{ef}\cdot F_{bg}H_{kcd}\nabla_h(\nabla H)^{kbef}\mathcal R_S^{ghcd}
\\
&\quad
+\nabla_g\hat F^{ac}\cdot\nabla^gF_b{}^dH_{kcd}H_{aef}\mathcal R_S^{efkb}
-\frac12F^{bk}\cdot\nabla_g\hat F_{ef}H_{kcd}\nabla_bH_h{}^{ef}\mathcal R_S^{ghcd}
\\
&\quad
+\frac12\hat F_{ef}\cdot F^{bk}H_{kcd}\nabla_bH_{gh}{}^f\nabla^e\mathcal R_S^{ghcd}
-\hat F_{ef}\cdot F_{bg}\nabla^fH_{kcd}\nabla_hH^{kbe}\mathcal R_S^{ghcd}
\\
&\quad
-\frac12\hat F_{ef}\cdot F_{bg}\nabla^hH_{kcd}\nabla_hH^{kbe}\mathcal R_S^{fgcd}
+\hat F_{ef}\cdot F^{bk}\nabla_gH_{kcd}\nabla_bH_h{}^{gf}\mathcal R_S^{hecd}
\\
&\quad
+(H\rightarrow-H,\, F\leftrightarrow\hat F)
+\mathcal O(H^3)
+\mathrm{H.O.}
+\mbox{e.o.m. terms of }\mathcal O(H^2)\,.
\end{split}
\label{eq:l7red}
\end{equation}

\subsection{\texorpdfstring{$l_8$--$l_{36}$}{l8-l36} terms}
Finally, the terms with at least two traces defined in (\ref{eq:l8to36}) give rise to the following $O(d,d)$-violating terms quadratic in the KK vectors upon dimensional reduction
{
\allowdisplaybreaks
\begin{align}
l_8&\rightarrow
\frac12\nabla^g\hat F^{ac}\cdot\nabla_gF_{ef}H^2_{abcd}\mathcal R_S^{bdef}
-\hat F_b{}^k\cdot\nabla^cF^{ef}H_{kcd}(\nabla H)_{ghef}\mathcal R_S^{bdgh}
\nonumber\\
&\quad
+(H\rightarrow-H,\, F\leftrightarrow\hat F)
+\mathcal O(H^3)
\\
l_9&\rightarrow
\hat F_{ab}\cdot F_{cd}(\nabla H)^{abgh}(\nabla H)_{efgh}\mathcal R_S^{efcd}
+\hat F^{bk}\cdot\nabla_bF^{ef}H_{kcd}(\nabla H)_{efgh}\mathcal R_S^{ghcd}
\nonumber\\
&\quad
+\frac12\nabla^g\hat F^{ab}\cdot\nabla_gF_{ef}H^2_{abcd}\mathcal R_S^{efcd}
+(H\rightarrow-H,\, F\leftrightarrow\hat F)
+\mathcal O(H^3)
\\
l_{10}
&\rightarrow
\hat F_{ab}\cdot F_{cd}(\nabla H)^{begh}(\nabla H)_{ef}{}^{cd}\mathcal R_S^{fa}{}_{gh}
-\frac12\hat F_{ka}\cdot\nabla^eF^{gh}H^{kcd}(\nabla H)_{efcd}\mathcal R_S^{fa}{}_{gh}
\nonumber\\
&\quad
-\hat F^{kc}\cdot\nabla_cF_{ef}H_{abk}(\nabla H)^{begh}\mathcal R_S^{fa}{}_{gh}
-\frac14\nabla^b\hat F^{gh}\cdot\nabla_fF^{cd}H^2_{abcd}\mathcal R_S^{fa}{}_{gh}
\nonumber\\
&\quad
+(H\rightarrow-H,\, F\leftrightarrow\hat F)
+\mathcal O(H^3)
\\
l_{11}
&\rightarrow
\hat F_{ab}\cdot F_{cd}\nabla^eH^{agh}\nabla_eH_{fgh}\mathcal R_S^{fbcd}
+\hat F_{bk}\cdot\nabla^eF^{gh}\nabla_eH_{fgh}H^{kcd}\mathcal R_S^{fb}{}_{cd}
\nonumber\\
&\quad
+2\nabla^e\hat F^{ha}\cdot\nabla_eF_{hf}H^2_{abcd}\mathcal R_S^{fbcd}
+(H\rightarrow-H,\, F\leftrightarrow\hat F)
+\mathcal O(H^3)
\\
l_{12}
&\rightarrow
\nabla^e(\hat F_{ab}\cdot F_{cd}H^{agh})\nabla_eH_{fgh}\mathcal R_S^{fbcd}
+\nabla^e(\hat F_{bk}\cdot F^{gh}H^{kcd})\nabla_eH_{fgh}\mathcal R_S^{fb}{}_{cd}
\nonumber\\
&\quad
+2\nabla_e\hat F_{hf}\cdot\nabla^e(F^{ha}H^2_{abcd})\mathcal R_S^{fbcd}
+(H\rightarrow-H,\, F\leftrightarrow\hat F)
+\mathcal O(H^3)
\\
l_{13}
&\rightarrow
2\hat F_{bf}\cdot\nabla^eF^{bk}H_{kcd}\nabla_eH^{fgh}\mathcal R_S^{cd}{}_{gh}
+\hat F_{ab}\cdot\nabla^eF^{gh}H_{kcd}\nabla_eH^{abk}\mathcal R_S^{cd}{}_{gh}
\nonumber\\
&\quad
+\hat F_{cd}\cdot\nabla^eF^{ab}H_{abf}\nabla_eH^{fgh}\mathcal R_S^{cd}{}_{gh}
+(H\rightarrow-H,\, F\leftrightarrow\hat F)
+\mathcal O(H^3)
\\
l_{14}
&\rightarrow
2\nabla^k(\hat F^{be}\cdot F_{bf})\nabla_k(H_{ecd}H^{fgh})\mathcal R_S^{cd}{}_{gh}
+2\nabla^k(H_{abf}\hat F^{ab})\cdot\nabla_k(\hat F_{cd}H^{fgh})\mathcal R_S^{cd}{}_{gh}
\nonumber\\
&\quad
+(H\rightarrow-H,\, F\leftrightarrow\hat F)
+\mathcal O(H^3)
\\
l_{15}
&\rightarrow
2\nabla^k\hat F^{be}\cdot\nabla_kF_{bf}H_{ecd}H^{fgh}\mathcal R_S^{cd}{}_{gh}
+2\hat F_{cd}\cdot\nabla^kF^{ab}H^{fgh}\nabla_kH_{abf}\mathcal R_S^{cd}{}_{gh}
\nonumber\\
&\quad
+(H\rightarrow-H,\, F\leftrightarrow\hat F)
+\mathcal O(H^3)
\\
l_{16}
&\rightarrow
\hat F_{ab}\cdot F_{cd}\nabla_eH^{cgh}\nabla_fH^d{}_{gh}\mathcal R_S^{abef}
+2\nabla_e\hat F^{gh}\cdot F^{dk}H_{abk}\nabla_fH_{dgh}\mathcal R_S^{abef}
\nonumber\\
&\quad
+2\nabla_e\hat F^{cg}\cdot\nabla_fF^d{}_gH^2_{abcd}\mathcal R_S^{abef}
+(H\rightarrow-H,\, F\leftrightarrow\hat F)
+\mathcal O(H^3)
\\
l_{17}
&\rightarrow
4\hat F_{bh}\cdot\nabla_eF^{bg}H_{kcd}\nabla_fH_g{}^{cd}\mathcal R_S^{efkh}
+\nabla_e\hat F^{ab}\cdot\nabla_fF^{cd}H_{abh}H_{kcd}\mathcal R_S^{efkh}
\nonumber\\
&\quad
+(H\rightarrow-H,\, F\leftrightarrow\hat F)
+\mathcal O(H^3)
\\
l_{18}
&\rightarrow
-\frac12\hat F_{ab}\cdot\nabla_fF_{gh}H_{kcd}(\nabla H)^{cdgh}\mathcal R_S^{fkab}
-\hat F^{dk}\cdot\nabla_dF_{gh}H_{abe}(\nabla H)^{efgh}\mathcal R_S^{ab}{}_{fk}
\nonumber\\
&\quad
+\frac12\nabla_gF_{ef}\cdot\nabla^g\hat F^{cd}H_{ab}{}^eH_{kcd}\mathcal R_S^{fkab}
+(H\rightarrow-H,\, F\leftrightarrow\hat F)
+\mathcal O(H^3)
\\
l_{19}
&\rightarrow
\hat F_{bc}\cdot\nabla_bF^{hk}H_{def}(\nabla H)_{ck}{}^{gd}\mathcal R_S^{ef}{}_{gh}
-\frac12\hat F_{ab}\cdot\nabla_kF^{gd}H_{def}(\nabla H)^{hkab}\mathcal R_S^{ef}{}_{gh}
\nonumber\\
&\quad
-\frac12\hat F_{ef}\cdot\nabla^gF_{ck}H^{abc}(\nabla H)^{hk}{}_{ab}\mathcal R_S^{ef}{}_{gh}
+\frac14\nabla_c\hat F^{gd}\cdot\nabla^hF_{ab}H^{abc}H_{def}\mathcal R_S^{ef}{}_{gh}
\nonumber\\
&\quad
+(H\rightarrow-H,\, F\leftrightarrow\hat F)
+\mathcal O(H^3)
\\
l_{20}
&\rightarrow
\hat F^{bc}\cdot\nabla_bF^{dk}H_{def}(\nabla H)_{ckgh}\mathcal R_S^{efgh}
-\frac12\hat F_{ab}\cdot\nabla_kF_{gh}H_{def}(\nabla H)^{dkab}\mathcal R_S^{efgh}
\nonumber\\
&\quad
-\frac12\hat F_{ef}\cdot\nabla^kF_{ab}H^{abc}(\nabla H)_{ckgh}\mathcal R_S^{efgh}
+\frac14\nabla^d\hat F_{ab}\cdot\nabla_cF_{gh}H^{abc}H_{def}\mathcal R_S^{efgh}
\nonumber\\
&\quad
+(H\rightarrow-H,\, F\leftrightarrow\hat F)
+\mathcal O(H^3)
\\
l_{21}
&\rightarrow
\hat F^{bc}\cdot\nabla_bF^{ef}H_{def}(\nabla H)_{ckgh}\mathcal R_S^{dkgh}
-\frac12\hat F_{ab}\cdot\nabla_kF_{gh}H_{def}(\nabla H)^{efab}\mathcal R_S^{dkgh}
\nonumber\\
&\quad
+\hat F_{de}\cdot\nabla^eF_{ab}H^{abc}(\nabla H)_{ckgh}\mathcal R_S^{dkgh}
+(H\rightarrow-H,\, F\leftrightarrow\hat F)
+\mathcal O(H^3)
\\
l_{22}
&\rightarrow
\hat F_{bc}\cdot\nabla^bF_{dk}H^{def}(\nabla H)_{efgh}\mathcal R_S^{ckgh}
+\hat F^{de}\cdot\nabla_eF_{gh}H_{abc}(\nabla H)_{dk}{}^{ab}\mathcal R_S^{ckgh}
\nonumber\\
&\quad
-\frac12\hat F^{ef}\cdot\nabla_kF^{ab}H_{abc}(\nabla H)_{efgh}\mathcal R_S^{ckgh}
+(H\rightarrow-H,\, F\leftrightarrow\hat F)
+\mathcal O(H^3)
\\
l_{23}
&\rightarrow
\hat F_{cd}\cdot\nabla^eF^{gh}H_{abe}(\nabla H)^c{}_{fgh}\mathcal R_S^{fdab}
-\frac12\hat F_{dk}\cdot\nabla_fF_{gh}H_{abe}\nabla^eH^{kgh}\mathcal R_S^{fdab}
\nonumber\\
&\quad
+\nabla^e\hat F^{hk}\cdot\nabla_hF^c{}_fH_{kcd}H_{abe}\mathcal R_S^{fdab}
+(H\rightarrow-H,\, F\leftrightarrow\hat F)
+\mathcal O(H^3)
\\
l_{24}
&\rightarrow
2\nabla^e(\hat F^{fb}\cdot F_{cb})H^{kcd}\nabla_kH_{fgh}\mathcal R_S^{gh}{}_{de}
+\nabla^e(H_{abc}\hat F^{ab})\cdot\nabla_kF^{gh}H^{kcd}\mathcal R_{Sdegh}
\nonumber\\
&\quad
+\nabla^e(H_{fab}\hat F^{ab})\cdot F^{dk}\nabla_kH^{fgh}\mathcal R_{Sdegh}
+(H\rightarrow-H,\, F\leftrightarrow\hat F)
+\mathcal O(H^3)
\\
l_{25}
&\rightarrow
2\nabla^c(\hat F^{ea}\cdot F_{fa})H_{kcd}\nabla_eH^d{}_{gh}\mathcal R_S^{fkgh}
+\hat F_{kc}\cdot\nabla^eF_{gh}\nabla^cH^2_{ef}\mathcal R_S^{fkgh}
\nonumber\\
&\quad
+(H\rightarrow-H,\, F\leftrightarrow\hat F)
+\mathcal O(H^3)
\\
l_{26}
&\rightarrow
2\hat F_{eb}\cdot F^{fb}(\nabla H)^{ekcd}(\nabla H)_{kfgh}\mathcal R_S^{gh}{}_{cd}
-\hat F_{ab}\cdot\nabla^kF_{cd}H^{fab}(\nabla H)_{kfgh}\mathcal R_S^{cdgh}
\nonumber\\
&\quad
-\frac14\nabla^e\hat F_{cd}\cdot\nabla^fF_{gh}H^2_{ef}\mathcal R_S^{cdgh}
+(H\rightarrow-H,\, F\leftrightarrow\hat F)
+\mathcal O(H^3)
\\
l_{27}
&\rightarrow
2\hat F^{ec}\cdot F_{fc}\nabla_kH_{eab}(\nabla H)^{abgh}\mathcal R_S^{kf}{}_{gh}
+\hat F^{cd}\cdot\nabla_kF^{ab}H_{fcd}(\nabla H)_{abgh}\mathcal R_S^{kfgh}
\nonumber\\
&\quad
+\nabla_k\hat F^{ea}\cdot\nabla_aF_{gh}H^2_{ef}\mathcal R_S^{kfgh}
+(H\rightarrow-H,\, F\leftrightarrow\hat F)
+\mathcal O(H^3)
\\
l_{28}
&\rightarrow
2\hat F^{eb}\cdot F_{fb}\nabla_eH_{kcd}\nabla^fH^{kgh}\mathcal R_S^{cd}{}_{gh}
+\nabla^e\hat F_{cd}\cdot\nabla^fF_{gh}H^2_{ef}\mathcal R_S^{cdgh}
\nonumber\\
&\quad
+(H\rightarrow-H,\, F\leftrightarrow\hat F)
+\mathcal O(H^3)
\\
l_{29}
&\rightarrow
2\nabla^e(\hat F^{fa}\cdot F_{ka})H^{kcd}(\nabla H)_{cdgh}\mathcal R_S^{gh}{}_{ef}
+\hat F^{cd}\cdot\nabla_e(F^{ab}H_{fab})(\nabla H)_{cdgh}\mathcal R_S^{ghef}
\nonumber\\
&\quad
-\hat F^{dk}\cdot\nabla_dF_{gh}\nabla_eH^2_{fk}\mathcal R_S^{ghef}
+(H\rightarrow-H,\, F\leftrightarrow\hat F)
+\mathcal O(H^3)
\\
l_{30}
&\rightarrow
2\nabla_k(\hat F_e{}^b\cdot F_{fb})H^{kcd}\nabla_hH_{gcd}\mathcal R_S^{eghf}
+2\hat F^{kc}\cdot\nabla_hF_{gc}\nabla_kH^2_{ef}\mathcal R_S^{eghf}
\nonumber\\
&\quad
+(H\rightarrow-H,\, F\leftrightarrow\hat F)
+\mathcal O(H^3)
\\
l_{31}
&\rightarrow
\nabla^e(\hat F_{ab}\cdot F_{cd})\nabla_eH^2\mathcal R_S^{abcd}
+3\nabla_g(\hat F_{ef}\cdot F^{ef})\nabla^gH^2_{abcd}\mathcal R_S^{abcd}
\nonumber\\
&\quad
+(H\rightarrow-H,\, F\leftrightarrow\hat F)
+\mathcal O(H^3)
\\
l_{32}
&\rightarrow
2\hat F_c{}^h\cdot F_{dh}\nabla_aH_{efg}\nabla_bH^{efg}\mathcal R_S^{acdb}
+3\nabla_a\hat F^{ef}\cdot\nabla_bF_{ef}H^2_{cd}\mathcal R_S^{acdb}
\nonumber\\
&\quad
+(H\rightarrow-H,\, F\leftrightarrow\hat F)
+\mathcal O(H^3)
\\
l_{33}
&\rightarrow
3\hat F^{ef}\cdot F_{ef}\nabla_cH_{agh}\nabla_dH_b{}^{gh}\mathcal R_S^{abcd}
+2\nabla_c\hat F_a{}^g\cdot\nabla_dF_{bg}H^2\mathcal R_S^{abcd}
\nonumber\\
&\quad
+(H\rightarrow-H,\, F\leftrightarrow\hat F)
+\mathcal O(H^3)
\\
l_{34}
&\rightarrow
3\hat F^{ef}\cdot F_{ef}\nabla_bH_{kcd}\nabla^bH^{kgh}\mathcal R_S^{cd}{}_{gh}
+\nabla_b\hat F_{cd}\cdot\nabla^bF_{gh}H^2\mathcal R_S^{cdgh}
\nonumber\\
&\quad
+(H\rightarrow-H,\, F\leftrightarrow\hat F)
+\mathcal O(H^3)
\\
l_{35}
&\rightarrow
3\nabla^h\hat F_{ef}\cdot\nabla^kF^{ef}H_{kcd}H_{abh}\mathcal R_S^{abcd}
+(H\rightarrow-H,\, F\leftrightarrow\hat F)
+\mathcal O(H^3)
\\
l_{36}
&\rightarrow
3\hat F^{ef}\cdot F_{ef}(\nabla H)_{abcd}(\nabla H)^{abgh}\mathcal R_S^{cd}{}_{gh}
+\frac12\nabla_a\hat F_{cd}\cdot\nabla_aF^{gh}H^2\mathcal R_S^{cd}{}_{gh}
\nonumber\\
&\quad
+(H\rightarrow-H,\, F\leftrightarrow\hat F)
+\mathcal O(H^3)\,.
\end{align}
}

\bibliographystyle{nb}
\bibliography{biblio}{}
\end{document}